\documentclass[journal]{IEEEtran}

\ifCLASSINFOpdf
\else
\fi
\usepackage{moreverb}
\usepackage{graphicx}
\usepackage{amsmath}
\usepackage{amssymb}
\usepackage{epsfig}
\usepackage{epstopdf}
\usepackage{float}
\usepackage{caption}
\usepackage[caption=false]{subfig}
\usepackage{multirow}
\usepackage{algorithm}
\usepackage{algpseudocode}
\usepackage{blkarray}
\usepackage{wrapfig}
\usepackage{cleveref}
\usepackage{cite}
\usepackage{dblfloatfix}
\usepackage{color}
\usepackage{tabularx,booktabs}
\newcolumntype{C}{>{\centering\arraybackslash}X} % centered version of "X" type
\renewcommand\IEEEkeywordsname{Keywords}
\usepackage{xcolor,colortbl}
\definecolor{my3}{rgb}{0.6 0.6196 .8}

\begin{document}
	
	% paper title
	\title{Grant-free Non-orthogonal Multiple Access for IoT: A Survey}
	
	\author{\IEEEauthorblockN{Muhammad Basit Shahab, Rana Abbas, Mahyar Shirvanimoghaddam, and Sarah J. Johnson}
		
		%\author{\IEEEauthorblockN{Author 1\IEEEauthorrefmark{1}, Author 2\IEEEauthorrefmark{1}, Author 3\IEEEauthorrefmark{2}, Author 4\IEEEauthorrefmark{3}, and Author 5\IEEEauthorrefmark{4}}
		
		%\thanks{Manuscript received XXXX; accepted XXXX. Date of publication XXXX. This research was supported by (Grant Number: XXXX)}
		%\thanks{Authors are with the Department X, University Y, Country Z, Email: A,B, etc.}
		\thanks{Muhammad Basit Shahab and Sarah J. Johnson are with the School of Electrical Engineering and Computing, The University of Newcastle, Australia (Email: basit.shahab@newcastle.edu.au, sarah.johnson@newcastle.edu.au).}
		\thanks{Rana Abbas and Mahyar Shirvanimoghaddam are with the School of Electrical and Information Engineering, University of Sydney, Australia (Email: rana.abbas@sydney.edu.au, mahyar.shirvanimoghaddam@sydney.edu.au).}	
	}

	\maketitle
	
	% As a general rule, do not put math, special symbols or citations
	% in the abstract or keywords.
	\begin{abstract}
		Massive machine-type communications (mMTC) is one of the main three focus areas in the 5th generation (5G) of mobile standards to enable connectivity of a massive number of internet of things (IoT) devices with little or no human intervention. In conventional human-type communications (HTC), due to the limited number of available radio resources and orthogonal/non-overlapping nature of existing resource allocation techniques, users need to compete for connectivity through a random access (RA) process, which may turn into a performance bottleneck in mMTC. In this context, non-orthogonal multiple access (NOMA) has emerged as a potential technology that allows overlapping of multiple users over a radio resource, thereby creating an opportunity to enable more autonomous and grant-free communication, where devices can transmit data whenever they need. The existing literature on NOMA schemes majorly considers centralized scheduling based HTC, where users are already connected, and various system parameters like spreading sequences, interleaving patterns, power control, etc., are predefined. Contrary to HTC, mMTC traffic is different with mostly uplink communication, small data size per device, diverse quality of service, autonomous nature, and massive number of devices. Hence, the signaling overhead and latency of centralized scheduling becomes a potential performance bottleneck. To tackle this, grant-free access is needed, where mMTC devices can autonomously transmit their data over randomly chosen radio resources. This article, in contrast to existing surveys, comprehensively discusses the recent advances in NOMA from a grant-free connectivity perspective. Moreover, related practical challenges and future directions are discussed. 
	\end{abstract}
	% Note that keywords are not normally used for peerreview papers.
	\begin{IEEEkeywords}
		Non-orthogonal multiple access (NOMA), massive machine-type communications (mMTC), internet of things (IoT), random access (RA), grant-free transmission.
	\end{IEEEkeywords}
	\IEEEpeerreviewmaketitle

	\section{Introduction}
	\label{sec1}
	\IEEEPARstart{T}{he} Internet of Things (IoT) in recent years has emerged as a revolutionary transformation, where almost every physical device is expected to be connected to a communication network through a wired/wireless channel \cite{al2015internet,zanella2014internet,da2014internet}. Emerging services such as remote monitoring and real-time multi-device control (e.g., connected cars/homes, moving robots, and sensors) represent some prominent examples of the IoT framework. A high proportion of these services demonstrate a common autonomous feature, where communication between various devices and with the underlying network takes place with little or no human intervention \cite{lien2011toward}.
	\subsection{IoT Traffic Framework}
	International telecommunication union (ITU) and third generation partnership project (3GPP) have defined three network usage scenarios by considering the variety of connected devices and their diverse quality of service (QoS) requirements. These include enhanced mobile broadband (eMBB), massive machine-type communications (mMTC) and ultra-reliable low-latency communications (URLLC) \cite{wp5d2015imt}. The eMBB use case typically refers to the human-type communications (HTC) or human-to-human (H2H) communications, where the number of devices is less, communication is majorly downlink (DL), and the data size per device is large. On the contrary, mMTC and URLLC use cases of the IoT framework exhibit very different features from the HTC.
	\par
	%The IoT framework can be broadly classified into two types on the basis of different use cases; massive IoT (i.e., mMTC) and critical IoT (i.e., URLLC). 
	%The data traffic in both use case typical HTC/H2H traffic, has the following characteristics: majorly uplink (UL), transmit-data size per device is very small, extremely high energy efficiency is required for long-life of devices, communication is partially or completely autonomous, devices have diverse QoS requirements, etc \cite{zheng2014challenges,zheng2012radio}. 
	In massive IoT or mMTC (e.g., devices reporting to cloud, smart buildings, logistics tracking, and smart agriculture), some key traffic characteristics are:  majorly uplink (UL), transmit-data size per device is very small, extremely high energy efficiency is required for long-life of devices, communication is partially or completely autonomous, devices have diverse QoS requirements, etc \cite{zheng2014challenges,zheng2012radio}. However, these massive IoT use cases may have some degree of tolerance on data reliability and latency constraints. On the contrary, in critical IoT or URLLC (e.g., tele-surgery, intelligent transportation, and industrial automation), highest priority is given to data reliability and low latency communication \cite{Ericsson2016cellular,3gpp2019medical,3gpp2019vehical}. For instance, in critical medical applications, end-to-end latency for robot aided and augmented reality assisted surgeries is targeted to be less than 2ms and 750$\mu$s respectively \cite{3gpp2019medical}. Considering these diverse requirements of the IoT devices, significant upgradation in the existing communication technologies is needed.
		\begin{table}[!t]
		\label{taba}
		\centering
		\caption{List of abbreviations}
		\setlength\extrarowheight{0.05cm}
		\resizebox{\columnwidth}{!}{
			\begin{tabular}{| l | l |} 
				\hline
				3GPP & Third generation partnership project \\ 
				\hline
				5G & Fifth generation wireless systems \\
				\hline
				BOMA & Building block sparse-constellation based orthogonal multiple access \\ 
				\hline
				BP & Belief propagation \\ 
				\hline
				BS & Base station \\
				\hline
				CDMA & Code division multiple access \\
				\hline
				CoF & Compute-and-forward \\
				\hline
				CoSaMP & Compressive sampling matching pursuit \\
				\hline
				CP & Cyclic prefix \\ 
				\hline
				CS & Compressive sensing \\ 
				\hline
				CS-MUD & Compressive sensing multi-user detection \\ 
				\hline
				DL & Downlink \\ 
				\hline
				eMBB & enhanced mobile broadband \\ 
				\hline
				eNB & evolved NodeB \\ 
				\hline
				ESE & Elementary signal estimator \\ 
				\hline
				FDS & Frequency domain spreading \\ 
				\hline
				FEC & Forward error correction \\ 
				\hline
				GA & Grant acquisition \\ 
				\hline
				GOCA & Group orthogonal coded access \\ 
				\hline
				HARQ & Hybrid automatic repeat request \\ 
				\hline
				HTC & Human-type communications \\ 
				\hline
				H2H & Human-to-human \\ 
				\hline
				IDMA & Interleave division multiple access \\ 
				\hline
				IGMA & Interleave-grid multiple access \\ 
				\hline
				IoT & Internet of things \\ 
				\hline
				JMPA & Joint message passing algorithm \\ 
				\hline
				LCR & Low complexity receiver \\ 
				\hline
				LCRS & Low code rate spreading \\ 
				\hline
				LDS & Low density spreading \\ 
				\hline
				LDS-CDMA & Low density spreading code division multiple access\\ 
				\hline
				LDS-OFDM & Low density spreading orthogonal frequency-division multiplexing \\ 
				\hline
				LDS-SVE & Low density spreading signature vector extension \\ 
				\hline
				LPMA & Lattice partition multiple access \\ 
				\hline
				LSSA & Low code rate and signature based shared access \\ 
				\hline
				LTE & Long term evolution \\ 
				\hline
				LTE-A & Long term evolution Advanced \\ 
				\hline
				MA & Multiple access \\ 
				\hline
				MAP & Maximum a posteriori probability \\ 
				\hline
				MCS & Modulation and coding scheme \\ 
				\hline
				MIMO & Multi-input multi-output \\ 
				\hline
				ML & Machine Learning \\ 
				\hline
				MPA & Message passing algorithm \\ 
				\hline
				MTC & Machine-type communications \\ 
				\hline
				mMTC & Massive machine-type communications \\ 
				\hline
				MTCD & Machine-type communications device \\ 
				\hline
				MUD & Multi-user detection \\ 
				\hline
				MUSA & Multi-user shared access \\ 
				\hline
				M2M & Machine-to-machine \\ 
				\hline
				NCMA & Non-orthogonal coded multiple access \\ 
				\hline
				NOCA & Non-orthogonal coded access \\ 
				\hline
				NOMA & Non-orthogonal multiple access \\ 
				\hline
				NR & New radio \\ 
				\hline
				OMA & Orthogonal multiple access \\ 
				\hline
				PDMA & Pattern division multiple access \\ 
				\hline
				PD-NOMA & Power domain non-orthogonal multiple access \\ 
				\hline
				PIC & Parallel interference cancellation \\ 
				\hline
				PRACH & Physical random access channel \\ 
				\hline
				QoS & Quality of service \\ 
				\hline
				RA & Random access \\ 
				\hline
				RACH & Random access channel \\ 
				\hline
				RAN & Radio access network \\ 
				\hline
				RAR & Random access response \\ 
				\hline
				RB & Resource block \\ 
				\hline
				RDMA & Repetition division multiple access \\ 
				\hline
				%			 RRC & Radio resource control \\ 
				%			 \hline
				RSMA & Resource spread multiple access \\ 
				\hline
				SAMA & Successive interference cancellation aided multiple access \\ 
				\hline
				SCMA & Sparse code multiple access \\ 
				\hline
				SDMA & Spatial division multiple access \\ 
				\hline
				SINR & Signal-to-interference-plus-noise ratio \\ 
				\hline
				UE & User equipment \\
				\hline
				UL & Uplink \\
				\hline
				URLLC & Ultra-reliable low-latency communication \\
				\hline
				%		\rowcolor{my1}
			\end{tabular}
		}
	\end{table}
	\begin{figure}[t]
		\centering
		\includegraphics[width=3.6in,height=3.35in]{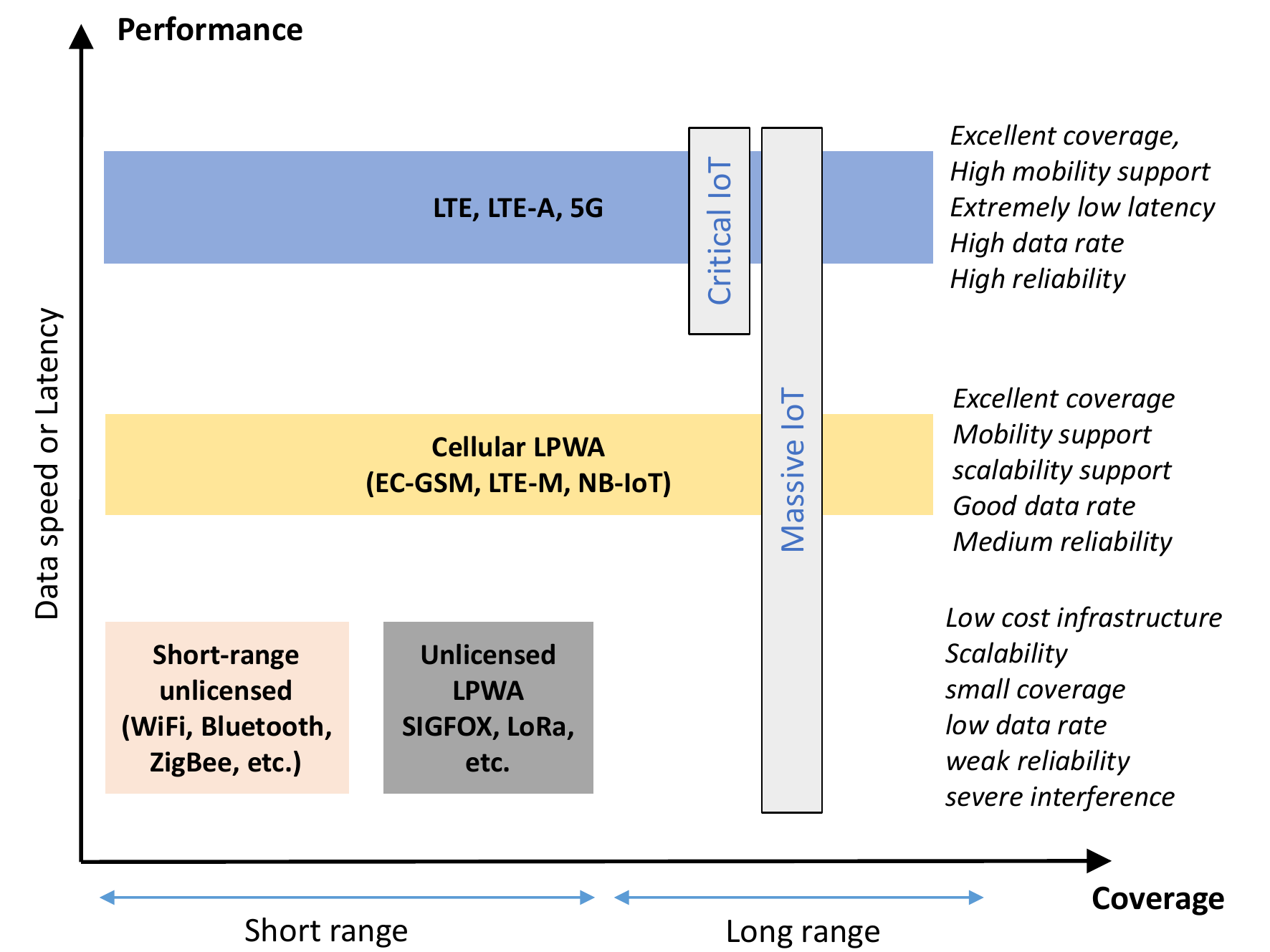}
		\caption{Wireless technologies for IoT applications}
		\label{fig1}
	\end{figure}
	\par
	The major focus of this work is on the mMTC use case, related requirements, and possible supporting solutions. mMTC is enabled through machine-type communications (MTC), also known as machine-to-machine (M2M) communication, where data transmission between various MTC devices (MTCDs) and with the underlying network takes place with little or no human intervention \cite{wu2011m2m,jain2012machine,etsi2011machine}. According to Cisco visual networking index and forecast, there will be around 28.5 billion connected devices by 2022, where the share of MTCDs will be above 51 percent \cite{cisco2018cisco}. Moreover, according to the global network connectivity/access trends, Cisco predicts a traffic contribution of wireless/mobile devices to be around 71 percent of total IP traffic by 2022. Similarly, Ericsson predicts that the number of devices connected to communication networks will reach 28 billion by 2021, out of which more than 53 percent will be MTCDs and consumer electronics devices \cite{Ericsson2016mobility}. Considering these massive devices with diverse QoS requirements, a dramatic shift from the current protocols, mostly designed for HTC, will be needed.
	\begin{table*}[!t]
		\caption{Summary of existing surveys on NOMA in a chronological order (some prominent schemes covered, grant-free access not discussed): $\checkmark \rightarrow$ covered in detail, $\bullet$ $\rightarrow$briefly introduced, X$\rightarrow$ not-covered, }
		\label{tab3}
		\resizebox{\textwidth}{!}{%
			\begin{tabular}{|l|c|c|c|c|c|l|c|c|c|c|c|c|c|c|c|c|c|}
				\hline
				\multicolumn{1}{|c|}{\multirow{2}{*}{\textbf{Survey}}} & \multicolumn{13}{c|}{\textbf{Uplink NOMA schemes}} & \multicolumn{4}{c|}{\textbf{Grant-free NOMA}}\\ 
				\cline{2-18} 
				%			\multicolumn{1}{|c|}{} & \multicolumn{3}{c|}{\textbf{Scrambling based}} & \multicolumn{10}{c|}{\textbf{Spreading based}} & \multicolumn{3}{c|}{\textbf{Interleaving based}} & \multirow{2}{*}{\textbf{\begin{tabular}[c]{@{}c@{}}Grant\\ Free \end{tabular}}} \\ \cline{2-17}
				\multicolumn{1}{|c|}{} & PD-NOMA & RSMA & LSSA & SCMA & \multicolumn{2}{c|}{PDMA} & LDS-SVE & MUSA & NOCA & NCMA &  IDMA & IGMA & RDMA & Signature & CS & CoF & ML\\ \hline
				\begin{tabular}[c]{@{}l@{}}Non-orthogonal multiple access for 5G: Solutions,\\ challenges, opportunities, and future directions,\\ Sep-2015 \cite{dai2015non}\end{tabular} & \checkmark & X & X & \checkmark & \multicolumn{2}{c|}{\textbullet} & X & \checkmark & X & X & X & X & X & X & X & X & X \\ \hline
				\begin{tabular}[c]{@{}l@{}}A survey: Several technologies of  non-orthogonal\\ transmission for 5G, Oct-2015 \cite{tao2015survey} \end{tabular} & \checkmark & X & X & \checkmark & \multicolumn{2}{c|}{\checkmark} & X & \checkmark & X & X & X & X & X & X & X & X & X \\ \hline
				\begin{tabular}[c]{@{}l@{}}Analysis of non-orthogonal multiple access for 5G,\\ 2016 \cite{wang2016analysis}\end{tabular} & \checkmark & X & X & \checkmark & \multicolumn{2}{c|}{\checkmark} & X & \checkmark & X & X & X & X & X & X & X & X & X \\ \hline
				\begin{tabular}[c]{@{}l@{}}Power-domain non-orthogonal multiple access \\ (NOMA) in 5G systems: Potentials and\\ challenges, Oct 2016 \cite{islam2016power}\end{tabular} & \checkmark & X & X & X & \multicolumn{2}{c|}{X} & X & X & X & X & X & X & X & X & X & X & X \\ \hline
				\begin{tabular}[c]{@{}l@{}}Uplink multiple access schemes for 5G:  A survey,\\ June-2017 \cite{yang2017uplink2} \end{tabular} & \checkmark & \checkmark & \checkmark & \checkmark & \multicolumn{2}{c|}{\checkmark} & \checkmark  & \checkmark & \checkmark & \checkmark & \checkmark & \checkmark & X & X & X & X & X\\ \hline
				\begin{tabular}[c]{@{}l@{}}A survey on non-orthogonal multiple access for \\5G networks: Research challenges and future \\ directions, Jul-2017 \cite{ding2017survey} \end{tabular} & \checkmark & X & X & \checkmark & \multicolumn{2}{c|}{\checkmark} & X & X & X & X & X & X & X & X & X & X & X \\ \hline
				\begin{tabular}[c]{@{}l@{}}Modulation and multiple access for 5G networks,\\ Oct-2017 \cite{cai2017modulation} \end{tabular} & \checkmark & X & X & \checkmark & \multicolumn{2}{c|}{\checkmark} & X & X & X & X & X & X & X & X & X & X & X \\ \hline
				\begin{tabular}[c]{@{}l@{}}Nonorthogonal multiple access for 5G and beyond,\\ Dec-2017 \cite{liu2017nonorthogonal} \end{tabular} & \checkmark & X & X & \textbullet & \multicolumn{2}{c|}{\textbullet} & X & X & X & X & \textbullet & X & X & X & X & X & X \\ \hline
				\begin{tabular}[c]{@{}l@{}}A survey and taxonomy on nonorthogonal \\ multiple‐access schemes for 5G networks,\\ Jan-2018 \cite{basharat2018survey} \end{tabular} & \checkmark  & X & X & \textbullet & \multicolumn{2}{c|}{\textbullet} & X & \textbullet & X & X & X & X & X & X & X & X & X\\ \hline
				\begin{tabular}[c]{@{}l@{}}Embracing non-orthogonal multiple access in\\ future wireless networks, Mar-2018 \cite{ding2018embracing} \end{tabular} & \checkmark  & X & X & X & \multicolumn{2}{c|}{X} & X & X & X & X & X & X & X & X & X & X & X\\ \hline
				\begin{tabular}[c]{@{}l@{}}A survey of non-orthogonal multiple access for \\ 5G, May-2018 \cite{dai2018survey} \end{tabular} & \checkmark  & \checkmark & \checkmark & \checkmark & \multicolumn{2}{c|}{\checkmark} & X & \checkmark & \checkmark & \checkmark & \checkmark & \checkmark & \checkmark & X & X& X& X\\ \hline
				\begin{tabular}[c]{@{}l@{}}Uplink nonorthogonal multiple access \\ technologies toward 5G: A survey, June-2018 \cite{ye2018uplink} \end{tabular}  & \checkmark  & \checkmark & \checkmark & \checkmark & \multicolumn{2}{c|}{\checkmark} & \checkmark & \checkmark & \checkmark & \checkmark  & \checkmark & \checkmark & \checkmark & \textbullet & X& X & X\\ \hline 
				\begin{tabular}[c]{@{}l@{}}A survey of rate-optimal power domain \\NOMA schemes for enabling technologies of \\future wireless networks, Sep-2019 \cite{maraqa2019survey} \end{tabular}  & \checkmark  & X & X & X & \multicolumn{2}{c|}{X} & X & X & X & X  & X & X & X & X & X& X & X\\ \hline 
				%\multicolumn{18}{l}{\begin{tabular}[c]{@{}l@{}}FDS: Frequency domain spreading  \hspace{4.8cm} LDS-SVE: Low density spreading signature vector extension      \hspace{2.8cm} PDMA: Pattern division multiple access  \\ GOCA: Group orthogonal coded access \hspace{4.2cm}  LSSA: Low code rate and signature based shared access \hspace{3.3cm} PD-NOMA: Power domain non-orthogonal multiple access \\          
				%IDMA: Interleave division multiple access \hspace{3.9cm}  MUSA: Multi-user shared access \hspace{6.05cm} RDMA: Repetition division multiple access \\ 
				%IGMA: Interleave grid multiple access  \hspace{4.35cm} NCMA: Non-orthogonal coded multiple acces   \hspace{4.55cm} RSMA: Resource spread multiple access  \\ 
				%LCRS: Low code rate spreading \hspace{5.05cm} NOCA: Non-orthogonal coded access  \hspace{5.55cm} SCMA: Sparse code multiple access                                                                                                                                                                                                                            \end{tabular}}
			\end{tabular}%
		}
	\end{table*}
	\subsection{Wireless Connectivity Options}
	Fig. \ref{fig1} depicts the variety of wireless connectivity options to support different use cases in the IoT framework. The basic classification of these technologies for different IoT use cases is carried out by taking network coverage and performance criterion into consideration \cite{Ericsson2016cellular}. It is expected that a large share of these devices will be facilitated by short-range radio technologies, such as WiFi, Bluetooth, and Zigbee, while a significant share will be enabled through wide area networks (WANs) that are majorly facilitated by cellular networks. Connectivity through cellular networks will be provided by the 3GPP technologies, including global system for mobile communications (GSM), wideband code division multiple access (WCDMA), long term evolution (LTE), LTE advanced (LTE-A), and the upcoming 5G. These technologies operate on the licensed spectrum and are primarily designed for high quality mobile voice and data services. An evolution of these technologies is being carried out for low-power IoT applications, where the key challenges are low device cost, long battery life, indoor connectivity and regional coverage, scalability, and diversity.
	\par
	3GPP has made significant improvements to meet the requirements of emerging massive IoT applications, which lead to a range of cellular low-power wide area solutions. They include a) extended coverage GSM (EC-GSM), which is achieved through new data and control channels mapped over legacy GSM, b) narrow-band IoT (NB-IoT), which is a self-contained carrier with a system bandwidth of 200 kHz and is enabled on an existing LTE network, and c) LTE for MTC (LTE-M), providing new power-saving functionalities to LTE. Overall, some noticeable improvements by 3GPP to enable massive IoT are: 
	\begin{itemize}
		\item Lower device cost by reducing peak data rate,
		memory requirements, and device complexity.
		\item Improved battery life using power saving mode and discontinuous reception.
		\item Better coverage (e.g., 15 dB and 20 dB
		in link budget on LTE-M and NB-IoT, respectively), allowing deeper indoor coverage.
	\end{itemize}
	While significant improvements focused on mMTC have been made, some major challenges still need to be tackled such as massive connectivity, which is explained next.
	\subsection{Massive Connectivity}
	The massive connectivity challenge to support mMTC can be broadly split into two categories.
	\subsubsection{Orthogonal/non-orthogonal Multiple Access}
	One primary challenge to provide connectivity to these devices is the limited number of available channel resources. The situation is exacerbated by the fact that radio resource allocation in existing multiple access (MA) techniques, orthogonal MA (OMA), is non-overlapping in nature i.e., a radio resource can be allocated to only a single device/user. Considering a large number of devices and limited available radio resources, such OMA based resource allocation becomes a performance bottleneck. In this context, non-orthogonal MA (NOMA) has been identified as a potential candidate technology to provide massive connectivity. NOMA works on the non-orthogonality principle, where multiple users can be overlapped over a single radio resource block (RB) \cite{dai2015non,tao2015survey,wang2016analysis,islam2016power,yang2017uplink2,ding2017survey,cai2017modulation,liu2017nonorthogonal,basharat2018survey,ding2018embracing,dai2018survey,ye2018uplink,maraqa2019survey,ding2017application,docomo2014docomo}, thereby creating new avenues for connectivity over limited radio resources. 
	\par
	In recent years, different NOMA techniques are proposed by academia and industry. However, most of these techniques are analyzed in literature from a HTC perspective, where a limited number of users are considered to be already connected to the base station (BS) or evolved NodeB (eNB), and capacity maximization is the key target goal. Moreover, considering centralized scheduling, the users and BS/eNB are assumed to know almost everything about each other i.e., number of users multiplexed over a RB, spreading sequences, power control, modulation and coding scheme (MCS), channel state information, etc. Furthermore, perfect system synchronization is assumed. These assumptions may not be applicable to mMTC scenarios, and need further investigation.
	\subsubsection{Grant-based/grant-free Transmission}
	Although NOMA can solve the massive connectivity issue by loading multiple users over a particular RB, another challenge is the way each device accesses a channel resource. In existing wireless networks, each device requests a data transmission slot via a contention-based (e.g., ALOHA \cite{abramson1970aloha}, slotted-ALOHA \cite{roberts1975aloha}, etc.) random access (RA) process followed by a multi-step handover process once granted a channel resource, which (in LTE/LTE-A) has been identified as a major performance bottleneck, and a source of excessive delay and signaling overhead \cite{3GPPGP100892,shirvanimoghaddam2015probabilistic,chen2018ultra}. Considering mMTC traffic, a step-wise shift towards grant-free/contention-based communication is inevitable, where devices can transmit data as per their needs without going through the RA / resource granting process, or by merging RA and data transmission. In this context, grant-free/contention-based transmission using NOMA is considered as a promising solution. 
	\subsection{Contributions and Organization}
	Table. \ref{tab3} provides a summary of how some prominent NOMA schemes (discussed later in Sec. \ref{sec3}) have been covered in recent surveys. While the centralized scheduling or grant-based NOMA schemes have been extensively covered through various surveys \cite{dai2015non,tao2015survey,wang2016analysis,islam2016power,yang2017uplink2,ding2017survey,cai2017modulation,liu2017nonorthogonal,basharat2018survey,ding2018embracing,dai2018survey,ye2018uplink,maraqa2019survey}, detailed information from a grant-free perspective for UL IoT communication is still lacking. Different from these works, this article primarily focuses on the grant-free access. In this context, the concept of grant-free NOMA, survey of different grant-free schemes, practical challenges, and future directions are comprehensively discussed.
	\par
\begin{figure}[!t]
		\centering
		\includegraphics[width=3.2in,height=4.85in]{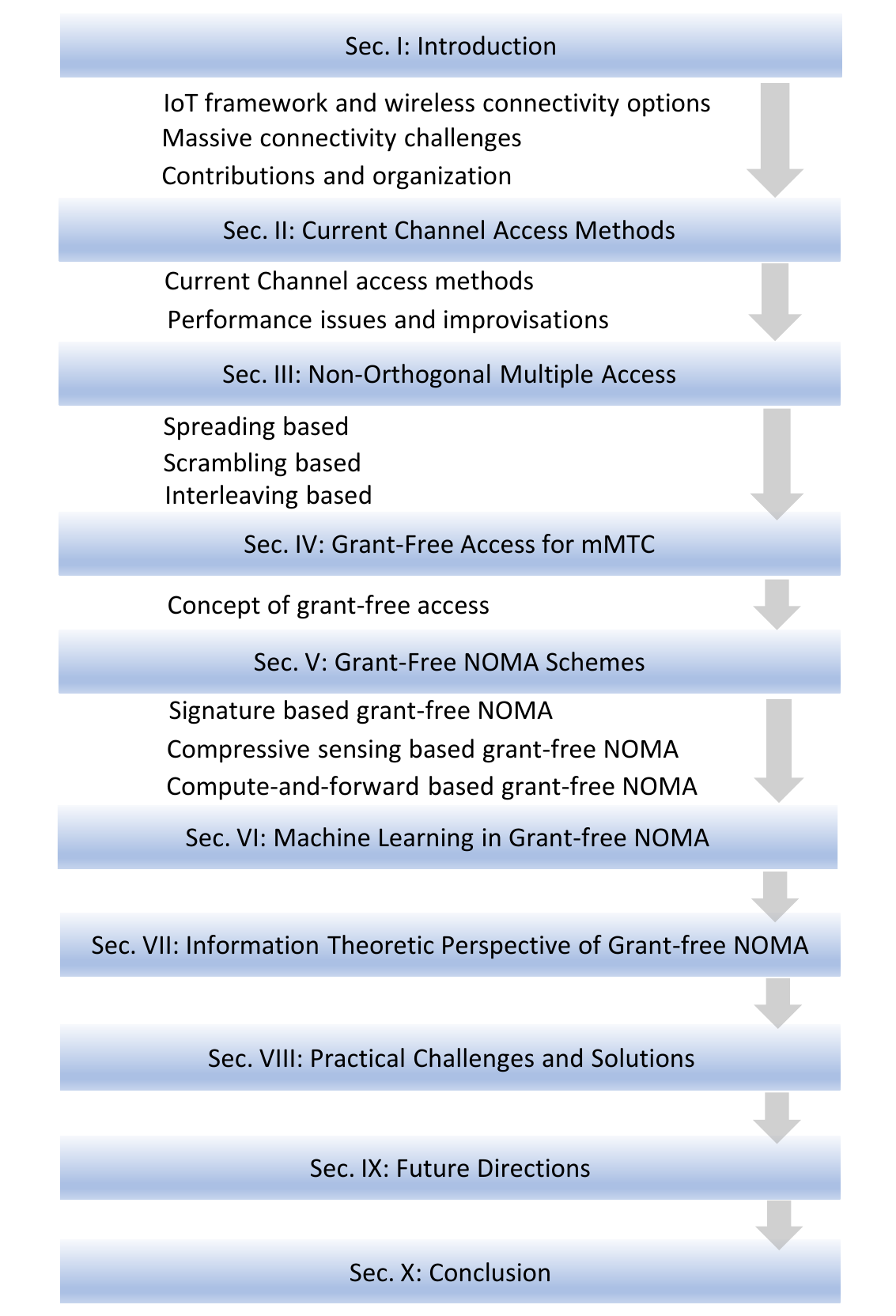}
		\caption{Organization of the Paper}
		\label{fig2}
	\end{figure}

	The major contributions of this article are highlighted as follows, whereas the organization is shown in Fig. \ref{fig2}.
	\begin{itemize}
		\item Firstly, different from the existing surveys, this article provides a comprehensive summary of the recent works on NOMA from a grant-free perspective.
		\item Initially, the existing scheduling based channel access methods and their related challenges are discussed including the basic NOMA principle. 
		\item The concept of grant-free access is then introduced, followed by a comprehensive overview of the recent research works on grant-free NOMA schemes designed for mMTC. 
		\item An information theoretic perspective of grant-free access is then discussed.
		\item In the later part, challenges related to the use of NOMA schemes for grant-free mMTC are highlighted.
		\item Finally, some possible future directions to deal with these challenges, and to design novel NOMA based grant-free schemes, are provided. 
	\end{itemize} 
	\section{Current Channel Access Methods}
	\label{sec2}
	3GPP has recently been working on the design of network architecture, services, specific signaling reduction, and optimization for mMTC. This is being done by considering a variety of mMTC services which, besides posing very diverse QoS requirements, may also need to handle a huge number of devices, provide ultra-low energy consumption, low cost solutions, and facilitate the coexistence of both mMTC and HTC applications. 
	\subsection{Channel Access Methods in LTE/LTE-A}
	\label{sec2:subsec1}
	In existing wireless networks, radio resources (e.g., time, frequency) are allocated orthogonally to connected devices. Therefore, in LTE/LTE-A, the entity requesting access to the cellular network has to first go through a coordination process over the physical random access channel (PRACH) to get aligned with the eNB. The process was originally intended to be used for multiple purposes i.e., initial access of a user not connected to the eNB, UL synchronization of users already connected, UL data transmission or acknowledgment of received data, handover management, etc \cite{sesia2011lte}. The LTE standard prescribes that PRACH access
	be performed using a four-way handshake, which is contention-based (the users can initiate the access process whenever they want). 
	\par
\begin{figure}[t]
		\centering
		\includegraphics[width=3.6in,height=2.8in]{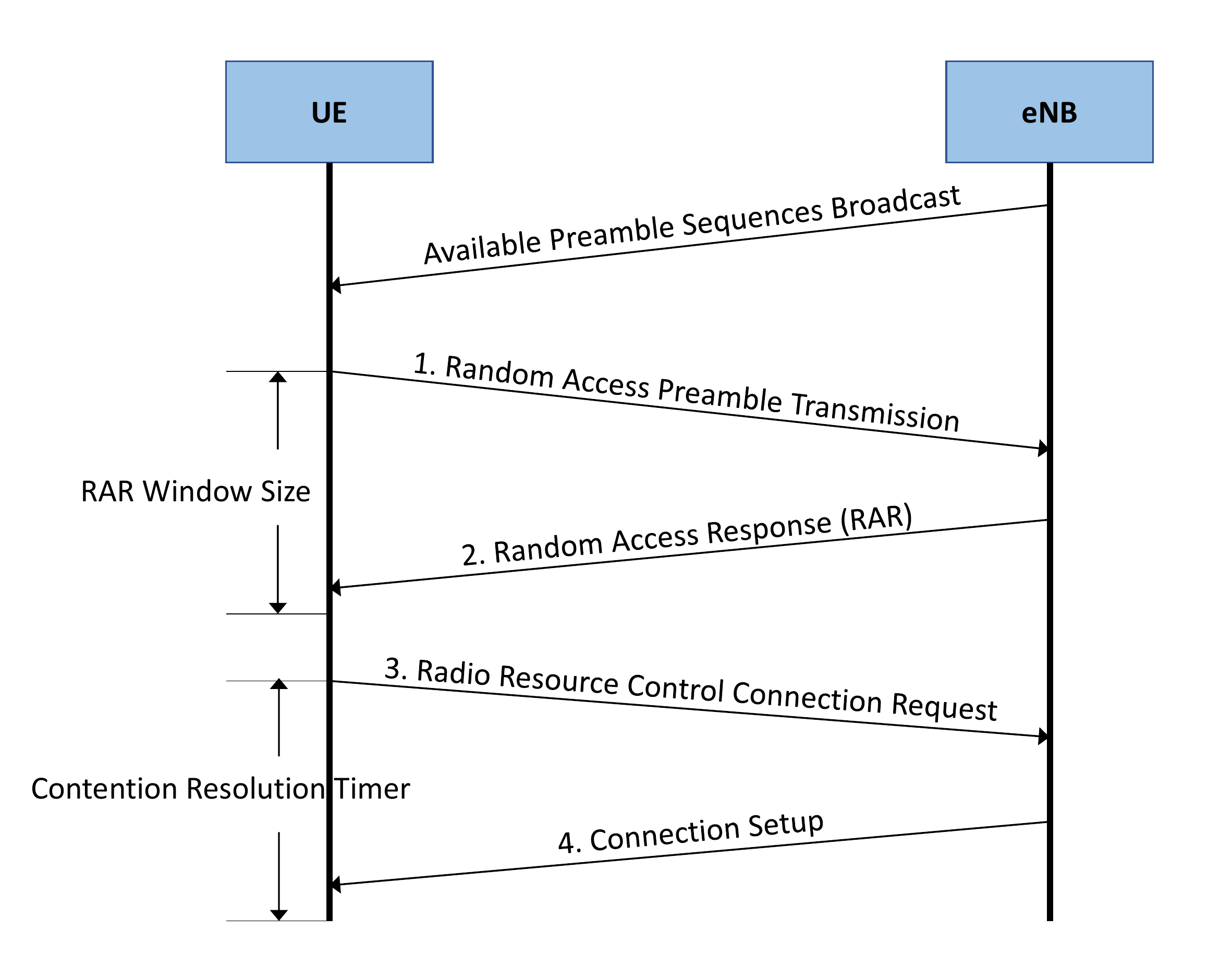}
		\caption{RACH of LTE/LTE-A networks}
		\label{fig3}
	\end{figure}

	Any device which is already aligned with the eNB can make a grant acquisition (GA) request to transmit its data when it needs. To enable PRACH access, devices are initially informed about available PRACH resources through a broadcast from eNB. This is followed by the four-step random access channel (RACH) handshake procedure. The RACH of LTE/LTE-A \cite{dahlman20103g} is shown in Fig. \ref{fig3}, and the steps are summarized below.
	\begin{enumerate}
		\item \textbf{Preamble Transmission:} Every device randomly chooses one out of 64 available preambles, and sends it to the eNB, which estimates the transmission time of each device from its detected preamble.
		\item \textbf{Random Access Response:} For each detected preamble, eNB sends a RA response (RAR) message with information about the radio resource allocated to the device and timing advance information for synchronization. If a device does not receive RAR within a predefined waiting time (RAR waiting window size) or gets a RAR with no information about its request, it postpones the access attempt to the next RACH opportunity.
		\item \textbf{Radio Resource Control Connection Request:} Each device that gets a successful RAR from the eNB makes a radio resource control (RRC) connection request by sending its temporary terminal identity to the eNB through the Physical UL Shared Channel (PUSCH). 
		\item \textbf{RRC Connection Setup:} eNB sends information allocating resources to all devices that have gained access by specifying their terminal identity.
	\end{enumerate}
	\subsection{Related Performance Issues}
	\label{sec2:subsec2}
\begin{figure}[t]
		\centering
		\includegraphics[width=3.25in,height=2.6in]{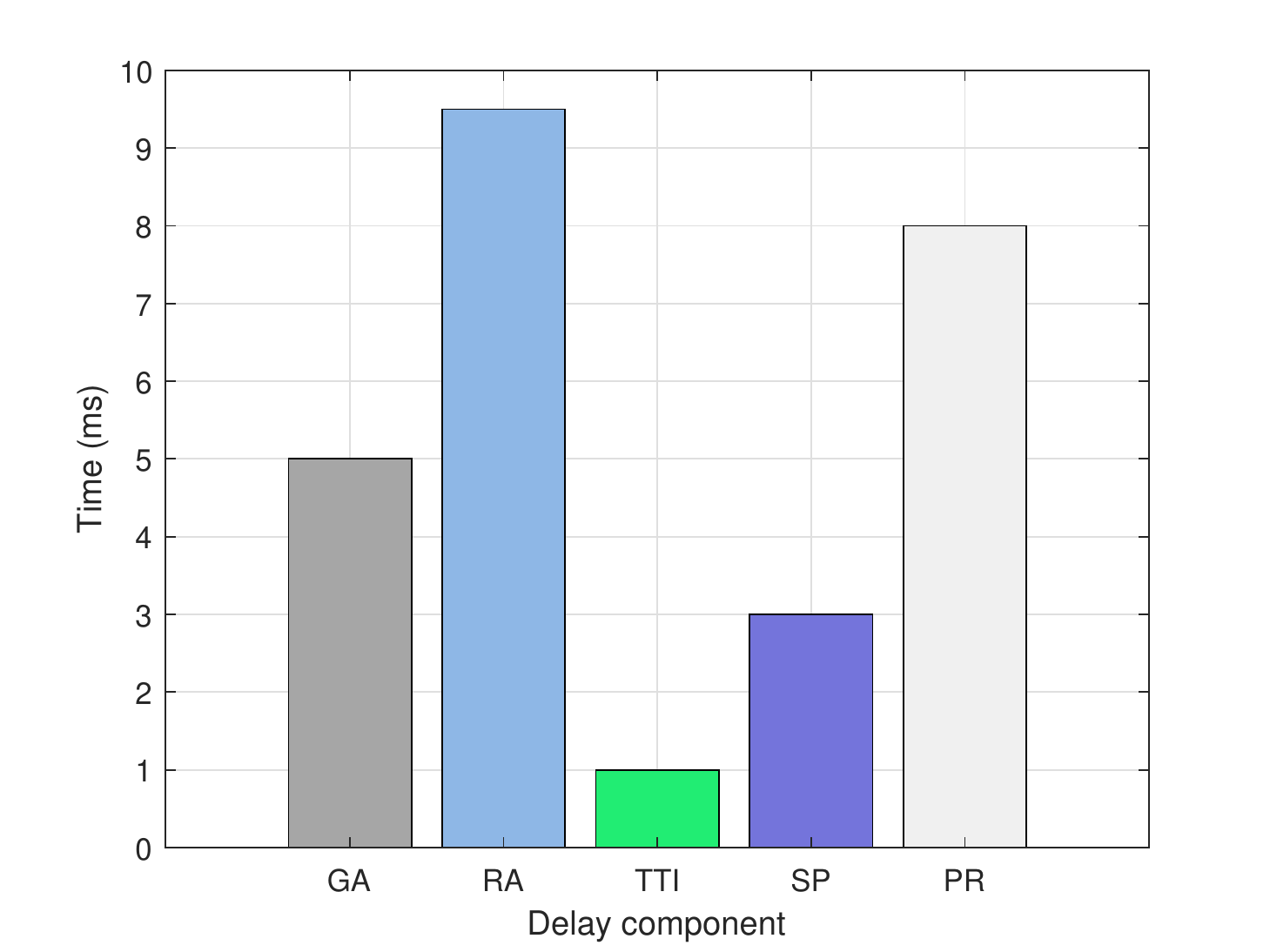}
		\caption{Delay sources of a LTE system (Release 8)}
		\label{fig4}
	\end{figure}

	The overall RA process (RACH and GA) comes with many problems, especially latency. In Fig. \ref{fig4}, some sources of delay in the LTE Release 8 are summarized, where GA means scheduling request of a connected user to transmit data, RA is the RACH plus GA for a user not connected/aligned with eNB, TTI is the transmit time interval of a data packet, SP is signal processing (e.g., encoding and decoding data), and PT represents packet retransmissions in an access network i.e., the UL hybrid automatic repeat request (HARQ) process delay for each retransmission. Furthermore, there can be other delays due to the core network such as queueing delay due to congestion, propagation delay, packet retransmission delay caused by upper layer, etc. It can be seen from Fig. \ref{fig4} that the RA process causes a latency of around 9.5 ms, which is too high for certain IoT use cases \cite{chen2018ultra}.
	\par
	The RA process is feasible for a small number of devices. However, if PRACH is congested due to a large number of MTC and/or HTC devices attempting access simultaneously, the signal-to-interference-plus-noise ratio (SINR) observed at the receiver (i.e., eNB) may be reduced to the extent that messages cannot be detected, and consequently, many of the access attempts
	fail; this is denoted as the outage condition \cite{khan2018priority}. Excessive preamble collisions and retransmissions result in problems like network congestion, unexpected delays, packet loss, radio resource wastage and high energy consumption. Excessive overhead is another major issue as significant resources are spent only to establish a connection for communicating very small sized mMTC traffic data \cite{wiriaatmadja2015hybrid}. For instance, to transmit 100 B of data, around 59 B of overhead in UL and 136 B of overhead in DL would be required for signal transmissions \cite{shariatmadari2015machine}. In LTE-A, the collision rate of the RA requests is around 10 percent and the signaling overhead is about 30-50 percent of the payload size \cite{durisi2016toward}. 
	\subsection{Proposed Access Method Modifications for mMTC Traffic}
	\label{sec2:subsec3}
	In the past several years, a number of techniques to improve the performance of mMTC traffic by alleviating outage due to congestion have been suggested. For instance, grouping the traffic into different classes for which RA process may be temporarily delayed or blocked, as done in access class barring technique \cite{lien2011cooperative}. Differentiation among traffic classes can be achieved by allowing different back-off windows for different classes of users \cite{cheng2011prioritized}. Similarly, predefined access attempt slots can be allocated based on the user grouping \cite{wu2013fasa}, or suitable polling scheme can be applied \cite{3GPP2011Studyx}. Moreover, PRACH resources (such as preamble sequences and RA slots) can be separated and dynamically allocated \cite{lo2011enhanced}.
	\par
	In addition to finding ways for reducing outage due to congestion, optimizing the RACH procedure has also been investigated in recent years by 3GPP radio access network (RAN) working group 1 (WG1) for the new radio (NR). For instance, it has been agreed that NR should support multiple RACH preamble formats with shorter/longer preamble lengths. Five different PRACH formats for the RA preamble parameters with different lengths are currently available in LTE \cite{3gpp2013accesse}. Details about preamble design, and corresponding RAR improvements, for NR PRACH can be found in \cite{Ericsson2016NRPrach,ZTE2016Considerations1,Huawei2016Evaluation1,Ericsson2016nrrandom}.
	\par
	Currently, there is only one RACH timeline applied to all scenarios and use cases in LTE. However, as NR supports various services and use cases with different latency requirements, one RACH timeline might not be efficient. Hence, the need for RACH timeline design with variable lengths for a variety of use cases is critical, and is addressed in \cite{Qualcomm2016RACH1}. In addition to these, 2-step RACH procedure is also proposed for consideration in 3GPP NR \cite{ZTE2016On2}. The use cases for such a RACH procedure are users with typically intermittent small packet transmissions, access in unlicensed spectrum, cases where UL timing advance is not needed, etc \cite{Nokia2016Random12}.
	\par
	While improvements in grant-based access procedures are ongoing, other solutions to reduce latency and signaling overhead due to scheduling requests by a massive number of IoT devices in the RA still need to be investigated. In order to overcome the limited resource issue and related challenges, grant-free NOMA based UL access for mMTC has been jointly agreed by both academia and industry. 
	\section{Non-orthogonal Multiple Access}
	\label{sec3}
	To enable massive connectivity and to deal with the limitations of existing MA schemes explained earlier, need for the design of new wireless technologies is inevitable. In this context, by breaking the orthogonality principle of conventional OMA schemes where a particular channel resource is used by a specific user, NOMA has emerged as a potential MA technique for 5G and beyond. The core idea of NOMA is to multiplex different data streams over the same radio resourse blocks and employ multi-user detection (MUD) algorithms at the receiver to recover users’ signals \cite{dai2015non,tao2015survey,wang2016analysis,islam2016power,yang2017uplink2,ding2017survey,cai2017modulation,liu2017nonorthogonal,basharat2018survey,ding2018embracing,dai2018survey,ye2018uplink,maraqa2019survey,ding2017application,docomo2014docomo,shafi20175g,li20145g,docomo2016initial}. Hence, NOMA provides massive connectivity and high spectral efficiency compared to existing OMA schemes, such as time division MA and orthogonal frequency-division MA \cite{ding2016impact,wang2006comparison}. The scheme has gained popularity and is recently used by the name layered-division-multiplexing in ATSC 3.0, a forthcoming digital TV standard \cite{zhang2016layered}. Moreover, it is considered for 3GPP LTE enhancements under the name multi-user superposition transmission. \cite{3gpp2015access}.
	\par
	Various NOMA schemes based on user-separation through spreading, scrambling, interleaving, or multiple domains are considered for 5G and beyond. The basic phenomenon behind all these schemes is same; multiplexing users over same time-frequency RBs by using some distinguishing parameter. In this context, some prominent NOMA schemes, and their basic working principles, are briefly summarized in Table. \ref{tab1}. The 3GPP RAN WG1 has also been studying some of these NOMA candidate solutions for 3GPP NR \cite{3GPP2016WF,China2016Classification}. 
	\par
	It has been agreed that NOMA schemes should be investigated for diverse 5G usage scenarios and use cases \cite{ETSI2016Final}, and 5G should target to support UL NOMA at least for mMTC \cite{ETSI2016Final2}. Some of the schemes which have so far been identified/agreed for investigation in UL mMTC use cases by 3GPP \cite{3GPP2016WF} are summarized in Table. \ref{tab2} in terms of their categorization, transmitter characteristics, and receiver types. It can be seen in Table. \ref{tab2} that the UL NOMA schemes can be divided into three broad categories i.e., scrambling, spreading, and interleaving based \cite{3GPP2016WF,yang2017uplink2,China2016Classification}. This section gives a brief summary of these prominent NOMA schemes agreed for investigation in UL mMTC use cases.
	\begin{table*}[]
		\caption{Various non-orthgonal multiple access schemes proposed by academia and industry}
		\label{tab1}
		\resizebox{\textwidth}{!}{%
			\begin{tabular}{ccclc}
				\hline
				\multicolumn{1}{|c|}{} & \multicolumn{2}{c|}{\textbf{NOMA Schemes}} & \multicolumn{1}{c|}{\textbf{Description}} & \multicolumn{1}{c|}{\textbf{\begin{tabular}[c]{@{}c@{}}Receiver \\ Type\end{tabular}}} 
				\\ 
				\hline
				\multicolumn{1}{|c|}{1} & \multicolumn{1}{c|}{\multirow{10}{*}{\begin{tabular}[c]{@{}c@{}}Spreading\\ based\end{tabular}}} & \multicolumn{1}{c|}{\begin{tabular}[c]{@{}c@{}}LDS-CDMA \cite{hoshyar2008novel,guo2008multiuser,van2009multiple} \end{tabular}} & \multicolumn{1}{l|}{\begin{tabular}[c]{@{}l@{}}\textbullet~Inspired by CDMA, where users can share a RB through unique orthogonal user-specific spreading sequences.\\ \textbullet~The basic difference is that LDS-CDMA uses LDS or sparse spreading sequences i.e., user data is spread over  \\ ~~~a small number of chips rather than all to limit interference on each chip in classic CDMA.\end{tabular}} & \multicolumn{1}{c|}{MPA} \\ \cline{1-1} \cline{3-5} 
				\multicolumn{1}{|c|}{2} & \multicolumn{1}{c|}{} & \multicolumn{1}{c|}{\begin{tabular}[c]{@{}c@{}}LDS-OFDM \cite{hoshyar2010lds,al2011subcarrier,wen2016non} \end{tabular}} & \multicolumn{1}{l|}{\begin{tabular}[c]{@{}l@{}}\textbullet~An integration of LDS-CDMA and conventional OFDM.\\ \textbullet~Symbols of users are first multiplied with LDS sequences, and then mapped onto different OFDM subcarriers.\\ \textbullet~More fit for wideband than LDS-CDMA, and achieves significant performance improvement.\end{tabular}} & \multicolumn{1}{c|}{MPA} \\ \cline{1-1} \cline{3-5} 
				\multicolumn{1}{|c|}{3} & \multicolumn{1}{c|}{} & \multicolumn{1}{c|}{\begin{tabular}[c]{@{}c@{}}SCMA \cite{nikopour2013sparse,zhang2014sparse,huawei2016sparse,taherzadeh2014scma,wu2015iterative,xiao2015iterative,du2016fast,mu2015fixed,bayesteh2015low}\end{tabular}} & \multicolumn{1}{l|}{\begin{tabular}[c]{@{}l@{}}\textbullet~Developed from basic LDS-CDMA. But, bit-to-constellation mapping and spreading are amalgamated in SCMA.\\ \textbullet~Each user has its own unique codebook for bit to codeword mapping.\\ \textbullet~These codebooks are built using multi-dimensional constellation mapping, that provides constellation shaping\\ ~~ gain and more diversity.\end{tabular}} & \multicolumn{1}{c|}{\begin{tabular}[c]{@{}c@{}}MPA, MPA \\ with SIC\end{tabular}} \\ \cline{1-1} \cline{3-5} 
				\multicolumn{1}{|c|}{4} & \multicolumn{1}{c|}{} & \multicolumn{1}{c|}{SAMA \cite{dai2014successive}} & \multicolumn{1}{l|}{\begin{tabular}[c]{@{}l@{}}\textbullet~Different from other LDS schemes, spreading sequences in SAMA have variable sparsity. \\ \textbullet~Data of different users can therefore be spread over different number of resources, providing more diversity. \end{tabular}} & \multicolumn{1}{c|}{MPA} \\ \cline{1-1} \cline{3-5}
				\multicolumn{1}{|c|}{5} & \multicolumn{1}{c|}{} & \multicolumn{1}{c|}{\begin{tabular}[c]{@{}c@{}}PDMA \cite{catt2016pdma,zeng2015pattern,kang2015pattern,chen2016pattern,dai2018pattern,ren2016advanced} \end{tabular}} & \multicolumn{1}{l|}{\begin{tabular}[c]{@{}l@{}}\textbullet~Similar to SAMA, sparsity of spreading sequences used by users is variable. \\ \textbullet~On top of it, users are multiplexed in multiple domains i.e., power, space, code, or their combination.\end{tabular}} & \multicolumn{1}{c|}{\begin{tabular}[c]{@{}c@{}} MPA/SIC, \\ BP \end{tabular}} \\ \cline{1-1} \cline{3-5} 
				\multicolumn{1}{|c|}{6} & \multicolumn{1}{c|}{} & \multicolumn{1}{c|}{\begin{tabular}[c]{@{}c@{}}LDS-SVE \cite{Fujitsu2016ldssve} \end{tabular}} & \multicolumn{1}{l|}{\begin{tabular}[c]{@{}l@{}}\textbullet~In spreading based schems, user-data bits are carried by vector form of signals spread on multiple resources.\\ \textbullet~Based on original LDS, the idea is to design larger user-specific signature (spreading) vectors.\\ \textbullet~For example, concatenating two element signature vectors of a user into a larger signature vector. Such\\ ~~~ signature-vector-extension can further exploit diversity gain.\end{tabular}} & \multicolumn{1}{c|}{MPA} \\ \cline{1-1} \cline{3-5} 
				\multicolumn{1}{|c|}{7} & \multicolumn{1}{c|}{} & \multicolumn{1}{c|}{MUSA \cite{yuan2016multi, yuan2015multi}} & \multicolumn{1}{l|}{\begin{tabular}[c]{@{}l@{}}\textbullet~Dense-spreading style scheme. Uses "complex" spreading codes with short length and advanced SIC receiver.\\ \textbullet~Increased pool of spreading codes. User-symbols are spread using same/different complex spreading sequences.\\ \textbullet~Importantly, each users's data after spreading is overlapped over same resources. \end{tabular}} & \multicolumn{1}{c|}{SIC} \\ \cline{1-1} \cline{3-5}  
				\multicolumn{1}{|c|}{8} & \multicolumn{1}{c|}{} & \multicolumn{1}{c|}{NCMA \cite{lg2014ncma,hu2013new}} & \multicolumn{1}{l|}{\textbullet~Spreading codes are obtained through Grassmannian line packaging problem.} & \multicolumn{1}{c|}{PIC} \\ \cline{1-1} \cline{3-5} 
				\multicolumn{1}{|c|}{9} & \multicolumn{1}{c|}{} & \multicolumn{1}{c|}{NOCA \cite{nokia2016noca}} & \multicolumn{1}{l|}{\textbullet~Based on LTE defined low correlation sequences as spreading codes.} & \multicolumn{1}{c|}{SIC} \\ \cline{1-1} \cline{3-5} 
				\multicolumn{1}{|c|}{10} & \multicolumn{1}{c|}{} & \multicolumn{1}{c|}{FDS \cite{intel2016fds}} & \multicolumn{1}{l|}{\textbullet~Directly spreads the modulation symbols with multiple
					orthogonal or quasi-orthogonal codes.} & \multicolumn{1}{c|}{SIC} \\ \cline{1-1} \cline{3-5} 
				\multicolumn{1}{|c|}{11} & \multicolumn{1}{c|}{} & \multicolumn{1}{c|}{LCRS \cite{intel2016fds}} & \multicolumn{1}{l|}{\begin{tabular}[c]{@{}l@{}}\textbullet~Direct spreading of modulation symbols with multiple orthogonal codes.\end{tabular}} & \multicolumn{1}{c|}{SIC} \\ \cline{1-1} \cline{3-5}
				\multicolumn{1}{|c|}{12} & \multicolumn{1}{c|}{\multirow{4}{*}{}} & \multicolumn{1}{c|}{GOCA \cite{MediaTek2016goca}} & \multicolumn{1}{l|}{\begin{tabular}[c]{@{}l@{}}\textbullet~Employs group orthogonal sequences to spread the modulation symbols over time-frequency resources.\\ \textbullet~Each group contains orthogonal sequences developed by localized frequency/time-domain repetitions.\\ \textbullet~Several non-orthogonal groups are then made by using non-orthogonal sequences in a second stage.\end{tabular}} & \multicolumn{1}{c|}{SIC} \\ 
				\hline
				\multicolumn{1}{|c|}{13} & \multicolumn{1}{c|}{\begin{tabular}[c]{@{}c@{}}\\Scrambling\end{tabular}} & \multicolumn{1}{c|}{\begin{tabular}[c]{@{}c@{}}PD-NOMA  \cite{benjebbour2013concept,higuchi2015non,benjebbour2015non1,benjebbour2015noma2,yan2015receiver,ding2014performance,shahab2016user,shahab2016power,shahab2018user,zhang2016uplink,al2014uplink,sheng2017novel,ali2016dynamic,yang2016general,ding2015cooperative,zhang2016full,zhong2016non,shahab2018time,ding2015application} \end{tabular}}  & \multicolumn{1}{l|}{\begin{tabular}[l]{@{}l@{}} \textbullet~Users are multiplexed in power domain. Users transmit their data with different power levels over same RB.\\ \textbullet~eNB exploits the received power difference to perform MUD using SIC receiver. \end{tabular}} & \multicolumn{1}{c|}{SIC} 
				\\ 
				\cline{1-1} \cline{3-5}
				\multicolumn{1}{|c|}{14} & \multicolumn{1}{c|}{\multirow{2}{*}{\begin{tabular}[c]{@{}c@{}}Based \end{tabular}}} & \multicolumn{1}{c|}{\begin{tabular}[c]{@{}c@{}}RSMA \cite{qualcomm2016RSMA,qualcom2016rsma1,qualcom2016rsma2} \end{tabular}} & \multicolumn{1}{l|}{\begin{tabular}[c]{@{}l@{}}\textbullet~Uses combination of low rate channel codes and scrambling codes (and optionally different interleavers) \\ ~~~with good correlation properties to separate users\end{tabular}} & \multicolumn{1}{c|}{SIC} \\ \cline{1-1} \cline{3-5} 
				\multicolumn{1}{|c|}{15} & \multicolumn{1}{c|}{} & \multicolumn{1}{c|}{LSSA \cite{etri2016lssa}} & \multicolumn{1}{l|}{\textbullet~Each user data is bit or symbol level multiplexed with user-specific signature pattern, unknown to others} & \multicolumn{1}{c|}{SIC} \\ \hline
				\multicolumn{1}{|c|}{16} & \multicolumn{1}{c|}{\multirow{3}{*}{\begin{tabular}[c]{@{}c@{}}Interleaving\\ Based\end{tabular}}} & \multicolumn{1}{c|}{\begin{tabular}[c]{@{}c@{}}IDMA \cite{ping2006interleave,nokia2016idma} \end{tabular}} & \multicolumn{1}{l|}{\textbullet~Unique user-specific bit or chip level interleaving used.} & \multicolumn{1}{c|}{ESE} \\ \cline{1-1} \cline{3-5} 
				\multicolumn{1}{|c|}{17} & \multicolumn{1}{c|}{} & \multicolumn{1}{c|}{IGMA \cite{samsung2016igma}} & \multicolumn{1}{l|}{\textbullet~Use bit level interleavers and/or grid mapping pattern to separate users} & \multicolumn{1}{c|}{\begin{tabular}[c]{@{}c@{}}ESE, or CCMUD\end{tabular}} \\ \cline{1-1} \cline{3-5} 
				\multicolumn{1}{|c|}{18} & \multicolumn{1}{c|}{} & \multicolumn{1}{c|}{\begin{tabular}[c]{@{}c@{}}RDMA  \cite{MediaTek2016goca} \end{tabular}} & \multicolumn{1}{l|}{\begin{tabular}[c]{@{}l@{}}\textbullet~Unlike IDMA, symbol-level interleaving based on cyclic-shift repetition of modulated symbols is used. \\
						\textbullet~RDMA transmitter is simpler than IDMA because no random interleave is required.\end{tabular}} & \multicolumn{1}{c|}{SIC} \\ \hline
				\multicolumn{1}{|c|}{19} & \multicolumn{1}{c|}{} &    \multicolumn{1}{c|}{\begin{tabular}[c]{@{}c@{}}SDMA \cite{bana2001space} \end{tabular}} & \multicolumn{1}{l|}{\begin{tabular}[c]{@{}l@{}}\textbullet~Uses unique user-specific channel impulse responses to multiplex users, instead of spreading sequences.\\ \textbullet~Due to potentially infinite variety of channel impulse responses, SDMA can support large number of users.\\ \textbullet~However, accurate channel impulse response estimation is needed at the base station.\end{tabular}} & \multicolumn{1}{c|}{\begin{tabular}[c]{@{}c@{}}PIC,\\NL-MUD\end{tabular}} \\ \cline{1-1} \cline{3-5} 
				\multicolumn{1}{|c|}{20} & \multicolumn{1}{c|}{Others} & \multicolumn{1}{c|}{\begin{tabular}[c]{@{}c@{}}LPMA \cite{fang2016lattice} \end{tabular}} & \multicolumn{1}{l|}{\begin{tabular}[c]{@{}l@{}}\textbullet~The power domain and code domain are combined to multiplex users.\\ \textbullet~Implements multilevel lattice superposition codes to allocate different code levels to different CSI users.\\ \textbullet~With two degrees of freedom in multiplexing, LPMA becomes more flexible than simple PD-NOMA.\end{tabular}} & \multicolumn{1}{c|}{SIC} \\ \cline{1-1} \cline{3-5} 
				\multicolumn{1}{|c|}{21} & \multicolumn{1}{c|}{} & \multicolumn{1}{c|}{\begin{tabular}[c]{@{}c@{}}BOMA \cite{naim2014building} \end{tabular}} & \multicolumn{1}{l|}{\begin{tabular}[c]{@{}l@{}}\textbullet~BOMA multiplexes users by attaching information from good channel user to symbols of bad channel user.\\ \textbullet~To achieve same bit error rate as good user, bad user applies coarse constellation with large minimum distance. \\ ~~~Hence, the small building block containing data of good user can be tiled in the constellation of bad user.\end{tabular}} & \multicolumn{1}{c|}{\begin{tabular}[c]{@{}c@{}}LCR \end{tabular}} \\  \hline
			\end{tabular}%
		}
	\end{table*}
	\begin{table*}[]
		\caption{Categorization of some prominent NOMA schemes proposed for the Rel-14 3GPP NR Study item}
		\label{tab2}
		\resizebox{\textwidth}{!}{%
			\begin{tabular}{|c|l|l|l|l|}
				\hline
				\multicolumn{1}{|l|}{} & \multicolumn{1}{c|}{\textbf{\begin{tabular}[c]{@{}c@{}}Category 1:\\ Scrambling based\end{tabular}}} & \multicolumn{2}{c|}{\textbf{\begin{tabular}[c]{@{}c@{}}Category 2:\\ Spreading based\end{tabular}}} & \multicolumn{1}{c|}{\textbf{\begin{tabular}[c]{@{}c@{}}Category 3:\\ Interleaving based\end{tabular}}} \\ \hline
				\multirow{4}{*}{\textbf{\begin{tabular}[c]{@{}c@{}}Transmitter\\ characteristics\end{tabular}}} & \multirow{4}{*}{\begin{tabular}[c]{@{}c@{}} \textbullet~Use different scrambling sequences \\~~~to distinguish different users\\ \textbullet~Can be used together with low code\\~~~rate FEC.\end{tabular}} & \multicolumn{2}{l|}{\multirow{2}{*}{\begin{tabular}[c]{@{}c@{}} \textbullet~Use different spreading sequences/codes\\~~~to distinguish different users\end{tabular}}} & \multirow{4}{*}{\begin{tabular}[c]{@{}c@{}}\textbullet~Use different interleavers to distinguish \\~~~different users.\\ \textbullet~Can be used together with low code\\~~~rate FEC.\end{tabular}} \\
				&  & \multicolumn{1}{l}{} &  \\ \cline{3-4}
				&  & ~~~~~~LDS code & ~~~~~~~Non-LDS code &  \\ 
				&  &  &  &  \\ \hline
				\textbf{\begin{tabular}[c]{@{}c@{}}Receiver\\ type\end{tabular}} & ~~~~~~~~~~~~~~~~~~~~~ SIC & ~~~~~MPA or BP & ~~~~~~~~~SIC or PIC & \begin{tabular}[c]{@{}c@{}}(iterative) ESE, or MAP/MPA if there exist\\ zero entries, SIC \end{tabular} \\ \hline
				\textbf{\begin{tabular}[c]{@{}c@{}}Candidate \\ schemes\end{tabular}} & \begin{tabular}[c]{@{}l@{}}~~~~~~~~~~~~~PD-NOMA \cite{benjebbour2013concept,higuchi2015non,benjebbour2015non1,benjebbour2015noma2,yan2015receiver,ding2014performance,shahab2016user,shahab2016power,shahab2018user,zhang2016uplink,al2014uplink,sheng2017novel,ali2016dynamic,yang2016general,ding2015cooperative,zhang2016full,zhong2016non,shahab2018time,ding2015application}\\ ~~~~~~~~~~~~~RSMA \cite{qualcomm2016RSMA,qualcom2016rsma1,qualcom2016rsma2}\\ ~~~~~~~~~~~~~LSSA \cite{etri2016lssa} \end{tabular} & \begin{tabular}[c]{@{}l@{}}~~~SCMA \cite{nikopour2013sparse,zhang2014sparse,huawei2016sparse,taherzadeh2014scma,wu2015iterative,xiao2015iterative,du2016fast,mu2015fixed,bayesteh2015low} \\ ~~~PDMA \cite{catt2016pdma,zeng2015pattern,kang2015pattern,chen2016pattern,dai2018pattern,ren2016advanced}\\ ~~~LDS-SVE \cite{Fujitsu2016ldssve} \end{tabular} & \begin{tabular}[c]{@{}l@{}}~MUSA \cite{yuan2016multi}~~~~~~LCRS\cite{intel2016fds}\\ ~NCMA \cite{lg2014ncma}~~~~~~GOCA\cite{MediaTek2016goca}\\ ~NOCA\cite{nokia2016noca}~~~~~FDS\cite{intel2016fds}\\ \end{tabular} & \begin{tabular}[c]{@{}l@{}}~~~~~~~~~~~~~~IDMA\cite{ping2006interleave,nokia2016idma}\\~~~~~~~~~~~~~~IGMA\cite{samsung2016igma} \\~~~~~~~~~~~~~~RDMA\cite{MediaTek2016goca} \end{tabular} \\ \hline
			\end{tabular}%
		}
	\end{table*}
	\par
	\subsection{Spreading Based}
	These schemes are majorly inspired by classic code division multiple access (CDMA), where multiple users share same time-frequency resources through unique user-specific spreading sequences. These schemes are further classified into low density spreading (LDS) based, and non-LDS based. The spreading sequences in LDS based schemes are sparse or non-orthogonal low cross-correlation sequences, which can be achieved by switching-off a large number of spreading signature chips in classic CDMA. In the receiver of LDS-based schemes, iterative algorithms, such as a message passing algorithm (MPA), can be applied for joint/simultaneous detection of multiple data streams with near maximum-likelihood performance. For non-LDS schemes, successive interference cancellation (SIC) or parallel interference cancellation (PIC) is is applied at the receiver. Some prominent LDS and non-LDS schemes are LDS-CDMA \cite{hoshyar2008novel,guo2008multiuser,van2009multiple}, LDS orthogonal frequency-division multiplexing (LDS-OFDM, \cite{hoshyar2010lds,al2011subcarrier,wen2016non}), sparse code multiple access (SCMA, \cite{nikopour2013sparse,zhang2014sparse,huawei2016sparse,taherzadeh2014scma,wu2015iterative,xiao2015iterative,du2016fast,mu2015fixed,bayesteh2015low}), pattern division multiple access (PDMA, \cite{catt2016pdma,zeng2015pattern,kang2015pattern,chen2016pattern,dai2018pattern,ren2016advanced}), LDS signature vector extension (LDS-SVE, \cite{Fujitsu2016ldssve}), multi-user shared access (MUSA, \cite{yuan2016multi, yuan2015multi}). 
	%The technique basically focuses on increasing the resource pool of spreading sequences by generating random non-orthogonal "complex" spreading codes with short length, from which each user can randomly choose one. 
	%Complex spreading sequences maintain lower cross-correlation than traditional pseudo random noise based sequences due to the additional freeedom of the imaginary part. Due to its advantages, MUSA is also being considered for grant-free UL in mMTC. 
	%An example of resource mapping in MUSA is shown in Fig. \ref{fig7}, where each user randomly chooses a spreading code from a pool, and transmits its data over the same radio resource. At the receiver, code level SIC is used for MUD.
	non-orthogonal coded multiple access (NCMA) \cite{lg2014ncma,hu2013new}, non-orthogonal coded access (NOCA, \cite{nokia2016noca}), low code rate spreading (LCRS, \cite{intel2016fds}), frequency domain spreading (FDS, \cite{intel2016fds}), group orthogonal coded access (GOCA \cite{MediaTek2016goca}), etc, as shown in Table. \ref{tab1}. 
	\subsection{Scrambling Based}
	Scrambling based NOMA schemes use different scrambling sequences/patterns to distinguish multiple users, with a SIC process applied at the receiver side for MUD. Moreover, these sequences can be used together with low code rate forward error correction (FEC) codes. Three important schemes that fall in this category are power domain NOMA (PD-NOMA \cite{benjebbour2013concept,higuchi2015non,benjebbour2015non1,benjebbour2015noma2,yan2015receiver,ding2014performance,shahab2016user,shahab2016power,shahab2018user,zhang2016uplink,al2014uplink,sheng2017novel,ali2016dynamic,yang2016general,ding2015cooperative,zhang2016full,zhong2016non,shahab2018time,ding2015application}), resource spread MA (RSMA \cite{qualcomm2016RSMA,qualcom2016rsma1,qualcom2016rsma2}) and low code rate and signature based shared access (LSSA \cite{etri2016lssa}). %\begin{figure}[t]
	In UL PD-NOMA, users can be separated by using different scrambling sequences and by creating received power differences among paired users through predefined transmit power allocation factors). SIC based MUD at the receiver is achieved by exploiting the received power difference \cite{zhang2016uplink,al2014uplink,sheng2017novel,ali2016dynamic,yang2016general}. 
	%A simple two-user UL PD-NOMA system model is shown in Fig. \ref{fig5a}. 
	%The SIC process at BS by exploiting the received power difference of both users' signals for MUD \cite{zhang2016uplink,al2014uplink,sheng2017novel,ali2016dynamic,yang2016general}, is shown in Fig. \ref{fig5b}. 
	%\par
	%The SIC process is performed in descending order of the received SINRs (signal to interference and noise ratio) of multiplexed users, where BS starts decoding the user with highest received SINR first.
	%(e.g., UE$_1$ in this case).
	%The signal detection block further consists of three sub-blocks. Firstly, a minimum mean-square error receiver detects the highest SINR signal by treating the other low SINR signals as interference/noise. This is followed by signal demodulation and FEC decoding.
	%Once highest SINR user is decoded, it's signal is then reconstructed and subtracted from the received superimposed signal to decode the next user, and so on.   
	%\begin{figure*}[!t]
	%	\centering
	%	\includegraphics[width=4.25in,height=1.75in]{SCMA/SCMA.png}
	%	\caption{Resource mapping of SCMA: 6 users, 4 RBs \cite{zte2016discussion1}}
	%	\label{fig8}
	%\end{figure*} 
	Other scrambling based schemes i.e., RSMA and LSSA, use a combination of low rate channel codes and scrambling codes (optionally different interleavers as well) to separate users' signals \cite{qualcomm2016RSMA,qualcom2016rsma1,qualcom2016rsma2,etri2016lssa}.
	\subsection{Interleaving Based}
	The key characteristic of interleaving based NOMA schemes is the use of different interleavers to distinguish between multiplexed users. Moreover, low code rate FEC can also be applied together with interleaving. For MUD, an elementary signal estimator (ESE) is used \cite{ping2006interleave}. Some prominent interleaving based schemes are interleave division multiple access (IDMA \cite{ping2006interleave,nokia2016idma}), 
	%that exhibits significant robustness against lack of synchronization among users. 
	%A basic interleaving based transmitter is shown in Fig. \ref{fig13}. 
	interleave-grid multiple access (IGMA, \cite{samsung2016igma}) and repetition division multiple access (RDMA, \cite{MediaTek2016goca}). 
	%In IGMA, interleaving is combined with grid mapping patterns to facilitate distinguishing between multiple user data streams at the receiver using ESE or chip-by-chip MUD (CCMUD). On the other hand RDMA focuses on symbol-level interleaving unlike IDMA. The interleaving process is designed using simple cyclic-shift repetitions. Moreover, SIC based receiver is used for MUD.
	\begin{figure*}[!t]
		\centering
		\includegraphics[width=6.25in,height=3.35in]{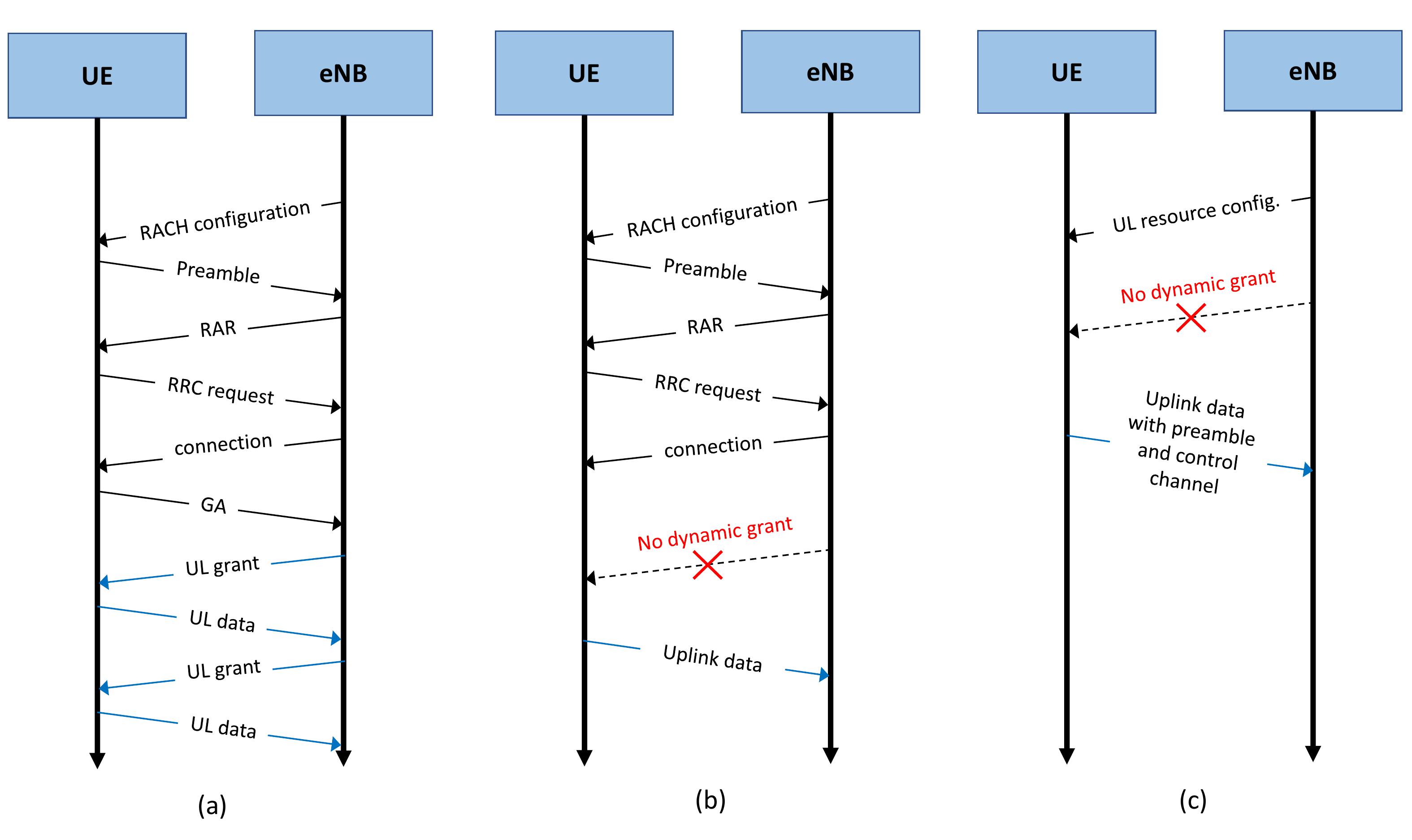}
		\caption{(a) RACH-based grant-based, (b) RACH-based grant-free, (c) RACH-less grant-free}
		\label{fig5}
	\end{figure*}
	\section{Grant-Free Access for mMTC}
	\label{sec4}
	As mentioned in Sec. \ref{sec3}, many variants of NOMA have been proposed by academia and industry in recent years. However, analysis of these schemes in existing literature is mainly done by considering centralized scheduling, where spreading sequences, interleaving patterns, and/or transmission powers of different users are predefined by the eNB. However, the major drawback of this is the excessive signaling overhead, which makes grant-free NOMA inevitable. 
	%In addition to these, NOMA schemes in other domains also exist in literature as detailed in Table. \ref{tab1}, e.g., lattice partition MA (LPMA \cite{fang2016lattice}) using lattice codes, or spatial domain MA (SDMA \cite{bana2001space,suard1998uplink,wolfgang2007parallel,zhang2013evolutionary,godara2018blind,zhang2012turbo,zhang2011joint}) using user-specific channel impulse responses. 
	%\par
	%As explained earlier in Sec. \ref{sec2}, conventional HTC is optimized for mobile broadband, where limited number of HTC devices use a diverse class of applications with high data rate demands. In such cases, signaling overhead due to RA and other control functions constitutes a small fraction of the high data traffic, and is not a major point of concern. On the contrary, MTC traffic is mostly UL, data size per device is very small, devices have a variety of QoS requirements, and high energy efficiency is needed. In MTC traffic, the control signaling overhead becomes significant part of the overall traffic, which is an undesirable situation. 
	\par
	A step-wise proposal towards grant-free communication using NOMA based user multiplexing for 3GPP NR was presented in \cite{docomo2016uplink}, with the following four steps. 
	\begin{enumerate}
		\item RACH-based grant-based OMA as the starting point or baseline UL transmission scheme. To deal with the problems of RA process discussed in Sec. \ref{sec2:subsec2}, and motivated by several improved RA strategies discussed in Sec. \ref{sec2:subsec3}, new RA methods can be designed to minimize collision and overload problems.
		\item RACH-based grant-based NOMA schemes. In \cite{shirvanimoghaddam2017massive2}, a novel RA strategy to enable multiple MTCDs to transmit over same RB is developed. The presence of preambles is detected using timing advance information. Moreover, through power control, eNB can detect the number of devices which have selected the same preamble. This enables eNB to perform RACH signaling for a group of devices instead of each individual, which significantly reduces the signaling overhead in mMTC. These devices can then be allowed to communicate using grant-based NOMA transmission over the same data channel.
		\item RACH-based (synchronous UL) grant-free NOMA to reduce the UL grant overhead. The MTCDs can perform the RACH, but once they get aligned with the BS, further transmissions of data need no GA, and are grant-free NOMA based.
		\item  RACH-less (asynchronous UL) grant-free NOMA. In this case, the MTCDs transmit their data without carrying out any RA or GA process.
	\end{enumerate}
	\begin{figure}[!t]
		\centering
		\includegraphics[width=3.55in,height=2.35in]{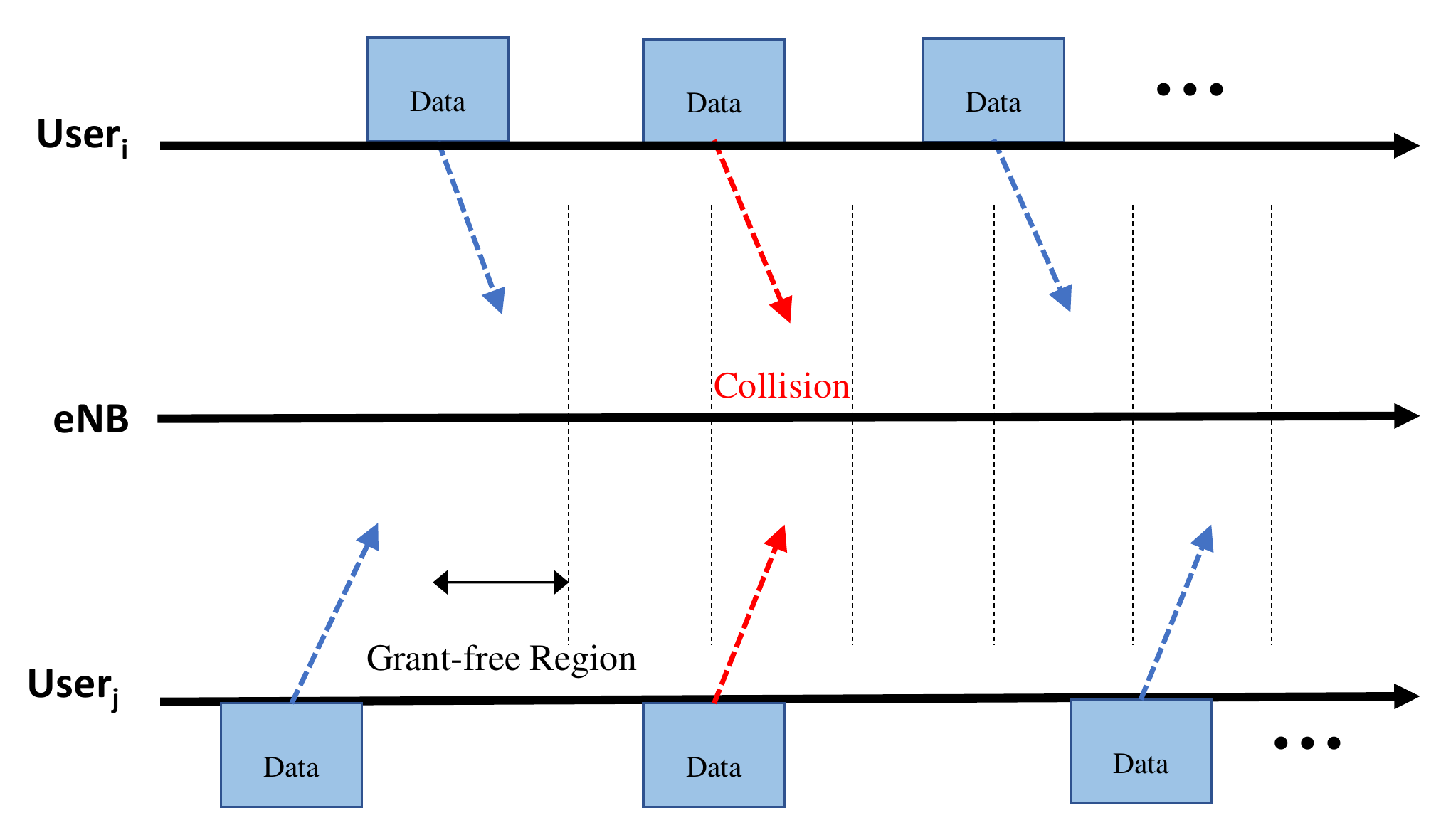}
		\caption{Grant-free contention based transmission}
		\label{fig6}
	\end{figure}
	Fig. \ref{fig5} shows an illustration of RACH-based grant-based, RACH-based grant-free, and RACH-less grant-free transmissions for UL of 3GPP NR \cite{docomo2016uplink}. It can be seen that grant-free achieves autonomous transmission without explicit dynamic grant. There are two possible options of resource allocation in grant-free transmissions. First option is that MTCD's radio resource is pre-configured by eNB or pre-determined, while the second option is that MTCD performs random resource selection itself \cite{Intel2016Grant}. In both cases, when the MTCD wants to transmit data, it takes no further scheduling grant or GA from the eNB, and is termed as a grant-free transmission. In this context, Fig. \ref{fig6} shows grant-free/contention-based arrive and go transmissions using OMA, where some $i^{th}$ and $j^{th}$ users transmit whenever they have data packets \cite{Huawei2016Discussion}. In case both users transmit in the same grant-free slot, a collision is said to have taken place due to OMA, and they re-transmit their data later using random back-off procedure.
	%\begin{figure}[!t]
	%	\centering	
	%	\subfloat[Grant-free UL transmission with or without control channel]{%
	%		\includegraphics[width=2.25in,height=0.85in]{TransmissionFormat/Transmissionformat1.pdf}%
	%		\label{fig13a}
	%	}	
	%	
	%	\subfloat[Grant-free UL transmission with or without control channel]{%
	%		\includegraphics[width=2.25in,height=0.75in]{TransmissionFormat/Transmissionformat2.pdf}%
	%		\label{fig13b}
	%	}
	%	\caption{Grant-free Transmission options}	
	%	\label{fig13}	
	%\end{figure}
	\par
\begin{figure}[!t]
		\centering
		\includegraphics[width=3.45in,height=2.75in]{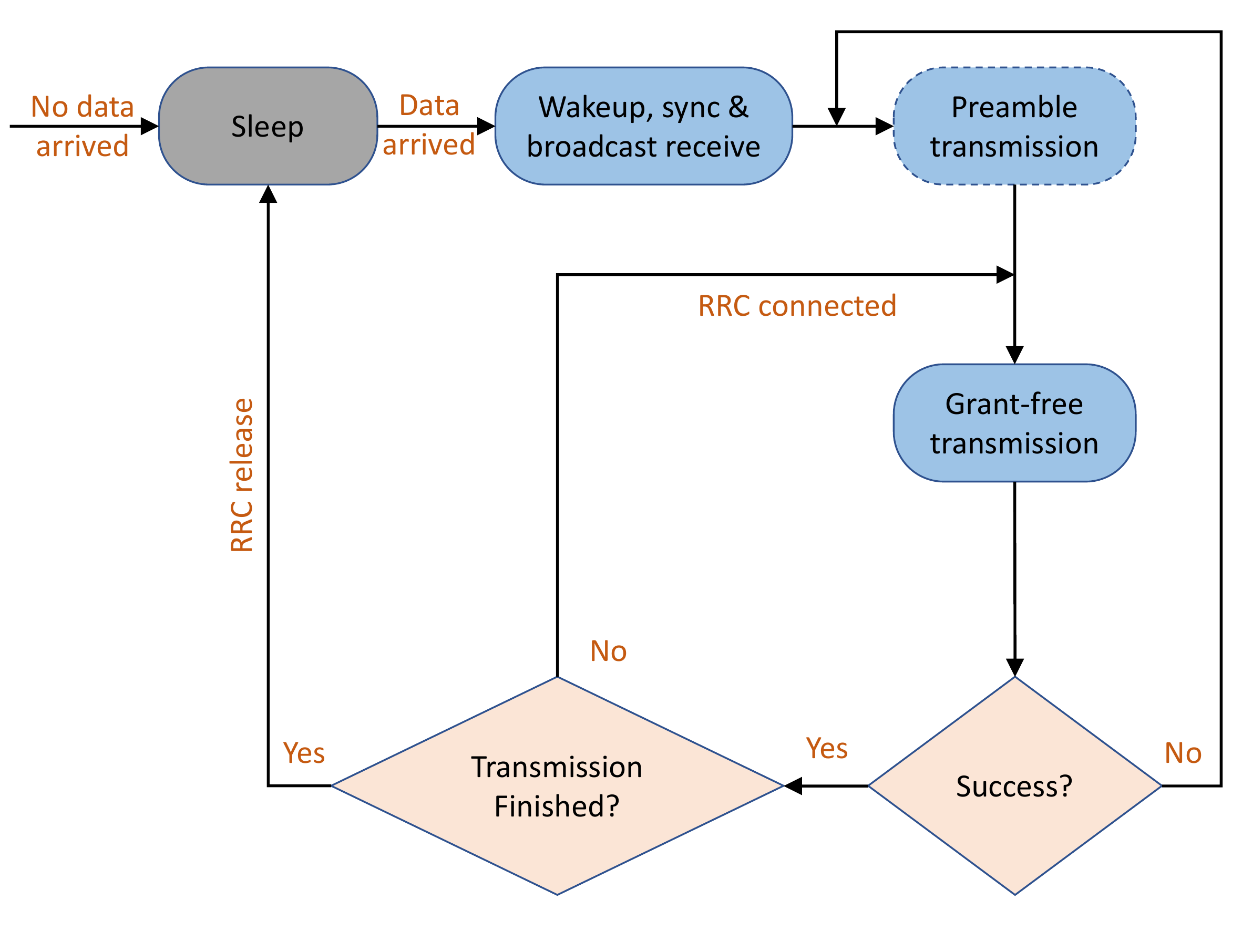}
		\caption{Illustration of Grant-free UL transmissions}
		\label{fig7}
	\end{figure}

	The state graph of a grant-free transmission is shown in Fig. \ref{fig7}. If user has no data to transmit in its buffer, it stays in a sleep state; otherwise, it would wake up, synchronize using reference signals, and acquire some necessary system information. Before direct grant-free transmission, preamble may be transmitted for UL synchronization to facilitate detection at receiver. Furthermore, some MA information could be implicitly indicated by the preamble, such as spreading signature, locations of radio resources, and the timing of retransmissions. With this information, collisions can be detected, and the blind detection complexity of eNB can also be greatly reduced. In general, grant-free UL transmission schemes need to ensure that transmission parameters, identification of the MTCD for purpose of decoding data (e.g., knowledge of spreading code, interleaver, etc.), synchronization, and channel estimation can be determined or detected by the eNB. 
	\par
	Grant-free transmission using OMA schemes will cause a tremendous amount of collisions due to limited available resources as depicted in Fig. \ref{fig6}, where two users transmitting in the same slot collide and need retransmissions. However, in case of NOMA, use of different signature patterns will avoid these collisions as BS can distinguish between multiple users transmitting over the same RB through their signature patterns. Hence, the basic features of grant-free/contention-based UL NOMA are a) transmission from a MTCD does not need the dynamic and explicit scheduling grant from eNB, and b) multiple MTCDs can share the same time-frequency resources through NOMA. It was agreed that grant-free NOMA schemes, where MTCDs can send their data without going into any explicit dynamic grant process, are well-suited for mMTC \cite{zte2016grant}. The devices can transmit their data whenever they want, which can reduce signaling overhead and end-to-end latency \cite{lenovo2016uplink}. Hence, it was decided that 3GPP NR should aim to support UL grant-free/contention-based transmissions at least for mMTC \cite{catt2016consideration,docomo2016discussion}.	\par
	As mentioned earlier, depending on weather RACH is present or not, we have RACH-based/RACH-less grant-free UL NOMA \cite{docomo2016discussion,zte2016discussion}. Collectively, they come under the umbrella of grant-free transmission. In RACH-based grant-free NOMA, once all users perform the RACH, grant-free transmission can occur in a more synchronized manner. In RACH-less grant-free NOMA, to reduce the signaling
	overhead, RACH can be completely eliminated, and the data transmission phase starts whenever a user has packets to transmit.  
	\begin{table*}[]
		\caption{Grant-free UL NOMA schemes}
		\label{tab4}
		\resizebox{\textwidth}{!}{%
			\begin{tabular}{|c|c|l|}
				\hline
				\multicolumn{2}{|c|}{\textbf{Grant-free NOMA schemes}} & \multicolumn{1}{c|}{\textbf{Description}} \\ \hline
				\multirow{6}{*}{\begin{tabular}[c]{@{}c@{}} MA Signature based\\ \cite{au2014uplink,nokia2016basic,qualcom2016RSMA,bayesteh2014blind,Huawei2016LLS,zte2016discussion1,zte2016discussion,zte2016contention,zte2016System2,zte2016System,zte2016receiver,zte2016receiver2,balevi2018aloha,choi2017nomaal,elkourdi2018enabling,qualcomm2016candidate,shirvanimoghaddam2017massive,shirvanimoghaddam2017fundamental,abbas2018novel,abbas2018multi} \end{tabular}} & \multirow{2}{*}{Spreading} & \multirow{6}{*}{\begin{tabular}[c]{@{}l@{}}\textbullet ~To enable grant-free access, a MA resource is defined consisting of a time-frequency block and an MA signature.\\ \textbullet ~The MA signature can be scrambling/spreading/interleaving based, and therefore includes at least one of the following; \\ ~~~codebook/codeword, sequence, interleaver and/or mapping patterns, power dimension, sptaial dimension, preamble, pilot, etc.\\ \textbullet ~Multiple users can transmit in a grant-free manner using any MA resource, where MUD at receiver is achieved by exploiting\\ ~~~the MA signatures.\end{tabular}} \\
				&  &  \\ \cline{2-2}
				& \multirow{2}{*}{Scrambling} &  \\
				&  &  \\ \cline{2-2}
				& \multicolumn{1}{l|}{\multirow{2}{*}{Interleaving}} &  \\
				& \multicolumn{1}{l|}{} &  \\ \hline
				\multicolumn{2}{|c|}{\begin{tabular}[c]{@{}c@{}}Compressive sensing based\\ \cite{hong2014sparsity,fazel2013random,alam2018survey,monsees2014reliable,monsees2015compressive,abebe2015compressive,wang2015compressive,needell2009cosamp,wang2016joint,tan2016compressive,shim2012multiuser,wang2016dynamic,vaswani2016recursive,chen2017sparsity,liu2017blind,abebe2017comprehensive}\end{tabular}} & \begin{tabular}[c]{@{}l@{}}\textbullet ~In UL grant-free mMTC, the inherent sparsity of user activities could be used to solve the MUD problem by using CS.\\ \textbullet ~Exploiting the low user activity ratio, CS techniques enable the eNB to handle more users.\\ \textbullet ~As blind MUD at eNB needs to jointly perform user activity and data detection in grant-free UL, this activity detection can\\ ~~~be achieved through CS-MUD by exploiting the sporadic nature of mMTC transmissions. \\ \textbullet ~Moreover, user-specific signature patterns enable the MUD to distinguish between these active users at the receiver.\end{tabular} \\ \hline
				\multicolumn{2}{|c|}{\begin{tabular}[c]{@{}c@{}}Compute-and-forward based\\ \cite{polyanskiy2017perspective,ordentlich2017low,nazer2011compute,bar1993forward,yang2017non,yang2016multiuser,goseling2015random}\end{tabular}} & \begin{tabular}[c]{@{}l@{}}\textbullet ~Users encode their messages with two concatenated channel codes; one for error correction, and one for user detection.\\ \textbullet ~The first, inner code, is to enable the receiver to decode the sum of all codewords.\\ \textbullet ~The second, outer code, is to enable the receiver to recover the individual messages of users that participated in the sum.\end{tabular} \\ \hline
				\multicolumn{2}{|c|}{\begin{tabular}[c]{@{}c@{}}Machine Learning based\\ \cite{du2018block,ye2019deep,ding2019sparsity}\end{tabular}} & \begin{tabular}[c]{@{}l@{}}\textbullet ~Inspired by powerful capabilities of ML to look for patterns in data for making best possible, nearly optimal, decisions. \end{tabular} \\ \hline
			\end{tabular}%
		}
	\end{table*}
	%\par
	%To facilitate this, two types of UL transmission procedures are shown in Figs. \ref{fig13a} and \ref{fig13b} respectively, where the basic difference is the control channel. In the first procedure of Fig. \ref{fig13a}, a dedicated control channel is transmitted with preamble and data channel to facilitate MUD. Contrary to this, the second procedure does not include a control channel. It is suggested in literature that both methods should be studied for 3GPP NR \cite{Intel2016Grant}. 
	\par
	%\begin{figure}[!t]
	%	\centering
	%	\includegraphics[width=2.5in,height=1.6in]{TransmissionFormat/Transmissionformat.pdf}
	%	\caption{Grant-free UL transmission with or without control channel}
	%	\label{fig11}
	%\end{figure}
	%\begin{figure*}[!t]
	%	\centering	
	%	\subfloat[Basic SCMA transmitter]{%
	%		\includegraphics[width=3.0in,height=1.55in]{SCMA/SCMAtransmitter.pdf}%
	%		\label{fig6a}
	%	}	
	%	\subfloat[Resource mapping of SCMA: 6 users, 4 subcarriers \cite{zte2016discussion1}]{%
	%		\includegraphics[width=4.0in,height=1.8in]{SCMA/SCMA.png}%
	%		\label{fig6b}
	%	}	
	%	\caption{Basic SCMA transmitter block diagram and resource mapping}
	%	\label{fig6}	
	%\end{figure*}
	\section{Grant-Free NOMA Schemes}
	We categorize the different
	grant-free NOMA approaches into three main classes: (1) MA signature based, (2) compressive
	sensing (CS) based, and (3) compute-and-forward (CoF)
	based. In what follows, we review the recent and
	significant works in each class. Moreover, the use of machine learning (ML) in grant-free NOMA is also discussed in next section. A summary of these schemes is provided in Table. \ref{tab4}. 
	\subsection{Signature based Grant-free NOMA Schemes}
		\begin{figure}[!t]
		\centering
		\includegraphics[width=3.5in,height=1.4in]{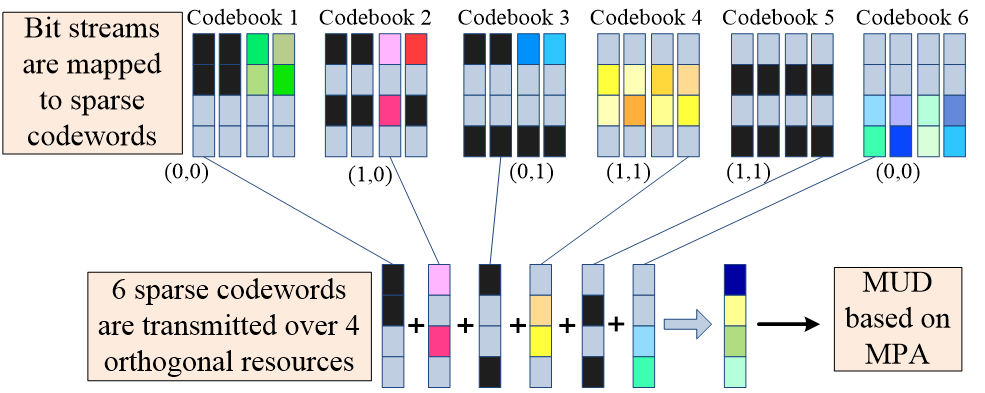}
		\caption{Resource mapping of SCMA: 6 users, 4 subcarriers \cite{zte2016discussion1}}
		\label{fig8}
	\end{figure}
	In grant-free UL NOMA, eNB may not have complete information about multiplexed users, with various other channel/user parameters also unknown/partially-known. To enable grant-free access, a MA resource is defined, which comprises of a physical resource (a time-frequency block) and a MA signature, which may include atleast one of the following; codebook/codeword, sequence, interleaver and/or mapping pattern, demodulation reference signal, power-dimension, spatial-dimension, preamble, etc \cite{catt2016consideration}. For MA signature selection, one option is that MTCD performs random selection and the other option is that MTCD's signature is pre-configured/pre-determined. Through these MA signatures, various grant-free NOMA transmissions can be enabled. Some prominent MA signature (i.e., spreading, scrambling, interleaving, and multiple domains) based grant-free UL strategies and MUD receivers are discussed in this section.  
	\begin{figure}[!t]
		\centering
		\includegraphics[width=3.5in,height=1.55in]{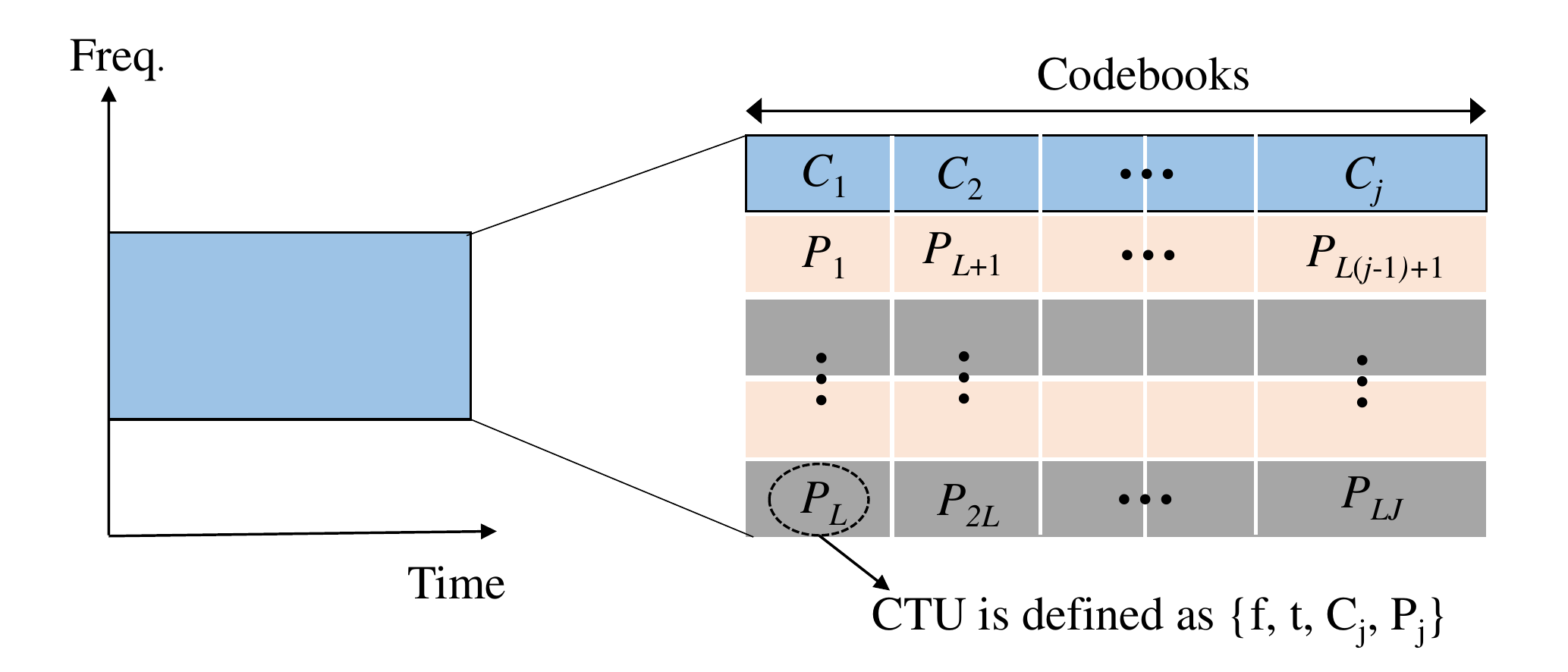}
		\caption{An illustration of CTU}
		\label{fig9}
	\end{figure}
	\subsubsection{Spreading based grant-free NOMA schemes:}
	Generally speaking, all MA signature-based NOMA schemes explained previously can be customized to achieve grant-free UL transmission. Consider spreading-based NOMA schemes for instance. The spreading code can be designed with either long or short sequence. Users randomly select the sequence from a predefined codebook set or a spreading sequence resource pool. To enable grant-free access, a contention-based unit (CTU) is defined. The CTU is a basic building block of a predefined region within the time-frequency plane for grant-free/contention-based transmissions, and may consist of several fields including radio resources, reference signals, and spreading sequences \cite{au2014uplink}. A CTU differs from others in any fields, and these differences can be exploited by receiver for efficient MUD. A MTCD, which has data to transmit, randomly selects a CTU and transmits its data packet accordingly. For mMTC, different MTCDs may choose the same radio resource, but different fields of the CTU still facilitate the eNB for efficient MUD. 
	\par
	To elaborate further, let us consider SCMA as an example. Developed using LDS-CDMA, the original bit stream of each user is directly mapped to a codeword chosen from its own dedicated codebook. SCMA codewords are sparse, i.e. only few of their entries are
	non-zero. The key difference between LDS-CDMA and SCMA is that SCMA relies on multidimensional constellations for generating
	its codebooks. All SCMA codewords have a
	unique location of non-zero entries. An illustration of resource mapping of SCMA is shown in Fig. \ref{fig8}, by considering 6 users, 4 resources, sparsity of 2 (each user transmits data over 2 out of 4 resources). The maximum number of codebooks $J$ that can be generated is a function of $N$ (non-zero entries in a codeword) and $K$ (codeword length). Selection of $N$ non-zero positions within $K$ elements is simply a combination problem. The maximum number of such combinations is given by the binomial coefficient $J={N\choose k}$. Hence, for the example shown in Fig \ref{fig8} with a 4-dimensional complex codebook ($K=4$) and 2 non-zero entries ($N=2$), a set of $J={4\choose 2}=6$ codebooks is generated, where each user selects one codeword from its codebook. Each user then maps its two bits ($b_1,b_2$) to that codeword. The data is then spread over the subcarriers. In this case, the data streams of multiple users are overlaid with codewords from different codebooks. As there are 6 possible codebooks, 6 users can be multiplexed over 4 subcarriers (150 percent loading). 
	\begin{figure}[!t]
		\centering
		\includegraphics[width=2.45in,height=2.75in]{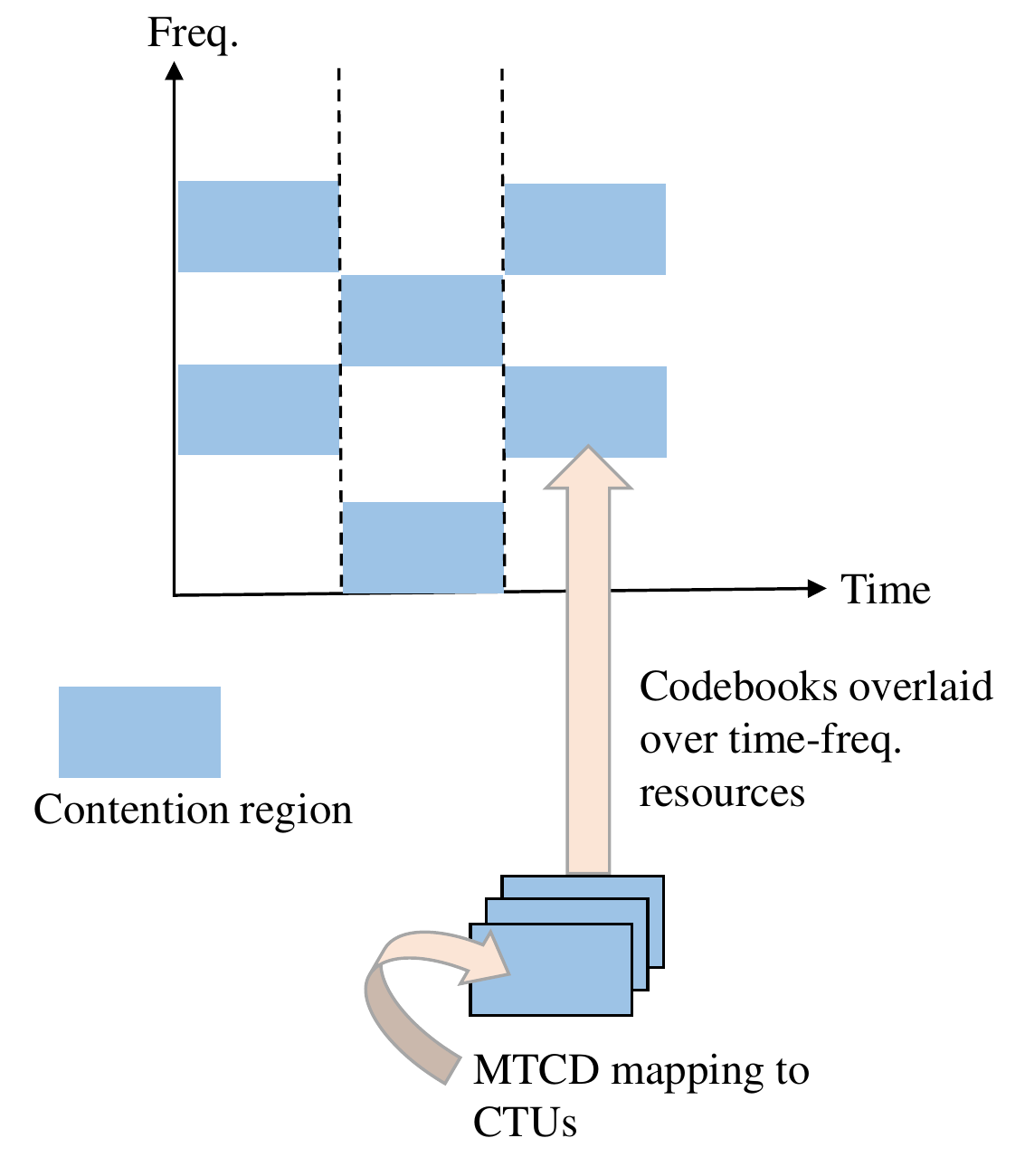}
		\caption{Uplink contention based regions}
		\label{fig10}
	\end{figure}
	\par
	In order to provide massive connectivity, a contention-based/grant-free SCMA scheme is proposed in \cite{au2014uplink}. In this context, a CTU is designed, as shown in Fig. \ref{fig9}. This CTU is a combination of time, frequency, SCMA codebook, and pilot sequence. There are $J$ unique codebooks defined over the time-frequency resources (RBs). For each codebook, there are $L$ associated pilot sequences, making it a total of $L \times J$ unique pilot sequences. In this way, there is a resource pool of $L \times J$ CTUs in the given time-frequency region. Multiple users/MTCDs may share the same codebook, and transmit at the same time. As codebooks go through different wireless channels, the MPA based detector can still detect these MTCDs data carried over the same codebook, as long as different pilot sequences are used. Moreover, the receiver can estimate channels of different MTCDs with different pilots. Therefore, the number of active users at a specific time slot can be potentially more than $J$. In this way, codebook reuse can help to increase the effective overloading factor and the number of connections to enable mMTC.
	\begin{figure}[!t]
		\centering
		\includegraphics[width=3.6in,height=2.75in]{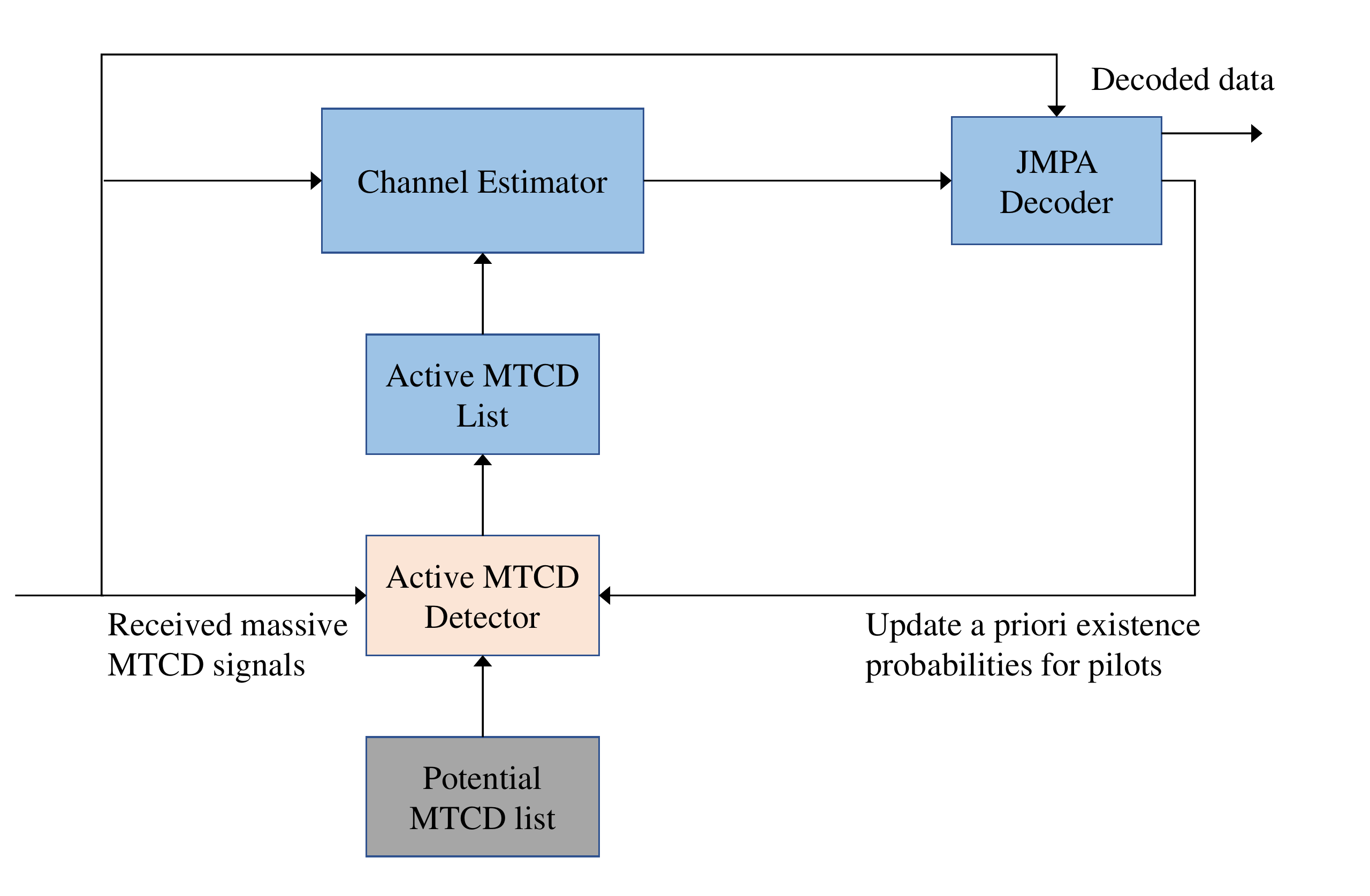}
		\caption{Uplink SCMA blind receiver}
		\label{fig11}
	\end{figure} 
	\par
	The user-to-CTU mapping rule can be defined, so that each MTCD can select the CTU itself. For instance, a MTCD can choose the CTU index as CTU$_{\text{index}}$=MTCD$_{\text{ID}}$ mod $N_{\text{CTU}}$; the CTU index that a MTCD chooses to transmits its data over is a function of the MTCD ID and the resource pool of CTUs $N_{\text{CTU}}$ \cite{au2014uplink}. In case different MTCDs choose the same CTU for transmission, a collision will occur, which can be resolved through random back-off procedure. 
	\par
	It is to note that the whole time-frequency region does not need to be contention supportive. For practical purposes, only a portion of UL bandwidth is configured as contention regions, while the other portion can be used for regular scheduled UL data transmissions. The coexistence of contention-based NOMA and scheduled access is supported by various studies in academia and industry. This is based on the fact, that in addition to the MTCDs, there are a comparatively smaller number of devices using eMBB services, where scheduled access has been shown to be quite efficient \cite{nokia2016basic}. The size and number of access regions are therefore dependent on many factors e.g., expected number of MTCDs and/or applications, etc. An example of CTU regions in time-frequency space is shown in Fig. \ref{fig10}. 
	\par
	In conventional SCMA, number of codebooks $J$ depends on spreading factor $K$ and non-zero entries $N$, i.e., $J={N\choose k}$. Hence, by varying the spreading factor and degree of sparsity, significant increase in the codebooks pool is possible. For instance, for $K=8$ and $N=4$, 70 codebooks can be generated. Correspondingly, the number of CTUs in the contention region increases manifold, thereby improving the performance of grant-free access.
	\par
	In order to perform MUD, different types of receivers are proposed in literature. In this context, performance comparison of SCMA with three types of receivers i.e., MPA, MPA with SIC, and MPA with Turbo decoder, is shown in \cite{qualcom2016RSMA}. Moreover, by considering the use of CTUs and massive number of MTCD transmissions using grant-free access, a blind detection based receiver is proposed in \cite{bayesteh2014blind}, as shown in Fig. \ref{fig11}. The receiver basically consists of two components, where the first one identifies active MTCDs to narrow down the list of potentially active MTCDs, and second one is joint data and active codebook (joint MPA - JMPA) detector to decode the active MTCDs with no knowledge of active codebooks. The first component (active MTCDs detector) acts as a pre-filter to narrow down the list of potential active MTCDs to control the complexity and efficiency of reception. This can be achieved through some efficient techniques detailed in the next section. Based on a short list created by active MTCD detector through pilot symbols, the channel estimator now needs to estimate only the channels of these identified active MTCDs. In addition to receiver, efficient design of codebooks for SCMA may also facilitate in enhancing the performance of grant-free transmission with efficient detection. In this context, a detailed analysis of variable overloading and robustness to codebook collision for SCMA is provided in \cite{Huawei2016LLS}.
	\par
	Similar to the case of SCMA, spreading sequences in the CTUs can reuse the sequences designed for other spreading based schemes such as LDS-SVE, PDMA, MUSA, etc., to facilitate grant-free transmissions. As it was mentioned earlier, a hard collision may occur if multiple MTCDs select exactly the same CTU. In these situations, these MTCDs may be distinguished and detected only if they have distinctive channel gains. However, from this perspective, an important solution is to enlarge the pool of spreading sequences, as done in MUSA, which is one of the candidate techniques for UL grant-free transmission \cite{zte2016discussion1}. Multiple random non-orthogonal complex spreading codes with short length constitute a pool in MUSA, from which each user/MTCD can randomly choose one. It is to be noted that for the same user, different spreading sequences may also be used for different symbols in order to improve the performance via interference averaging. All spreading symbols of MTCDs are transmitted over the same time-frequency resources. At the receiver, codeword level SIC is used to separate data from different users. Complex spreading sequences maintain lower cross-correlation than traditional pseudo random noise based sequences due to the additional freedom of the imaginary part. It should be noted that spreading sequences of MUSA are different from LDS and do not have low density property. 
	\par
\begin{figure}[!t]
		\centering
		\includegraphics[width=2.05in,height=3.10in]{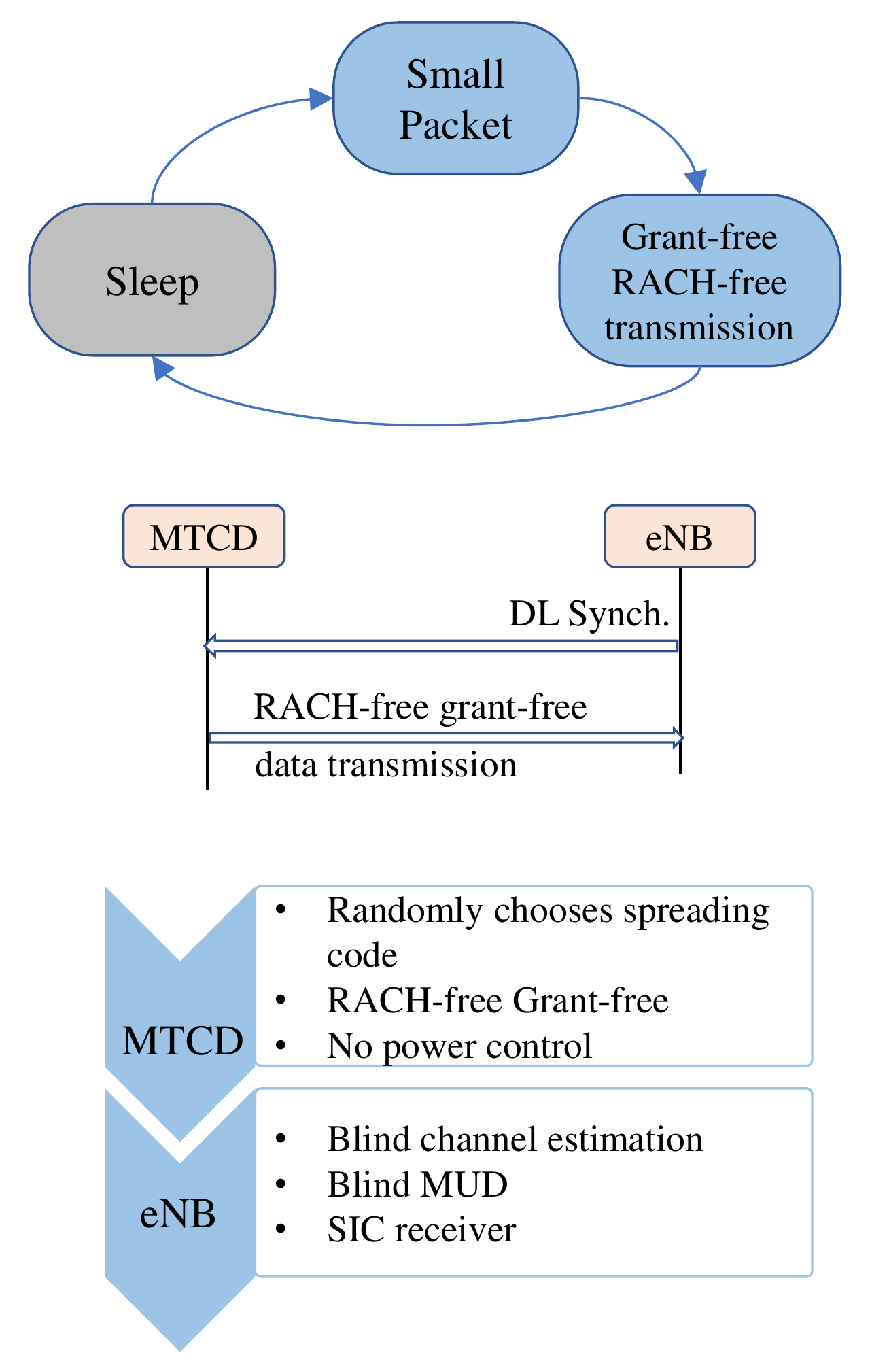}
		\caption{RACH-free Grant-free transmission based on MUSA}
		\label{fig12}
	\end{figure}

	As long spreading sequences used in traditional CDMA have relatively low cross-correlation, these codes combined with SIC for grant-free communication would cause processing complexity, delay and the error propagation in the receiver. Hence, short spread codes with relatively low cross-correlation are suitable for grant-free UL MUSA \cite{zte2016contention}. In this context, the family of complex spreading codes is a suitable option, as they are short due to the design freedom with real and imaginary parts. These spreading sequences are specifically designed to cope with heavy overloading of users, and to facilitate simple SIC on the receiver side. Moreover, in order to enable grant-free transmission and minimize the overhead of control signaling, users choose their spreading sequences locally/autonomously, without coordination by eNB \cite{zte2016discussion}. A RACH-free grant-free MUSA transmission model is shown in Fig. \ref{fig12}. Whenever a MTCD has data to transmit, it randomly chooses a spreading code, and transmits data without any RACH, GA, or power control. At the eNB, blind channel estimation and MUD using SIC receiver is performed.
	\par
	Overall, some practical challenges like collisions and receiver computational complexity are well addressed in MUSA. As the elements of spreading sequence are not binary, codewords do not need to be sparse, and more elements can be used by the spreading code in MUSA. In this way, a large pool of spreading codes can be generated, reducing the collision probability compared to SCMA and PDMA. Meanwhile, the complexity of blind MUD increases significantly as the pool size grows. Therefore, the pool size should be set to reasonable values in order to achieve massive connections while limiting the complexity of blind MUD \cite{zte2016discussion}. Hence, the design of spreading sequence is crucial to MUSA since it determines the interference between different users and system performance. Details about MUSA, its spreading process, and grant-free transmissions are provided in \cite{zte2016contention,zte2016System,zte2016System2}. 
	\par
	While most of these works consider an ideal SIC-based MUD to show the performance bounds of various spreading codes, a realistic receiver for grant-free MUSA is discussed in \cite{zte2016receiver,zte2016receiver2}. As mentioned earlier, the important assumptions for ideal receiver, such as the active users’ spreading codes and their fading channel may not be easily known at the BS before decoding. Fortunately, taking full advantage of the characteristics of grant-free accessing, e.g. inherent random near-far phenomenon and MUSA’s special features, e.g. short spreading code, blind MUD with very high performance is proposed and analyzed in \cite{zte2016receiver,zte2016receiver2}. The blind MUD investigated for MUSA consists of the following components. A SIC receiver is used to take full advantage of the near-far phenomenon usually observed in grant-free systems. Blind estimation is suggested by making full use of the characteristics of the spreading codes and the received signal, as conventional minimum mean-square error requires the much information which cannot be a prior known before the estimation in grant-free access. With blind estimation for the user with the highest post-SINR in current SIC stage, the decoding performance of the detected user can be guaranteed. Moreover, blind advanced channel estimation using pilot/reference signal is also suggested, where long preambles are not needed in the channel estimation to save the overhead. With the help of the advanced channel estimation and SIC, the post-SINR of the next detected user signal is enhanced and can be successfully decoded. It was shown through simulation results that the block error rate of MUSA, with blind SIC receiver, is smaller than 0.1 even with 300 percent user loading.
	\par
	Considering the coverage requirements for mMTC, repetition or secondary spreading can also be considered on the basis of short spreading. The secondary spreading could be orthogonal, and the spreading code can also be randomly selected to group users, similar to GOCA. Multiple receiving antennas can further improve the performance of MUSA due to the incremental freedom from spatial domain.
	%and includes the following three components; 1) SIC type receiver to take the full advantage of near-far phenomenon, 2) blind estimation by making full use of the characteristics of the spreading codes and the received signal, and 3) 
	\subsubsection{Scrambling/interleaving based grant-free NOMA schemes}
	The popular PD-NOMA with SIC receiver is one possible solution. However, the performance of PD-NOMA particularly relies on the power difference of multiplexed users/MTCDs over the same RB. The effectiveness of such a scheme is limited for the grant-free solutions in the absence of close-loop power control. The power differences of multiple users/MTCDs are out of control, since all (distant/close) near-far randomly distributed users/MTCDs can decide to transmit data at any time. In short, PD-NOMA generally needs a deterministic near-far situation. But in grant-free access, random near-far problem would make the SIC receiver for PD-NOMA difficult \cite{zte2016grant,shahab2016user,shahab2016power}. A detailed analysis of the effects of channel gains and power allocations of multiplexed users on their bit error rate and capacity for a PD-NOMA scenario is performed in \cite{shahab2018user}. It has been shown that the channel gains of users and their power allocation is critical for the performance of SIC receiver. However, some PD-NOMA based UL grant-free schemes do exist in literature such as integration of ALOHA or slotted-ALOHA protocols with PD-NOMA \cite{balevi2018aloha,choi2017nomaal,elkourdi2018enabling}. The eNB adaptively learns about the number of active devices using multi-hypothesis testing, and a novel procedure enables the transmitters to independently select distinct power levels. 
	\par
	For other scrambling based schemes, grant-free transmission and potentially asynchronous access using RSMA is proposed for UL in mMTC \cite{zte2016contention,qualcomm2016candidate}. Similarly, interleaving based NOMA schemes, and the schemes in multiple domains summarized in Table. \ref{tab1} and \ref{tab2} can also be customized to support grant-free communication. 
	\subsubsection{Other signature based grant-free NOMA schemes}
	Some other grant-free transmission schemes motivated by signature based NOMA are also proposed in academia. For instance, in \cite{shirvanimoghaddam2017massive}, a Raptor code based NOMA for mMTC is proposed with random data packet arrivals. In the considered system, MTCDs do not need to perform RA to get a transmission slot or network access. Instead, the RA and data transmission phases are combined over randomly selected subbands to minimize overhead. The eNB, being unaware of the number of MTCDs multiplexed over a subband performs load estimation and SIC for MUD. 
	\begin{figure}[!t]
		\centering	
		\subfloat[MTCDs transmitting over randomly chosen subbands/RBs]{%
			\includegraphics[width=3.0in,height=1.8in]{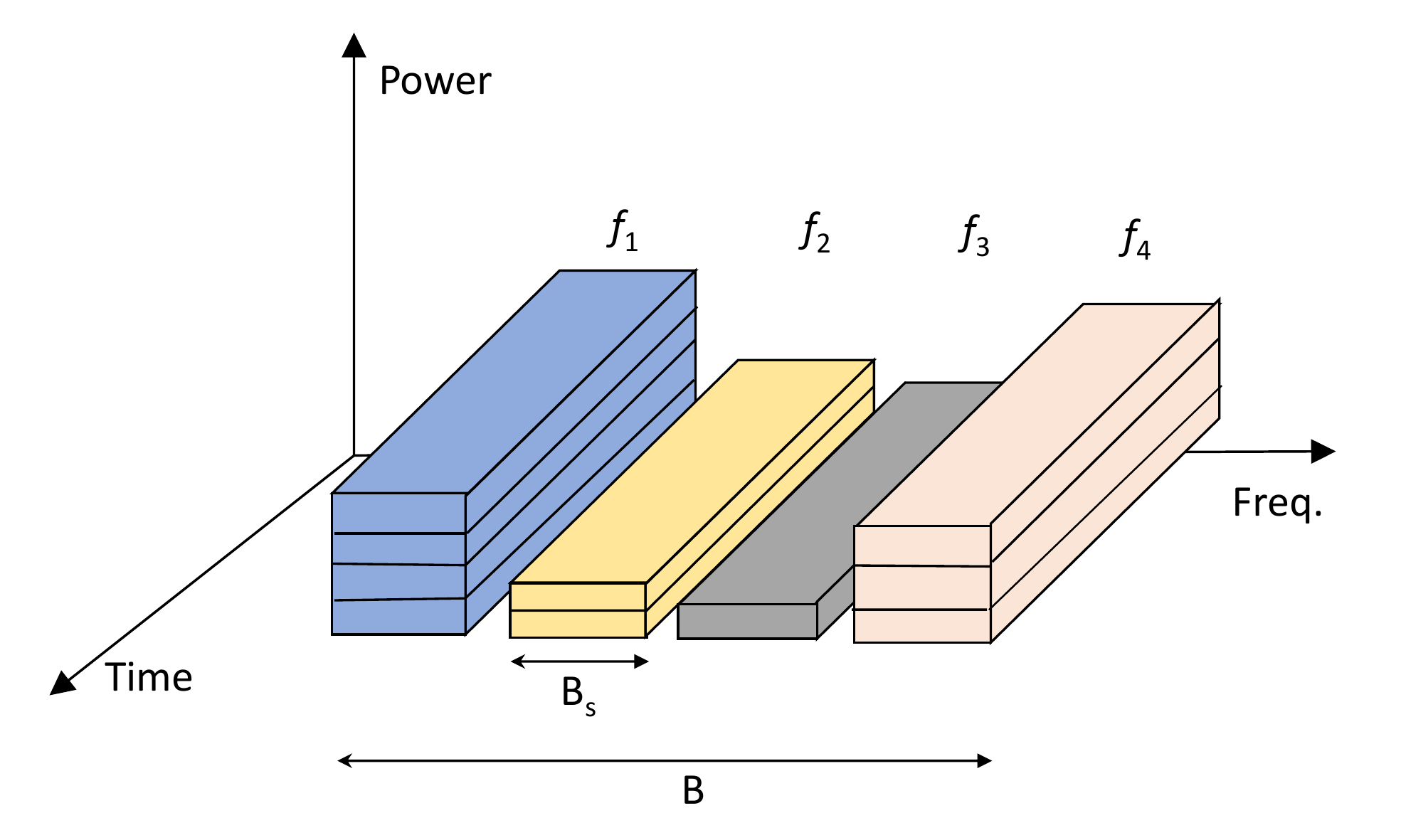}%
			\label{fig13a}
		}	
		
		\subfloat[Load estimation and SIC at BS for each subband]{%
			\includegraphics[width=2.6in,height=2.0in]{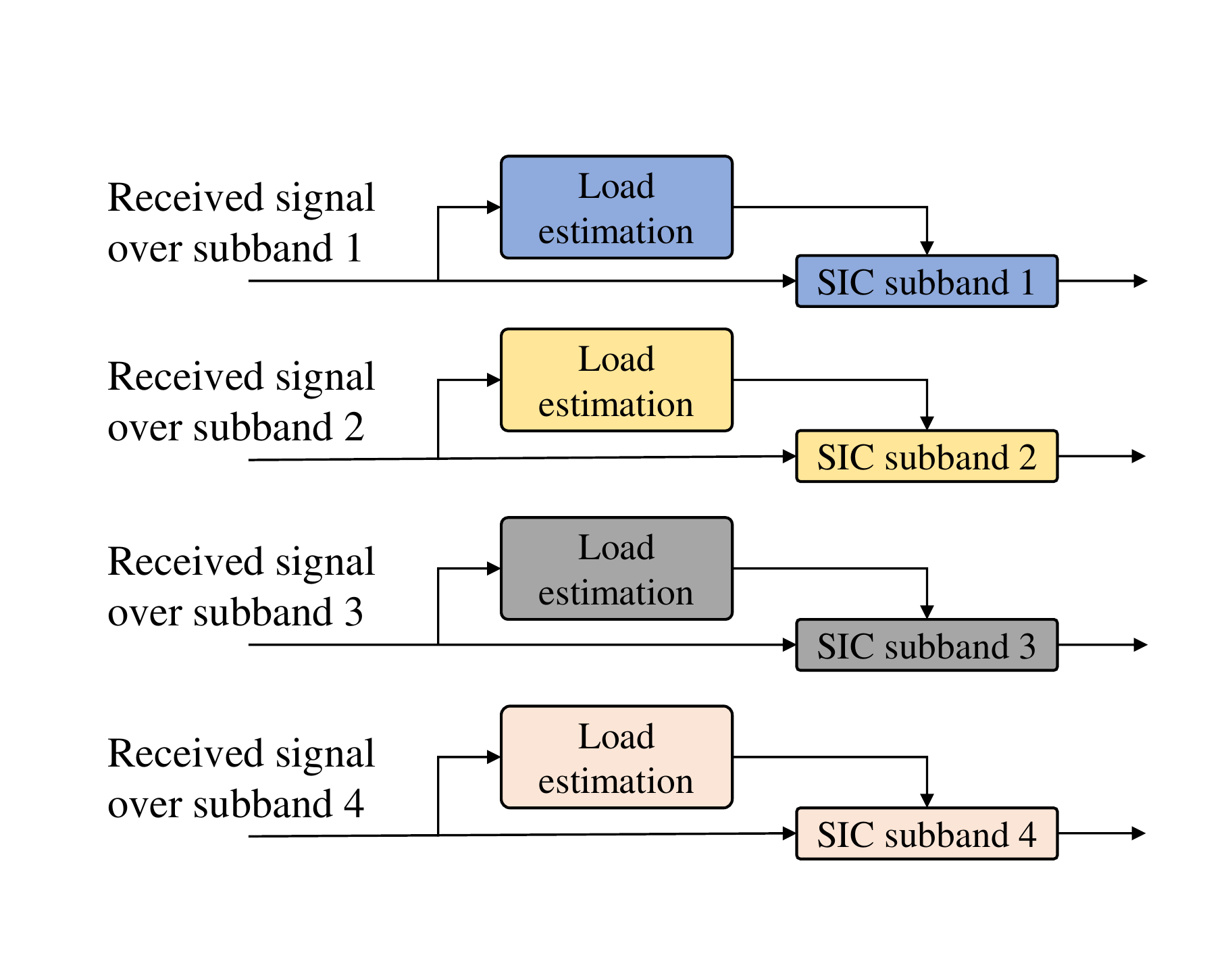}%
			\label{fig13b}
		}
		
		\subfloat[SIC process for users over a subband]{%
			\includegraphics[width=3.0in,height=2.05in]{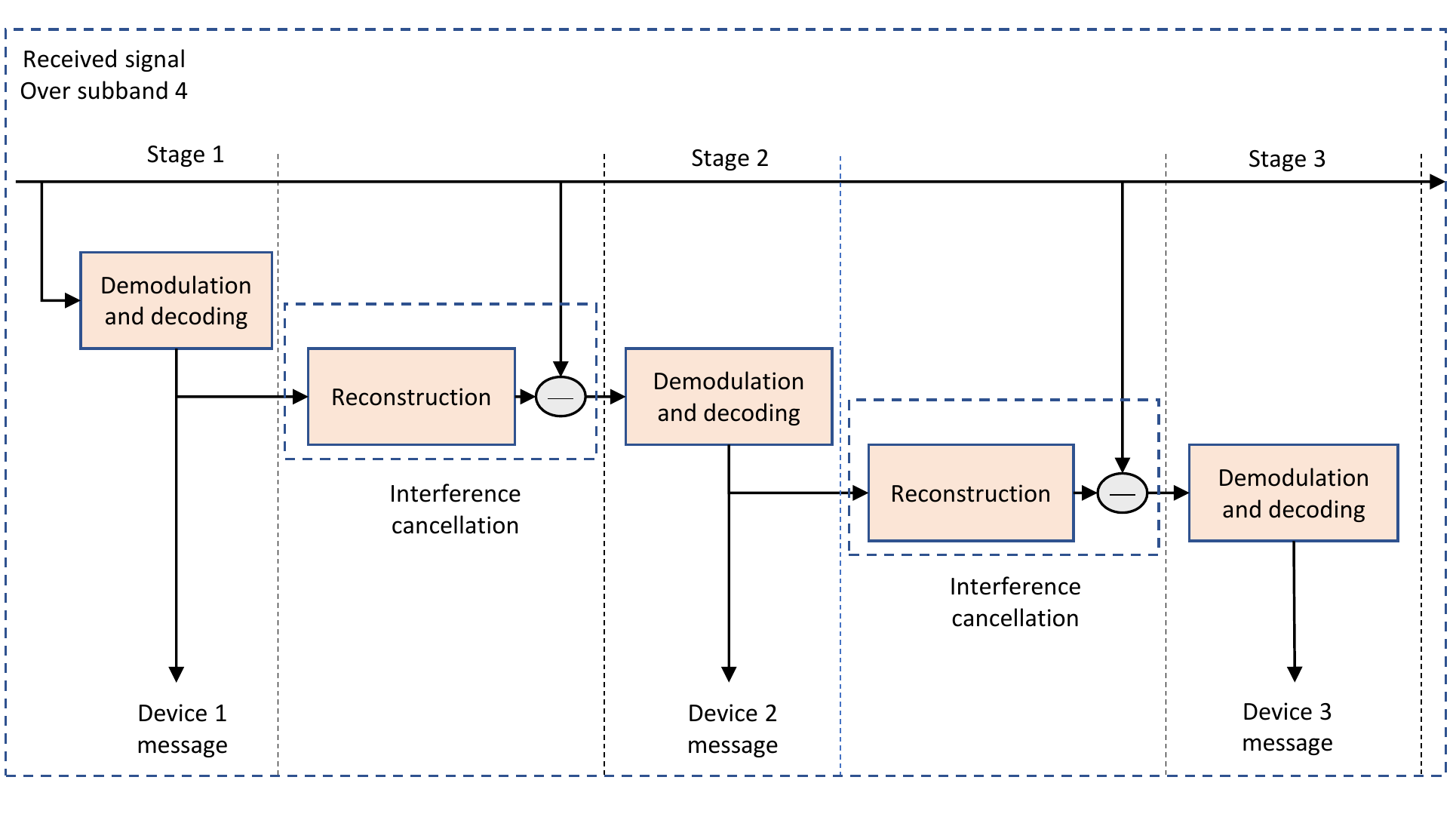}%
			\label{fig13c}
		}
		\caption{Illustration of grant-free Random NOMA for MTC}
		\label{fig13}	
	\end{figure}
	\par
	The steps and details of Random NOMA strategy are summarized as follows.
	\begin{enumerate}
		\item Initially, eNB broadcasts a pilot signal over each subband at the beginning of a time slot. 
		\item Each MTCD with data for transmission randomly chooses a subband from a set of available subbands, and listens to the pilot signal transmitted by eNB over that subband. Correspondingly, the MTCD estimates channel over that subband. Moreover, the MTCD also randomly chooses a seed for its random number generator from a set of $M_s$ available seeds.
		\item Each active MTCD, after randomly selecting a subband and seed, attaches its unique terminal ID to its message, and encodes using a Raptor code constructed from the selected seed, and transmit the codeword over the selected subband.  
		\item  When eNB receives superimposed message signals of various users over a subband, it performs load estimation followed by SIC for MUD. The SIC order is such that eNB starts decoding with the first seed and removes interference of it, followed by second seed and so on.
	\end{enumerate}
	An example scenario is shown in Fig. \ref{fig13a}, where bandwidth is divided into 4 subbands, and users randomly choose a subband for data transmission. The data retrieval at eNB is shown in Figs. \ref{fig13b} and \ref{fig13c}, where eNB performs load estimation (number of multiplexed users) for each subband, followed by SIC process to separate and recover the data of each user over that subband. All MTCDs are assumed to perform power control such that their received power at the eNB is same. In this way, eNB can effectively estimate the number of MTCDs multiplexed over each subband by calculating the total received power, as it would be proportional to the number of overlapped MTCDs. Further details about load estimation algorithms can be found in \cite{shirvanimoghaddam2017massive2}. 
		\begin{figure}[!t]
		\centering	
		\subfloat[Over a single subband]{%
			\includegraphics[width=2.25in,height=0.85in]{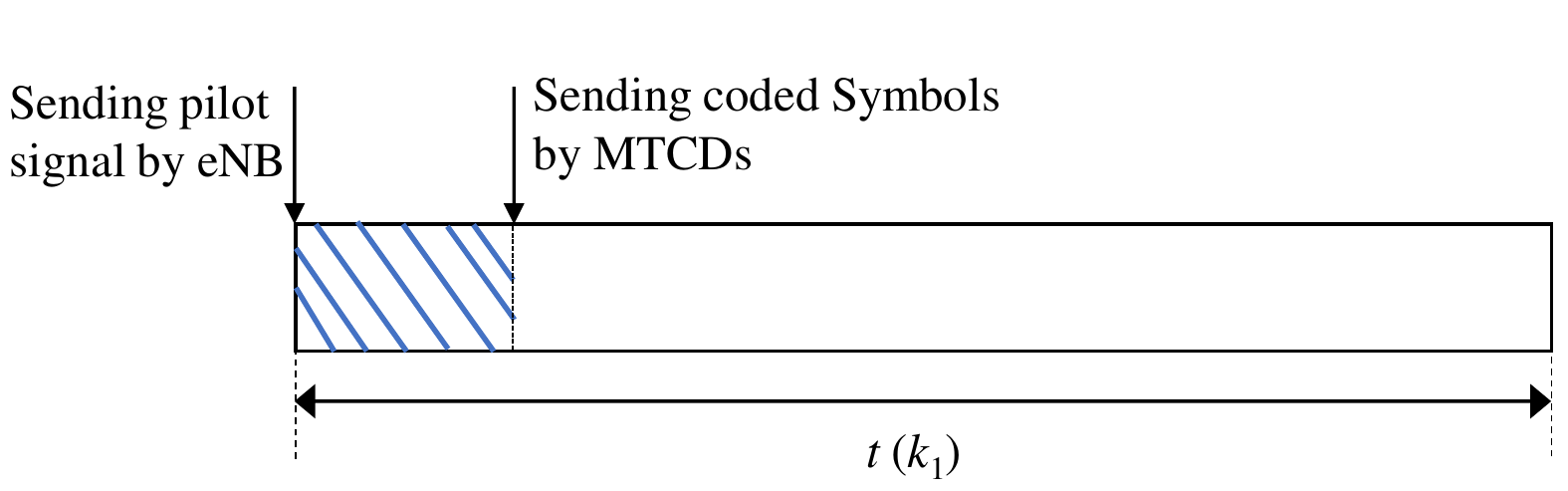}%
			\label{fig14a}
		}	
		
		\subfloat[Over two consecutive time slots]{%
			\includegraphics[width=3.5in,height=2.15in]{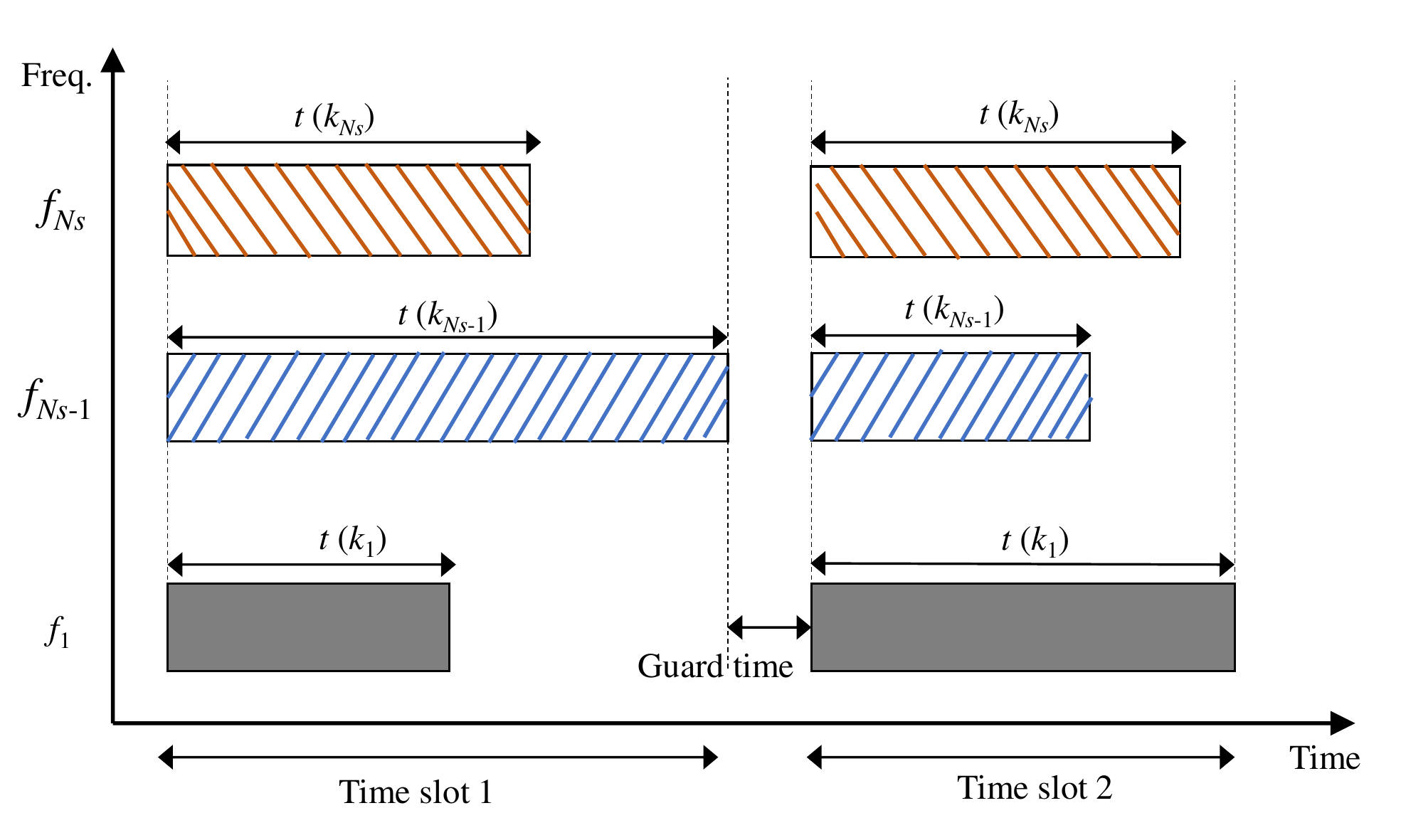}%
			\label{fig14b}
		}	
		\caption{Time slot duration of grant-free Random NOMA}
		\label{fig14}	
	\end{figure} 
	\par
	Due to random subband selection by MTCDs, the achievable rate over each subband is not fixed and depends on the number of these randomly overlapped MTCDs. Therefore, the number of coded symbols to be transmitted over each subband is also random. To elaborate this, Fig. \ref{fig14} shows the variable length of each sub-band in two consecutive time slots. It is important to note that each time slot duration will be mainly determined by the subband with highest number of active MTCDs, as its maximum achievable rate would be lower than the rest. This means that a fixed-rate code cannot be used in all time instances. Accordingly, use of Raptor codes is proposed in \cite{shirvanimoghaddam2017massive}.
	\par
	Raptor codes are rateless and can generate as many coded symbols as needed by eNB. They have a random structure represented by a bipartite graph, which depends upon a pseudo random generator's seed. By using the same seed, eNB can reproduce the same bipartite graph as that of the MTCD and decode the message. In case multiple MTCDs select the same seed while transmitting over the same subband, their transmitted code structure will be exactly the same. In this case, a collision is said to have taken place, as the eNB cannot differentiate between these users due to same structure of received codewords. However, the probability of such collision is still far less than the conventional RA collision. The work in \cite{shirvanimoghaddam2017massive} compared the average number of successful supported MTCDs by the random NOMA scheme with access class barring. It was shown that the proposed scheme supported significantly larger number of devices compared to access class barring. 
	\par
	Improvisations in random NOMA may completely avoid collisions. For instance, assume that each device always uses a unique user-specific seed determined in advance, then the collisions can be completely eliminated. However, this may still be complicated when number of MTCDs is incredibly large; need for massive seed pool and complex detection process at eNB by trying such massive seeds for data recovery of multiplexed users over each subband. The performance limits of random NOMA scheme for massive cellular IoT are comprehensively discussed in \cite{shirvanimoghaddam2017fundamental}. In \cite{abbas2018novel}, a novel framework for grant-free NOMA is developed, where collisions between any number of users over a radio resource does not entail for all simultaneously transmitting multiplexed users, and is treated as interference to the remaining received signals. Moreover, as PD-NOMA is less discussed from a grant-free UL perspective, a novel multi-level grant-free scheme is proposed in \cite{abbas2018multi}. The layers correspond to different codebooks and different power levels. The users first choose a layer randomly, which corresponds to a codebook. Users in each layer utilize the same codebook and perform power control such that they have the same received power level at the eNB. To achieve this, users choose between a predetermined set of power levels. At the receiver, the sum of codewords over each layer
	are first separated from remaining layers, symbol by symbol.
	Then, the codewords of the same layer are jointly decoded at
	the eNB.
	\begin{figure}[!t]
	\centering
	\includegraphics[width=2.3in,height=1.9in]{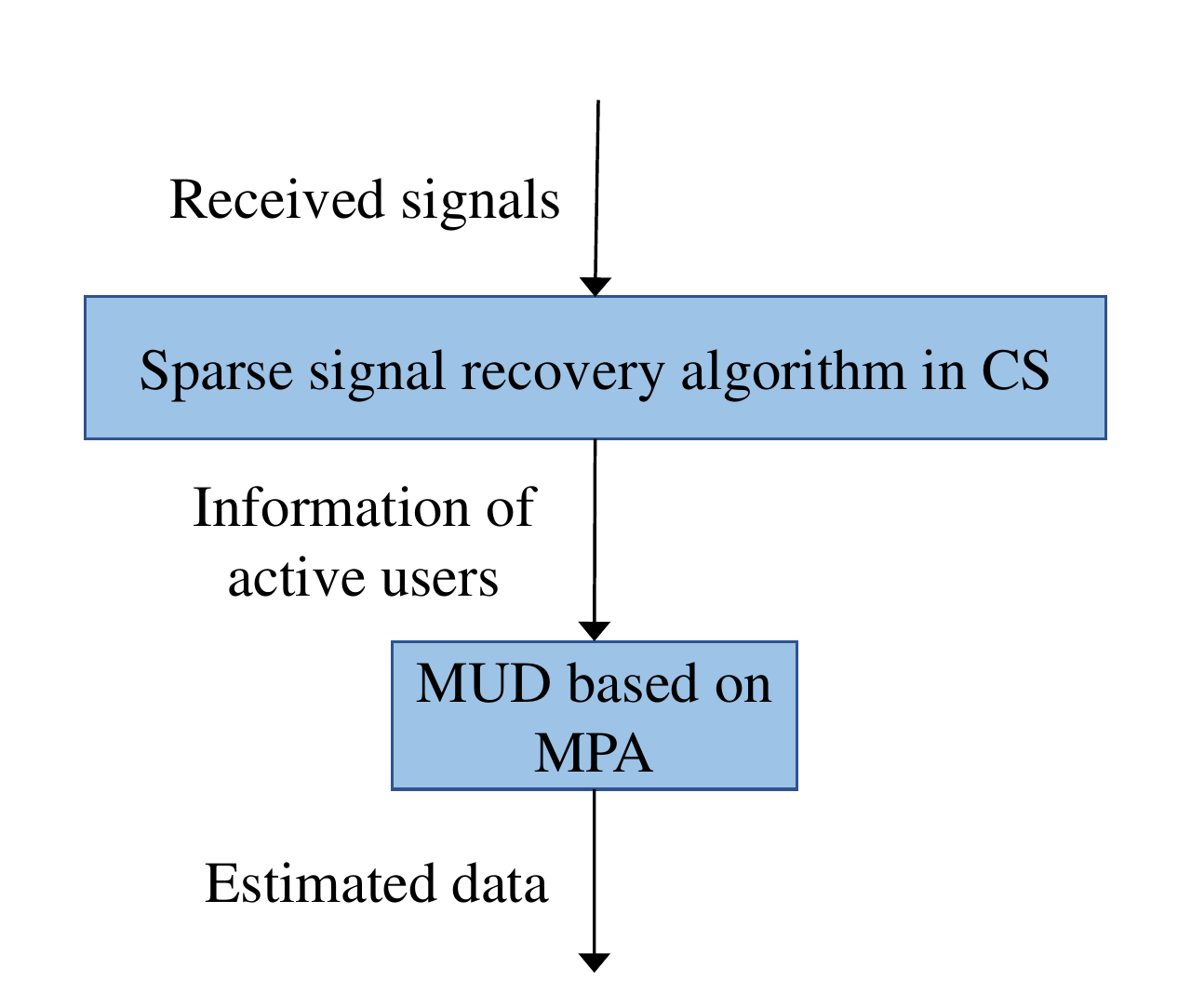}
	\caption{CS-based blind MPA detector}
	\label{fig15}
\end{figure}
	\subsection{Compressive Sensing based Grant-free NOMA Schemes}
	\label{sec5sub2}
	%\begin{figure}[!t]
	%	\centering
	%	\includegraphics[width=3.25in,height=2.5in]{CS-MUD/MassiveTransmission.pdf}
	%	\caption{mMTC scenario considered for CS-MUD}
	%	\label{fig20a}
	%\end{figure}
	%\begin{figure}[!t]
	%	\centering
	%	\includegraphics[width=3.25in,height=2.5in]{CS-MUD/SV-CS.png}
	%	\caption{Compressed sensing based blind MPA detector}
	%	\label{fig20b}
	%\end{figure}
		\begin{figure*}[!t]
		\centering
		\includegraphics[width=6.4in,height=3.25in]{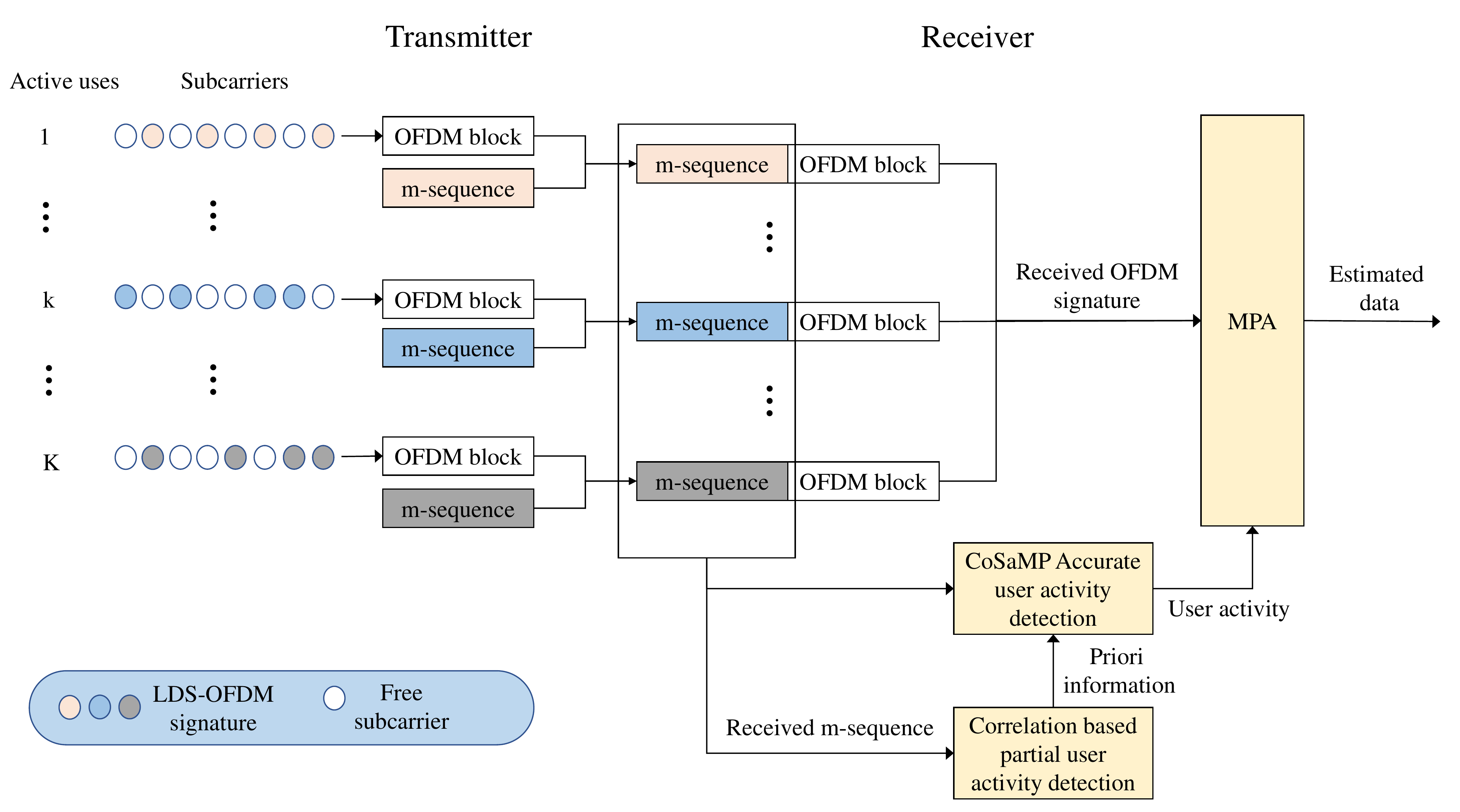}
		\caption{CS-based time-frequency joint NOMA transceiver}
		\label{fig16}
	\end{figure*}
	CS is an efficient signal processing technique that exploits the sparsity of a signal to recover it from far fewer samples than required by the Nyquist criteria. 
	In current wireless systems, the number of active users is usually much smaller than the total number of available users in the system, even during busy hours \cite{hong2014sparsity}. This characteristic also applies to mMTC scenarios. Thus, in UL grant-free NOMA systems, the inherent sparsity of user activities could be used to solve the MUD
	problem by using CS algorithms. Exploiting the low user activity ratio, CS techniques enable the eNB to handle more users \cite{fazel2013random}, because CS makes it possible to recover the desired signals from far fewer measurements than the total signal dimensions if signal sparsity is assumed.
	\par
	NOMA with CS-MUD is considered as a promising candidate to enable grant-free UL NOMA for mMTC. As blind MUD at the eNB needs to jointly perform user activity and data detection in grant-free UL, this activity detection can be achieved through CS-MUD by exploiting the sporadic nature of mMTC transmission, allowing the CS based algorithms to detect user activity. Moreover, user-specific signature patterns enable the MUD to distinguish these active users at the receiver. CS-MUD based NOMA schemes allow the users/MTCDs to transmit their data whenever they want without any control signaling. 
	\par
	Considering a fewer number of active users compared to the actual number, two models are used in the literature to formulate NOMA as a CS problem. The first model is termed as a single measurement vector based CS (SMV-CS), where a one shot transmission is considered by taking the received signals as a vector $\textbf{y}$, which consists of superimposed signals of active users. The received vector $\textbf{y}$ is product of a vector $\textbf{d}$ consisting of data symbols of the users, and a sensing matrix $\textbf{A}$, which contains the influences of channel and spreading matrices. In SMV-CS, when the number of users increases, the size of this sensing matrix $\textbf{A}$ becomes huge, which leads to poor sampling matrix properties and hence limits the scalability of the system in terms of detection speed. To deal with this, multiple measurement vector based CS (MMV-CS) considers the received signal as a matrix $\textbf{Y}$, which is a product of two matrices, i.e, a sensing matrix $\textbf{A}$ and a data symbols matrix $\textbf{D}$, where each row of $\textbf{D}$ contains data frame of a single user and each column represents the symbols from all users at a time instant. This is done to reduce the size of the sensing matrix $\textbf{A}$. Hence, compared with the SMV-CS model, MMV-CS can better mitigate higher complexity due to the growing number of users. These models have been actively used to represent the sparse spreading-based NOMA in UL mMTC scenarios \cite{monsees2014reliable,monsees2015compressive,abebe2015compressive}. Moreover, in order to perform CS-MUD, various algorithms exist in literature. These algorithms can be categorized as maximum a posteriori probability (MAP) based algorithms and greedy algorithms \cite{alam2018survey}, and are used frequently for CS-MUD.
	%\par
	%Some of the MAP based algorithms are sparsity aware MAP, sphere decoding based MUD, and K-best detector for sphere decoding. Similarly, some of the greedy algorithms are orthogonal least square and orthogonal matching pursuit, iterative order recursive least square, and structured matching pursuit.
	\par
	In the UL grant-free NOMA systems, the current near-optimal MUD based on MPA was discussed earlier to approximate the optimal MAP detection. The receiver assumes that the user activity information is exactly known, which is impractical yet challenging due to any of the massive users randomly entering or leaving the system. In this context, a joint use of CS and MPA for designing a CS-MPA detector to realize both user activity and data detection for LDS based UL grant-free NOMA is proposed in \cite{wang2015compressive}. The sparse signal recovery algorithms in CS can be used to realize user activity detection by identifying the positions of non-zero elements, for which a compressive sampling matching pursuit (CoSaMP, \cite{needell2009cosamp}) algorithm is used due to its low complexity and excellent robustness to noise. CoSaMP is one of the many CS algorithm, and is a common method applied to detect nonzero elements in a sparse signal by using the intrinsic sparsity of signals. After the activity detection through CoSaMP, by making full use of the sparsity of LDS structure, low complexity MPA based receiver is employed. A basic working principle of CS-MPA is shown in Fig. \ref{fig15}. Similarly, by observing a structured
	sparsity of user activity in mMTC, a low-complexity MUD based on structured compressive sensing for NOMA to further improve the signal detection performance is proposed in \cite{wang2016joint}. 
	\par
	In \cite{tan2016compressive}, the CS-MPA of \cite{wang2015compressive} is optimized by incorporating a two stage CS based activity detection for a LDS-OFDM UL grant-free NOMA, as shown in Fig. \ref{fig16}. In the first stage, a correlation-based activity detection is carried out. The approximated support obtained at the first stage is then fed into the second stage which executes the CoSaMP algorithm. In addition to these MUD receivers, by considering the variations in the sparsity level of the multi-user signals, a switching mechanism between the CS-MUD and the classical MUD can be adopted, as proposed in \cite{shim2012multiuser}. Similarly, as the sparsity of active users varies from time to time, a low complexity dynamic CS based MUD is proposed in \cite{wang2016dynamic}. This is based on the idea that although users
	can randomly access/leave the system, some users generally transmit their information in adjacent time slots with a high probability, which leads to the temporal correlation of active user sets in several continuous time slots \cite{vaswani2016recursive}. By exploiting this temporal correlation, the estimated active user set in a particular time slot is used as the initial set to estimate the transmitted signal in the next time slot in the dynamic CS-based MUD. 
	\par
	In addition to CS, the user activity detection could be realized by different algorithms and schemes. Three different algorithms are developed for active pilot detection in this context i.e., channel estimation-based algorithm, focal underdetermined system solver (FOCUSS), and expectation maximization (EM) \cite{bayesteh2014blind}. These algorithms can also be combined with blind detection, as shown earlier in the case of JMPA receiver. Furthermore, a sparsity-inspired sphere decoding (SI-SD) based blind detection algorithm for grant-free UL SCMA is proposed in \cite{chen2017sparsity}. By introducing one additional all-zero codeword, each user's status and data can be jointly detected, thus avoiding the redundant pilot overhead, and achieving the MAP detection. Similarly, an improved detection-based group orthogonal matching pursuit (DGOMP) MUD is proposed in \cite{liu2017blind} to facilitate massive grant-free UL SCMA transmission and reception. Moreover, in \cite{abebe2017comprehensive}, a comprehensive study is provided where synchronization, channel estimation, user detection and data decoding are performed in one-shot. 
	%It is evident from this discussion that CS based NOMA schemes are very promising candidates for enabling grant-free UL mMTC with efficient blind MUD.
	\subsection{Compute-and-forward based Grant-free NOMA Schemes}
	Another approach towards grant-free NOMA was proposed in \cite{polyanskiy2017perspective,ordentlich2017low}, where the principle of CoF \cite{nazer2011compute} is employed. 
	%As the name indicates, the CoF principle was actually proposed in a relaying network for enabling the relays to decode linear equations of the transmitted messages using the noisy linear combinations provided by the channel \cite{polyanskiy2017perspective}. A destination device, given sufficiently many linear combinations, could solve for its desired messages.
	The scheme relies on codes with a linear structure, specifically nested lattice codes. The linearity of the codebook ensures that integer combinations of codewords are themselves codewords. A destination is free to determine which linear equation to recover \cite{polyanskiy2017perspective}. The concept of network coding in CoF can be interpreted as a conversion of a network into a set of reliable linear equations. Inspired by this, in CoF based grant-free NOMA, users encode their messages with two different channel codes where one is used for error correction and the other is used for user detection. The latter is chosen such that the sum of $K$ or less distinct codewords is unique. Correspondingly, at the receiver side, the eNB has to first decode the sum of the received codewords. In situations when the eNB cannot recover the sum correctly, all the transmitted data is lost, which is similar to the existing MA signature-based grant-free NOMA schemes discussed earlier. 
	\par
	A channel use is divided into multiple sub-blocks, where each active user randomly chooses a sub-block, over which it transmits. All users encode their messages using the same codebook $C$, which are then modulated. The codebook $C$ is developed as a concatenation of two codes. The first one is an inner binary linear code, whose purpose is to enable the receiver to decode the modulo-2 sum of all codewords transmitted within the same sub-block, which can be referred as the CoF phase. The second code is an outer code, whose purpose is to enable the receiver to recover the individual messages of users that participated in the modulo-2 sum. This recovering of the individual messages from their modulo2 sum can be referred as the binary adder channel (BAC) phase. The success probability of the CoF phase in \cite{ordentlich2017low} is independent of the actual number of users that transmitted within the same sub-block. However, the outer code in \cite{ordentlich2017low} is designed such that if at most $M$ users use the channel over a particular sub-block, it is possible to determine the individual messages from their modulo-2 sum, essentially with zero error probability.
	\par
	The design of an inner code to be used in the CoF phase reduces to that of finding codes that perform well over a binary input memoryless output-symmetric channel, for which
	many off-the-shelf codes can be used. For the outer codes used in the second phase/stage, they can be constructed from the columns of $T$-error correcting Bose-Chaudhuri-Hocquenghen (BCH) codes \cite{bar1993forward}. As mentioned, at the receiver side, the eNB has to first decode the sum of the received codewords. When the eNB cannot recover the sum correctly, all the transmitted data is lost which is similar to the existing works on signature-based grant-free NOMA schemes. Moreover, in CoF based grant-free NOMA \cite{polyanskiy2017perspective,ordentlich2017low,nazer2011compute}, a closed loop power control is used with the extension to open-loop power control being not straight-forward. A similar approach with similar challenges is considered in \cite{yang2017non,yang2016multiuser,goseling2015random} using physical layer network coding where users transmit to a multi-antenna eNB.
	\section{Machine Learning in Grant-free NOMA}
	Recent research continues to confirm the powerful capabilities of ML technologies \cite{bacstanlar2014introduction} in enhancing the efficiency of transmitter/receiver designs in wireless communications. ML can solve NP-hard optimization problems in a faster, more accurate, and more robust manner than traditional approaches. Instead of relying on models and equations, ML algorithms look for patterns in the data to make the best possible, nearly optimal, decisions. The robustness of ML algorithms is especially desirable in wireless communications due to the dynamic nature of the networks, whether it is the fast-changing channel states, the dynamic network traffic, or even the network topology and scheduling. For NOMA systems, ML can be applied to several of its NP-hard problems that include (1) attaining channel state information, (2) resource allocation, (3) power allocation, (4) clustering, (5) complex joint decoding and (6) the fundamental trade-offs among them. This is especially useful in massive NOMA, as the complexity of these processes grows exponentially with the number of users.
	\par
	Before we proceed to survey the literature, it is of value to briefly explain some important terminology:
	\begin{itemize}
		\item \textit{Supervised/Unsupervised learning:} Supervised learning aims to learn the mapping function between input data and its respective output (called its label) by minimizing the function approximation error. On the other hand, unsupervised learning aims to extract the inner features of unlabelled data. One example of the latter would be the autoencoder.
		\item \textit{Reinforcement learning:} Reinforcement learning is based on a reward system where the learner is trained through multiple trial and error attempts to maximise rewards. Video games are a popular example of this paradigm.
		\item \textit{Deep learning:} Deep learning is a subset of ML that is capable of unsupervised learning from data that is unstructured or unlabelled. It is also known in the literature as deep neural networks.
		\item \textit{Online/Offline learning:} Offline learning is often also referred to as batch learning. In batch learning, the parameters are updated after consuming a whole batch of data. On the other hand, in online learning, the parameters are updated after each training data learnt.  
	\end{itemize}
	Although the concept of leveraging ML algorithms to solve communications-related problems can be dated back to almost 20 years ago, it seemed to have little, and possibly no, impact on the way communications was designed and implemented until now. We suspect that the main reason is that information theory, statistics and signal processing offered very accurate and often flexible models, that allowed for designs with reliable, analytically proven, performance guarantees. Nonetheless, with the development and wide spread of specialized software libraries and as well as cheap processing chips, training and testing machine learning models has become much more attractive.
	\par
	For the physical layer, learning algorithms have been applied to demodulation \cite{o2016convolutional}, channel codes \cite{o2016learning}, BP for channel decoding \cite{nachmani2016learning}, one-shot channel decoding \cite{gruber2017deep}, CS \cite{borgerding2016onsager}, coherent \cite{samuel2017deep}, and blind \cite{jeon2017blind} detection. For higher layers, learning algorithms have been applied to traffic load prediction, control and channel access schemes \cite{farsad2017detection,o2016unsupervised,goodfellow2016deep,pan2009survey,mitzenmacher2009survey}. Learning algorithms change the way we fundamentally solve communication problems. Traditionally, to find optimal solutions, our transmitters and receivers are divided into several independent blocks e.g. coding, modulation etc. Similarly, the overall communication system is divided into layers that perform tasks independently, e.g. routing, scheduling, resource allocation etc. Such a division, though sub-optimal, allows for the separate analysis and optimisation of tasks and eventually to stable systems. However, the approach of learning technologies disrupts this entire process by attempting to accomplish more than one task at once. For example, authors in \cite{dorner2017deep} went as far as demonstrating that it is possible to build a point-to-point communications system whose entire physical layer processing is carried out by neural networks. The work in \cite{liu2019uav} is another example of this, where UAV deployment design and trajectory amongst other aspects were optimized jointly through ML. 
	\begin{figure*}[!t]
		\centering
		\includegraphics[width=7.0in,height=2.5in]{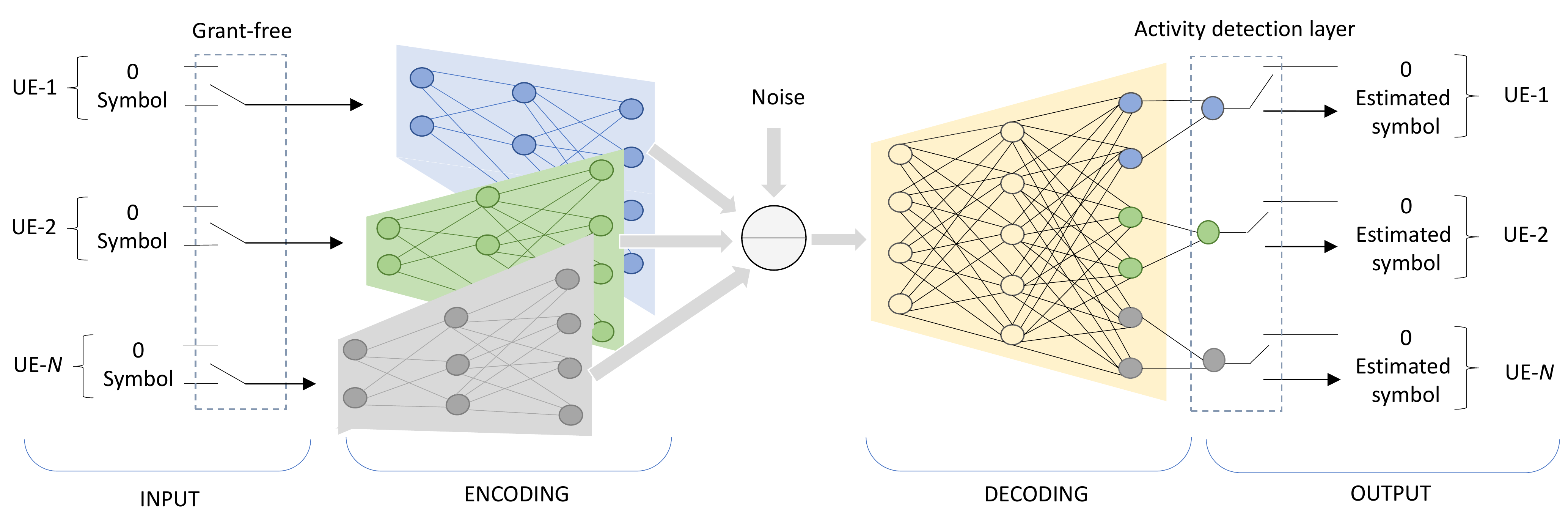}
		\caption{Detailed network structure for deep learning aided grant-free NOMA, as in \cite{ye2019deep}}
		\label{fig17}
	\end{figure*}
	\par
	Here, we review the few, yet growing number, of ML based papers that can be found in the literature on NOMA thus far. We cover both grant-based as well as grant-free NOMA.
	\par
	\subsection{Machine learning in grant-based NOMA:} 
	In \cite{gui2018deep}, authors use deep learning to solve the high computational complexity of traditional channel estimation and detection algorithms that is mainly due to the fast-changing wireless channel. Authors use long short-term memory (LSTM), a branch of recurrent neural networks commonly used for natural language processing, to perform automatic encoding, decoding and channel detection in a DL NOMA system with randomly deployed users. LSTM is made to learn the channel characteristics between the base station and each user for better power allocation. Results show performance enhancements in terms of sum rates as well as block error rates.
	\par
	In \cite{kim2018deep}, the authors apply deep learning to SCMA to facilitate codebook design which is notoriously difficult due to its high-dimension. The performance of SCMA heavily depends on the codebook chosen. These codebooks are usually hand-crafted in a case by case manner. The devised deep neural network can construct codebooks that minimize the block error rate and can adapt to the available set of resources. The authors also use deep neural networks to perform single shot decoding. Results show that the performance is very close to MPA and scores an order of magnitude improvement over conventional SCMA at high signal-to-noise ratio.
	\par
	In \cite{awan2018detection}, authors proposed an online learning detection algorithm for UL NOMA with SIC, where devices are clustered. Traditionally,  non-linear beamformers are highly sensitive to variations in the environment and their performance may deteriorate in dynamic networks where devices may join sporadically. Here authors propose a partially linear beamforming filter design. ML is more robust to cluster sizes even with perfect channel state information due to its ability to detect error propagation in SIC. In \cite{ liu2018deep}, authors use deep recurrent neural networks learning to obtain energy efficient resource allocation for heterogeneous cognitive radio networks between IoT and mobile users. In \cite{nguyen2018power}, deep learning was used for power allocation of caching based NOMA in three phases: exploration, training, and exploitation. In \cite{luo2019deep}, deep learning was used for the joint DL resource allocation problem for a multi-carrier NOMA system. Lastly, an unsupervised ML approach is  proposed for NOMA in \cite{cui2018unsupervised} to solve the clustering problem. 
	\subsection{Machine learning in grant-free NOMA:}
	The work in \cite{du2018block} aims at solving the MUD problem of UL grant-free NOMA through CS without relying on any a priori knowledge, namely the user sparsity level or noise level. The latter two are usually required for the correct termination of the recovery process. Instead, authors adopt statistical and ML mechanism cross validation (CV) to determine the user sparsity level, mathematically referred to as the model order, and eventually to decide when recovery needs to be terminated. Results show that the proposed algorithm well avoids overfitting and underfitting.
	\par
	In \cite{ye2019deep}, deep learning is used to solve a variational optimization problem grant-free NOMA scheme. The neural network model covers encoding, user activity, signature sequence generation and decoding. Simulation results show that the process can be of very low latency to suit tactile IoT applications. The detailed network structure for deep learning aided grant-free NOMA is shown in Fig. \ref{fig17}.
	\par
	In \cite{ding2019sparsity}, authors propose two MUD schemes, namely, random sparsity learning MUD and structured sparsity learning for synchronous and asynchronous transmissions, respectively. Authors show that even when users do not use pilot signals, the proposed algorithms demonstrates low error rates. Thus, the proposed algorithms can significantly reduce overhead in grant-free access scenarios.
	\par
	\subsection{The Way Forward}
	In general, the research work presented thus far in the literature demonstrates very promising performance enhancements to existing systems in terms of faster processing and near-optimal solutions. However, the research work is still too independent to give a clear picture on how things compare, especially when comparing different learning solutions to the same problem. Below, we put forth some issues and concerns related to the line of work, some of which are re-iterated from other authors. 
	\begin{itemize}
		\item \textit{Poor choice of benchmarks:} Benchmarks taken from the traditional line of work are often too basic and very old. Researchers need to be comparing against recent cutting-edge solutions proposed in the literature. 
		\item \textit{Poor choice of system models:} If research in this area focuses on simple yet practical system models, we have a chance at comparing our results to the theoretical limits of the system to better understand just how “near” we are to the optimal scenario. This is crucial point, as it is very difficult to asses what is the best we can get with ML. In other words, performance will almost never be analytically verified which leads us to believe that research on finding theoretical fundamental limits of systems will forever continue to evolve and should not be undervalued and marginalized.
		\item \textit{Practical channel statistics:} Systems with unknown channel models need to be better addressed by these approaches \cite{dorner2017deep} in order for learning algorithms to retain their reputation of ability to tune parameters on the fly. So far all algorithms rely heavily on channel models.
		\item \textit{User detection solutions need to incorporate collisions (for grant-free NOMA):} All work focuses on user detection and correct estimation of parameters. However, thus far assume that users have unique signature sequences and thus collisions are not an issue. However, in massive user settings, assigning unique signature sequences is not practical and collisions are the bottleneck of the performance. Thus, collision detection and resolution need to be better addressed in grant-free settings.
	\end{itemize}
	\section{Information Theoretic Perspective of Grant-Free NOMA}
	The capacity bounds of the classical multi access channel (MAC) with a fixed number of users and an infinitely large block length is well understood. However, for many years, information theorists have been hinting and even trying at deriving capacity bounds for grant-free channels. For instance, over 30 years ago, \cite{gallager1985perspective} sought a coding technology that is applicable for a large set of transmitters of which a small, but variable, subset simultaneously uses the channel. Over 10 years ago, authors in \cite{cover2012elements} concluded that when the total number of senders is very large, so that there is a lot of interference, we can still send a total amount of information that is arbitrary large even though the rate per individual sender goes to 0 \cite[pp. 546, 547]{cover2012elements}. With the fast evolution of mMTC, information theorists have amplified their efforts in deriving insightful and easy ways to evaluate capacity bounds that suit these new and pertinent scenarios. The main challenges faced for grant-free NOMA are with the finite block length regime, the random activity of users, and the growth of the number of users with the block length. Below, we provide the main information theoretical studies to date related to grant-free NOMA.
	\subsection{Capacity of Gaussian MAC in Finite Block-Length }
	The works in \cite{molavianjazi2015second,molavianjazi2012random,jazi2012simpler} investigated simple bounds on the $K$-user MAC, where $K$ is fixed and block length is finite. It was shown that, unlike the asymptotic case, OMA is strictly bounded below the capacity. It was further showed that the capacity can only be achieved with NOMA and joint decoding. In general, these bounds involve the evaluation of probabilities in 2$K$-dimensional spaces, and thus can only be evaluated for small values of $K$. Similar results have been found in \cite{truong2017gaussian,huang2012finite,scarlett2014second}. Other works aimed at investigating the penalty on the performance bounds introduced by the random activity of users and collisions \cite{abbas2018novel}. The work showed that by considering collisions as interference, the throughput can be significantly increased. Moreover, it was also shown that there is a significant gap in performance between joint decoding and successive decoding. \subsection{The Gaussian Many Access Channel}
	The work in \cite{chen2017capacity} introduced a new paradigm to this area of research, namely the many access paradigm. A case was considered where both the number of users and the block length go to infinity; however, the number of users can scale as fast as linearly with the block length. This assumption renders classical theory inapplicable as the number of users can grow larger than the block length which is typical for most if not all mMTC scenarios. When user activity is unknown at the receiver (random), authors show that the capacity is achievable by having transmitters concatenate a unique signature sequence to the payload and performing a two-stage decoding scheme at the receiver: (1) user detection, followed by (2) data decoding. The derived capacity for RA is the same as that with known access, minus a penalty factor characterised by the minimum length of signature sequences needed to obtain an arbitrarily low and even vanishing error probability in active user detection. An error in active user detection can be either a missed detection or a false alarm. Similar to the results in \cite{molavianjazi2015second}, authors show that successive decoding cannot achieve the sum capacity in this scenario unlike classical multiple access capacity regions. Overall, the derivations in \cite{chen2017capacity} are heavily rooted in it being a CS problem. According to our knowledge, no attempts to date have been shown to achieve or approach this capacity bound.
	\subsection{Random Coding Bound}
	Unlike the work in \cite{chen2017capacity}, \cite{polyanskiy2017perspective} sought a capacity bound for the case where users utilize the same encoders/codebook (no unique signature sequences) which is more practical when the total number of users in the network is large. This is also known as symmetrical encoding. The sole task of recovering the unordered list transmitted messages is assigned to the decoder, thus decoupling the role of user identification from data recovery, reasoning that the identification is part of the payload. Provided that this permutation invariance holds (i.e. the conditional channel output distributions are independent of the channel input permutations), \cite{polyanskiy2017perspective} derives the random coding bound that dictates the limits on RA code lengths that can achieve an arbitrarily small error probability. The error probability is defined as the number of average fractions of correctly recovered messages. Thus, the error is defined on user level rather than a network level. For a $K$-user MAC, the RA codes should be such that the sum of any $K$ or less unique codewords can be decoded reliably.
	\subsection{Achievability of the Random Coding Bound}
		\begin{figure}[t]
		\centering	\includegraphics[width=3.5in,height=3.2in]{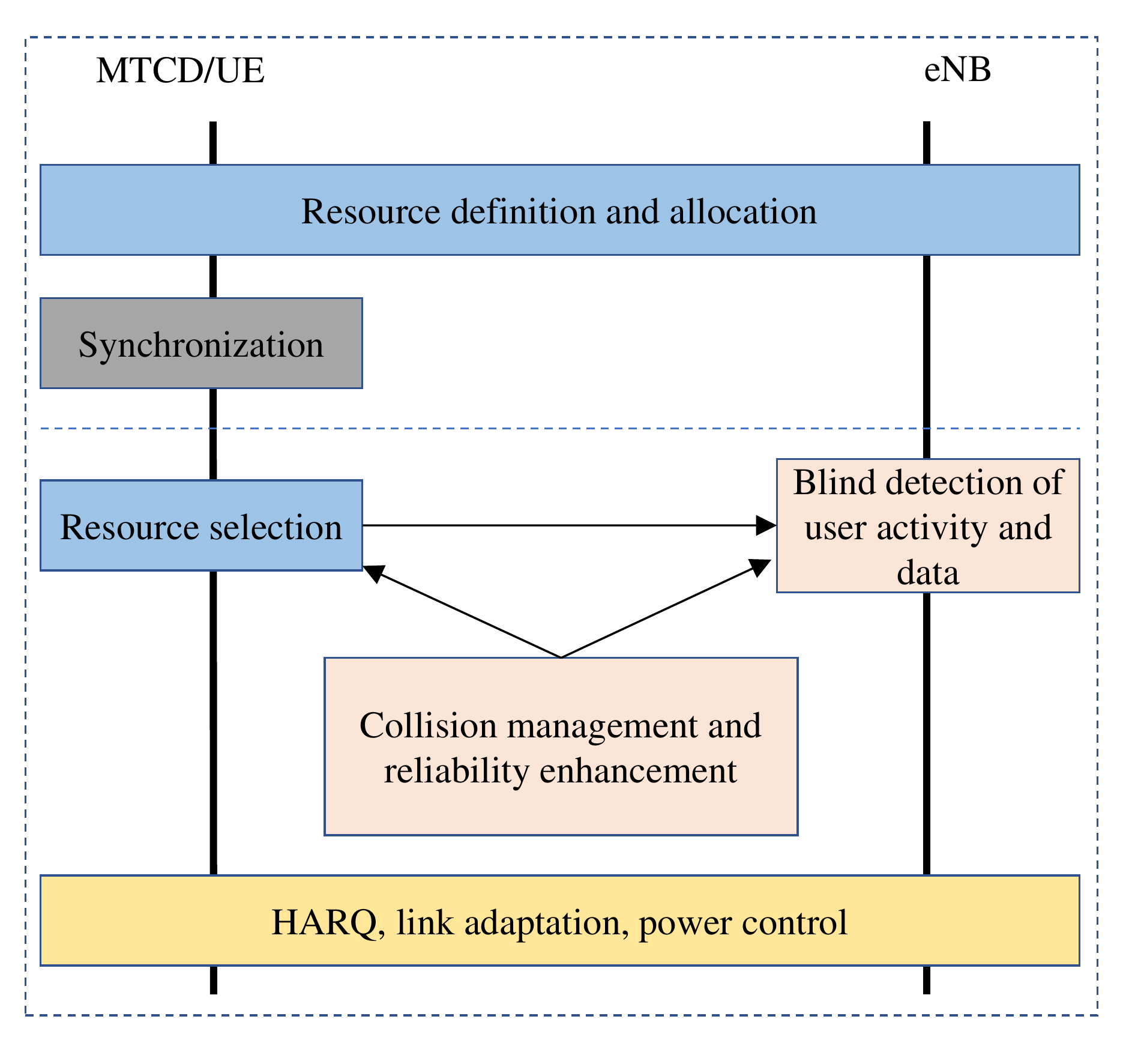}
		\caption{Overview of technicalities in grant-free transmission}
		\label{fig18}
	\end{figure}
	Following on from \cite{polyanskiy2017perspective} RA channel setup, \cite{effros2018random} demonstrated the achievability using rateless codes. Unlike \cite{chen2017capacity}, where random user activity was shown to introduce a penalty on the capacity, \cite{effros2018random} showed that the performance is the same in first and second order, whether user activity is known or not. Moreover, for a symmetric multiple access channel, there is a single-threshold decoding rule rather than the more familiar $2K-1$ simultaneous threshold rules. The considered class of rateless codes differs slightly from that in the literature. Here, codes are not rateless in that codewords vary in length, but rather that decoding varies in time. Moreover, traditional rateless codes, such as Raptor codes, assume arbitrary decoding times and single-bit feedback to terminate transmission, the codes in \cite{effros2018random} considered that a single-bit feedback is transmitted at every time step indicating whether to continue or terminate transmission. In \cite{effros2018random}, users listen at finite and predetermined set of times thus allowing the feedback rate to vanish as the block length grows. The decoding times are fixed to $n_1$, $n_2$, such that $n_i$ denotes the time at which the decoder believes there are $i$ active users. After every coding epoch $n_i$, the receiver sends a feedback bit that indicates whether it is ready or not to decode. Other works inspired by the random coding bound and the system model introduced by it can be found in \cite{ordentlich2017low,abbas2018multi}. However, the work therein focuses on proposing practical and low complexity system architectures rather than achieving or approaching the capacity bound.
	\section{Practical Challenges and Solutions for Grant-free UL NOMA}
	NOMA has demonstrated significant potential to provide massive connectivity and large spectral gains. However, to support/optimize grant-free transmission, resource definition, allocation, and selection is important. Moreover user synchronization, active user and data detection, potential collision management, HARQ, link adaptation, and power control procedures would need to be investigated \cite{Huawei2016Discussion}. Some of these practical challenges are highlighted in Fig. \ref{fig18}, and are discussed as follows.
	%\begin{figure*}[!t]
	%	\centering
	%	\includegraphics[width=2.5in,height=1.5in]{MAB/MAB.pdf}
	%	\caption{An example configuration of MAB}
	%	\label{fig26}
	%\end{figure*}
	\subsection{Resource definition, allocation and selection}
	In order to enable grant-free access, the radio resource for transmission should be defined before any grant-free transmission starts, and be known to both the UE and BS. The pre-defined resource for grant-free transmission can be similar to the CTU defined earlier in Fig. \ref{fig9}. Each CTU may include time-frequency resources, and may combine with a set of pilots for channel estimation and/or UE activity detection, and a set of MA signatures (e.g., codebooks/sequences/interleavers) for robust signal transmission and interference whitening, etc. Moreover, the size and location of time/frequency resources, as well as the pilot/signature patterns associated with it should be pre-defined, as shown earlier in Fig. \ref{fig10}. Once resources are defined, the resource allocation problem is to study how to allocate the CTUs to different UEs. It is possible for the BS to allocate the CTUs to users. However, to enable more autonomous transmission, the users can select a CTU i.e., some specific pilot and signature pattern. The selection of the specific pilot and signature can either be done randomly from the resource pool, or according to some pre-defined rule. 
	\subsection{Synchronization among devices}
	Grant-free/contention-based UL transmission is expected to be supported by transmission without close-loop time alignment signaling or a RA process (RACH-less grant-free). In such scenarios, if the timing offset between randomly transmitting MTCDs is larger than the cyclic prefix (CP), it is referred to as asynchronous transmission. This asynchronous transmission will cause a tremendous complexity increase for MTCD detection and decoding at the receiver side. In this case, well-designed preambles can assist the
	active user identification, timing-offset/frequency-offset estimation, and channel estimation \cite{Huawei2016Discussion}. Some preamble transmission procedures for the mMTC UL are provided in \cite{Nokia2016Preamble}.
	\begin{figure}[!t]
		\centering	
		\subfloat[Code/interleave pattern based NOMA]{%
			\includegraphics[width=3.6in,height=1.725in]{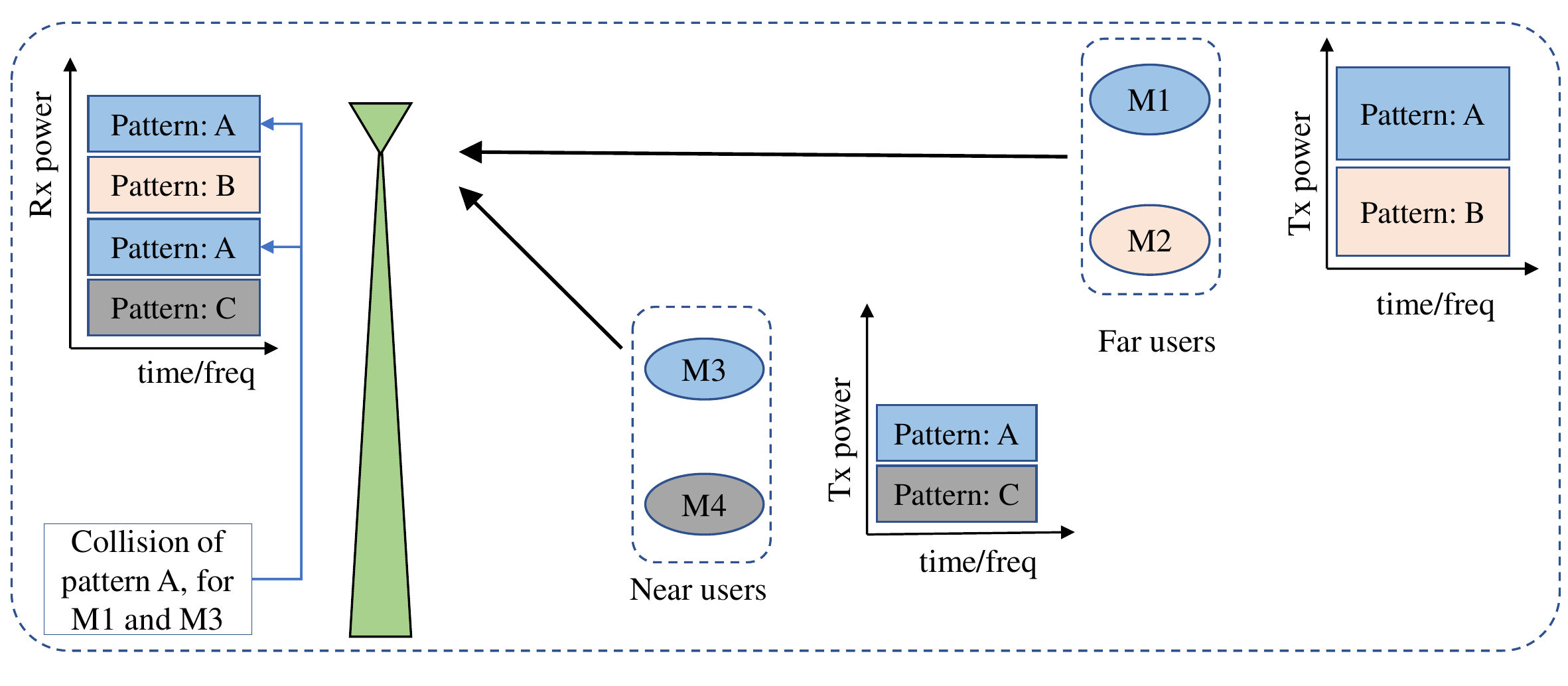}%
			\label{fig19a}
		}	
	
		\subfloat[Combination of code/interleave pattern and PD-NOMA]{%
			\includegraphics[width=3.6in,height=1.725in]{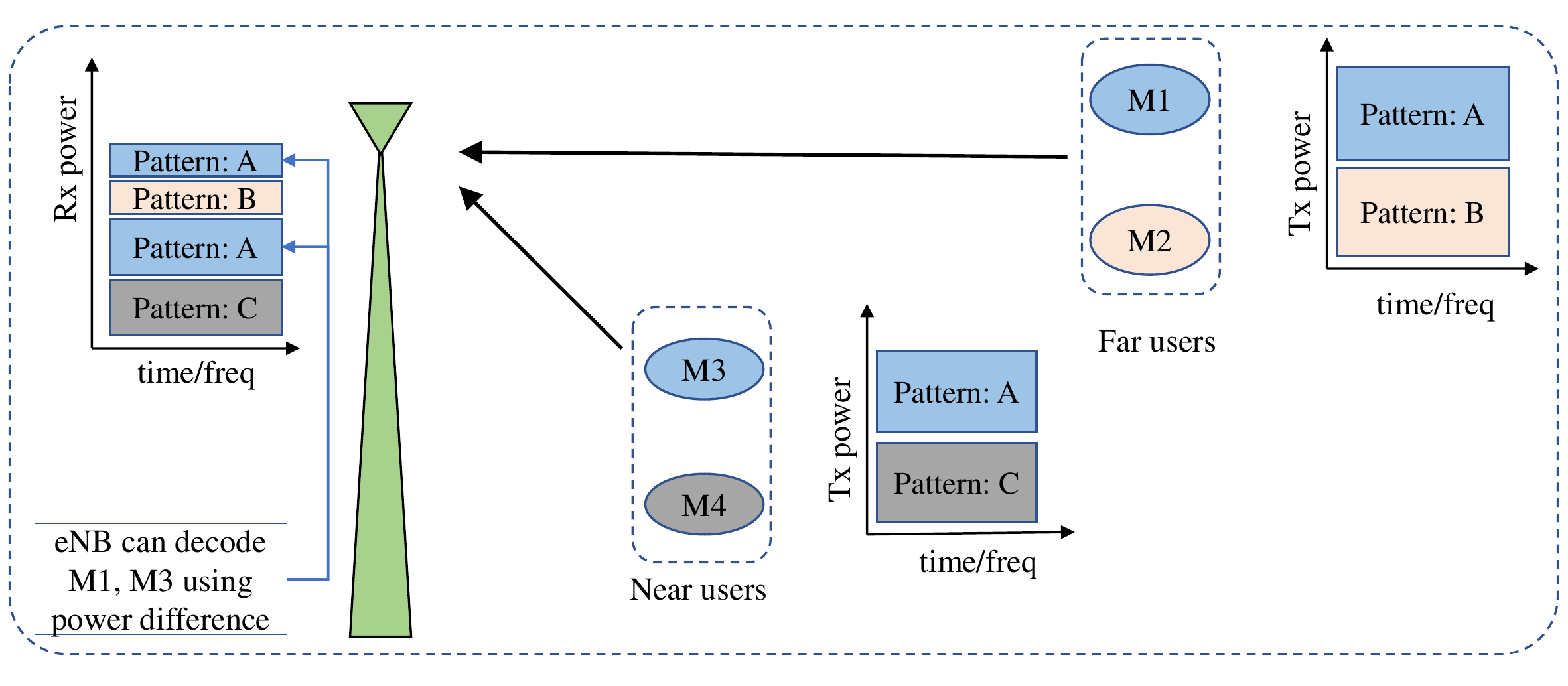}%
			\label{fig19b}
		}	
		\caption{Use of multiple MA signatures for collision handling}
		\label{fig19}	
	\end{figure} 
	\begin{figure*}[!t]
		\centering
		\includegraphics[width=4.65in,height=1.0in]{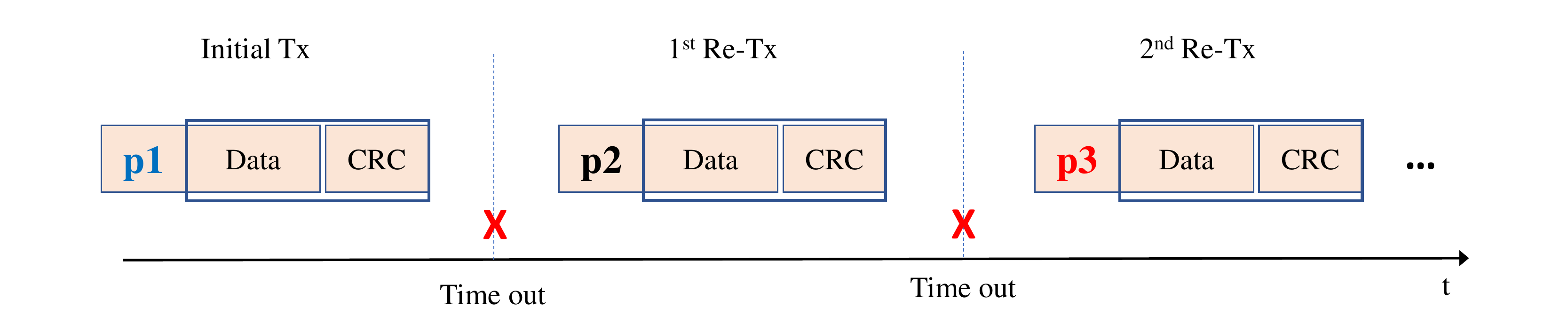}
		\caption{Initial transmission and re-transmission of one user with mapped pilots}
		\label{fig20}
	\end{figure*}
	\par
	Moreover, a MTCD may still need to detect synchronization signals and system information blocks for DL synchronization \cite{Samsung2016Discussion}. Given DL synchronization, a UE can adjust its UL transmit timing and, in many cases \cite{Samsung2016Non}, achieve UL synchronization (time-offsets between UEs within CP length) without close-loop timing advance command. Hence, synchronous UL transmissions within CP can often be feasible for mMTC. To avoid increasing the reception complexity and operation difficulty, time misalignment can be limited by using proper CP length and symbol duration.
	\subsection{Blind detection of MTCD activity and data}
	In grant-free transmission, since eNB has no prior information of when a MTCD may initiate transmission, it has to detect on each CTU which MTCDs have transmitted. Such MTCD activity detection is normally preferred to be done jointly with data decoding to reduce latency and overhead. This blind detection is crucial to the performance of grant-free UL communication. However, how to do the blind detection and based on what to detect is the major problem that needs to be investigated. One better option is to use pilots as was done in grant-free UL SCMA. In this case, pilots may serve the purpose of both the MTCD activity detection and channel estimation. In this context, efficient pilot designs for joint user activity detection and channel estimation should be studied. Moreover, performance of MUD is also a major concern. For example, the SIC receiver may suffer from imperfections, and the SIC error is then propagated and effects other overlapped MTCDs detection. A comprehensive discussion and performance evaluation of different advanced MUD receivers for grant-free transmissions is provided in \cite{Huawei2016Advanced}. Some non-coherent detection techniques have recently been proposed in literature for massive NOMA in grant-free access \cite{chen2019noncoherent,liu2018sparse,senel2018joint}. Moreover, the CS-MUD and ML techniques described earlier can also be used for efficient data recovery. 
	%\begin{figure*}[t]
	%	\centering	\includegraphics[width=7.0in,height=4.0in]{PracticalChallenges/PracticalChallenges.pdf}
	%	\caption{Summary of practical challenges, their description, and possible solutions}
	%	\label{fig21}
	%\end{figure*}
	\subsection{Collision management and reliability enhancement}
	In grant-free transmission, it is likely that more than one MTCD would end up using the same time-frequency resource. In the code/interleave domain NOMA, which includes spreading, interleaving or codebook-based signatures, colliding MTCDs are still separable through differences in their signatures or pattern vectors. However, if some MTCDs happen to use the same pattern vector, a hard collision is said to have taken place and these MTCDs will suffer from each other's signal interference. In this context, the impact of such collision should be studied \cite{Nokia2016Collision}. Some solutions to deal with MA signature collisions are efficient MA signature design, larger resource pool, detection optimization, MA resource management, and mode switching mechanism (between grant-based and grant-free) \cite{Huawei2016Solutions}. 
			\begin{table*}[!t]
		\caption{Summary of practical challenges, their description, and possible solutions}
		\label{tab5}
		\resizebox{\textwidth}{!}{%
			\begin{tabular}{|c|l|l|}
				\hline
				\textbf{Practical Challenge} & \textbf{Description} & \textbf{Possible Solutions} \\ 
				\hline
				\begin{tabular}[c]{@{}c@{}}Resource definition, \\ allocation, and selection\end{tabular} & \begin{tabular}[c]{@{}l@{}} \textbullet ~ How to define a grant-free resource?\\ \textbullet ~ How to allocate resources to users?\end{tabular} & \begin{tabular}[c]{@{}l@{}} \textbullet ~CTU based resource; containing MA signature, pilots etc.\\ \textbullet ~Predefined/pre-allocated CTU or randomly selected.\\ \textbullet ~Random pilot and MA signature selection by MTCD.\end{tabular} \\ 
				\hline
				UL synchronization & \begin{tabular}[c]{@{}l@{}}\textbullet ~RACH-less random transmission.\\ \textbullet ~Large timing offsets cause synchronization issues.\\ \textbullet ~Detection complexity/inefficiency of MUD.\end{tabular} & \begin{tabular}[c]{@{}l@{}}\textbullet ~MTCD adjusts UL timing based on DL synchronization.\\ \textbullet ~Well-designed preambles.\\ \textbullet ~Proper CP length and/or symbol duration.\end{tabular} \\ 
				\hline
				Blind detection & \begin{tabular}[c]{@{}l@{}}\textbullet ~eNB has no prior information of active MTCDs.\\ \textbullet ~Joint MTCD activity and data detection needed.\\ \textbullet ~How and based on what?\end{tabular} & \begin{tabular}[c]{@{}l@{}}\textbullet ~Use of pilot symbols for activity detection and channel estimation.\\ \textbullet ~CTU design: may contain MCS and other MUD information.\\ \textbullet ~CS and ML based activity/data detection.\end{tabular} \\ 
				\hline
				Collision management & \begin{tabular}[c]{@{}l@{}}\textbullet ~How to minimize collisions?\\ \textbullet ~How to manage once a collision happens?\end{tabular} & \begin{tabular}[c]{@{}l@{}}\textbullet ~Efficient MA signature design and increasing resource pool.\\ \textbullet ~Mixing multiple domains e.g., spreading and power.\\ \textbullet ~Efficient ACK/NACK feedback and retransmissions.\end{tabular} \\ 
				\hline
				\begin{tabular}[c]{@{}c@{}}Link adaptation and \\ power control\end{tabular} & \begin{tabular}[c]{@{}l@{}}\textbullet ~How to choose MCS?\\ \textbullet ~How to achieve power control?\end{tabular} & \begin{tabular}[c]{@{}l@{}}\textbullet ~Different CTUs linked with different MCS and power values.\\ \textbullet ~Use DL channel estimation for choosing UL MCS, power, etc.\end{tabular} \\ 
				\hline
				HARQ & \begin{tabular}[c]{@{}l@{}}\textbullet ~How to know a failed transmission?\\ \textbullet ~How to merge original and HARQ retransmissions?\\ \textbullet ~Distinguishing transmissions/retransmissions at eNB.\end{tabular} & \begin{tabular}[c]{@{}l@{}}\textbullet ~Efficient design of ACK/NACK feedback to users.\\ \textbullet ~Retransmissions from a user with different pilot values.\end{tabular} \\
				\hline
			\end{tabular}%
		}
	\end{table*}
	\par
	In the context of resource pool increase, Fig. \ref{fig19a} shows an example of code/interleave domain NOMA with 4 MTCDs, where a near-to-eNB MTCD M3 and far-from-eNB MTCD M1 choose the same pattern vector A, and are not separable at the eNB. As one of the countermeasures, PD-NOMA is introduced in the system as shown in Fig. \ref{fig19b}. The system model is same as previous, but the users adjust their transmission power, so that a power difference is achieved at the eNB for signal separation \cite{Sony2016Non}. Therefore, by using PD-NOMA, MA signature patterns can be reused. Hence, design of UL NOMA schemes should consider how to deal with collisions of NOMA patterns for grant-free/contention-based UL transmission. Moreover, similar to the example in Fig. \ref{fig19}, MA signature combination-based solutions to resolve collisions should be considered \cite{Sony2016Non}.
	\subsection{Link adaptation and power control}
	Link adaptation is a term used in wireless communications to denote the matching of the modulation, coding and other signal and protocol parameters to the conditions of the radio link. Efficient link adaptation results in better utilization of the network resources, detection error rate reduction, energy efficiency, reduced latency etc. Generally, link adaptation is based on the channel state information. However, in grant-free transmissions, MTCDs might not have the exact UL channel status. A solution is to consider the channel between MTCDs and eNB to be reciprocal in each direction (a case in time division duplexing). Hence, the MTCDs can estimate their channel to the eNB using the periodically received pilot/reference signals from eNB, and correspondingly adjust their transmission parameters to facilitate data recovery at the eNB. For instance, in the grant-free Random NOMA scheme explained earlier, the MTCDs after estimating their channel can adjust their transmission power so that their received power at the eNB is always the same fixed value, which helps the eNB in load estimation over a sub-band \cite{shirvanimoghaddam2017massive}. 
	\par
	Another solution is to divide the radio resources or CTUs into orthogonal MA blocks (MABs). Different MABs occupy different radio resources and may adopt different transmit parameter settings i.e., transmission block sizes, MCSs, transmit power, etc. Configurations of the MABs can be broadcasted by
	the eNB. During data transmission phase, any active MTCD first selects one MAB followed by MA signature or pattern vector. At eNB, MUD can be parallelly performed on each MAB. In addition, each MAB may be assigned with a limited number of signatures, thereby reducing the computational complexity of blind detection at the receiver.
	\subsection{Hybrid automatic repeat request (HARQ)}
	In UL grant-free transmission, a user waits for a fixed time period to figure out whether its previous transmission is successful or a retransmission is needed. This is usually achieved through an ACK/NACK feedback from eNB. However, unlike grant-based transmission, the eNB is not aware of which UEs are transmitting information in advance. Therefore, eNB has to perform user activity and data detection, followed by the ACK/NACK feedback. In addition to potentially poor channel conditions, collisions among users and increased interference level by contention based NOMA for mMTC may also lead to incorrect data detection at eNB, which results in a NACK. In this case, HARQ retransmissions are of prime importance to guarantee the reliability of data.
	\par
	For a failed UL transmission, MTCDs need to re-transmit their data one or more times. HARQ can efficiently merge new transmissions with the previous one. However, one issue is how to identify the first transmission and the retransmissions for a HARQ process. One potential solution is that the eNB can explicitly schedule retransmissions via DL control signaling. Another efficient mechanism to address this problem is to use different pilots mapped to transmission and re-transmissions to identify the transmissions of a same packet by a particular user. An example of this is shown in Fig. \ref{fig20}, where a user does one transmission and two retransmissions. Pilots p1, p2 and p3 are mapped on initial transmission, 1$^{\text{st}}$ retransmission and 2$^{\text{nd}}$ re-transmission, respectively. If the eNB successfully identified all pilots during the transmission, it can still potentially decode the signal by combing all the uncoded packets. Furthermore, due to the grant-free nature of transmission, the HARQ procedure can be different from LTE scheduled HARQ \cite{Samsung2016Discussion}. A detailed insight into the HARQ process, and potential HARQ techniques for grant-free transmissions is provided in \cite{Huawei2016The}.
	\par
	Based on the discussion in this section, the identified practical challenges, their description, and possible solutions are summarized in Table. \ref{tab5}.
	\section{Future Directions}
	For scheduling based NOMA schemes, comprehensive analysis of various MA signature (spreading, interleaving, scrambling, multiple domains) based schemes exists in literature. However, for grant-free/contention-based UL NOMA schemes, there is a need for further investigations, new designs, and integration of multiple technologies to deal with the challenges of different IoT use cases. In 3GPP RAN1 86th meeting, it was agreed that, parallel to new designs, the following should be continuously studied \cite{ZTE2016Discussion3}.
	\begin{figure*}[!t]
	\centering
	\includegraphics[width=5.2in,height=2.8in]{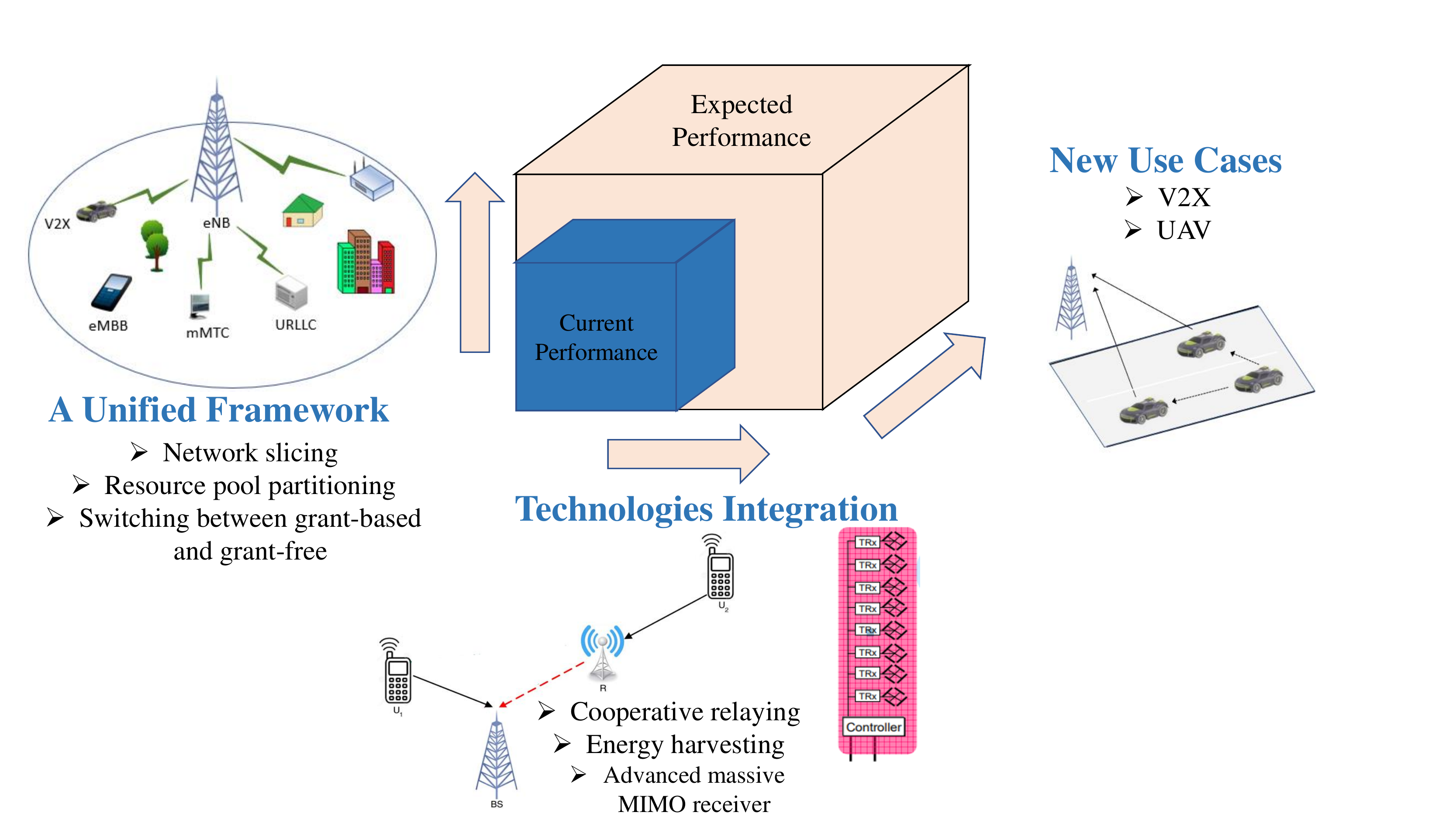}
	\caption{Future directions}
	\label{fig22}
\end{figure*}
	\begin{itemize}
		\item Resource allocation/selection options; a) randomly by user, b) pre-configured by eNB.
		\item UL synchronization (DL synchronization assumed) by considering two cases; timing offsets between users are within or greater than CP.
		\item Collision handling of time/frequency resources and MA signatures (e.g., code, sequence, interleaver pattern).
		\item Retransmission/repetition of failed transmissions and potential combining, e.g. HARQ.
		\item Link adaptation, e.g. MCS/signature assigning.
		\item Relationship between grant-free and grant-based transmissions and associated user behavior.
		\item Advanced receiver capabilities and complexity analysis.
		\item Requirement for power control.
	\end{itemize}
	These study items are basically a continuity in pursuit of solutions to the practical challenges explained earlier. In addition to these study items, other innovative options need to be explored for facilitating/improving grant-free communication for mMTC and other use cases. In this context, we discuss some other future directions to improve the performance of grant-free  transmissions, as shown in Fig. \ref{fig22}.
	\subsection{A unified framework for IoT use cases}
	Some major identified IoT use cases are eMBB (majorly HTC), mMTC, and URLLC. Considering the coexistence of these use cases with extremely diverse QoS requirements, a unified framework is required, where each of these use cases can be supported using a single backbone/core system. Some potential directions to achieve these goals are highlighted in Fig. \ref{fig22}, and are explained as follows.
	\subsubsection{Network slicing with grant-free UL NOMA}
	Considering a variety of IoT use cases, and a diverse range of QoS requirements, using the concepts of network slicing may provide additional flexibility of resource allocation to different use cases. The objective is to allow a physical mobile network operator to partition its network resources to allow for very different users, so-called tenants, to multiplex over a single physical infrastructure. The most commonly cited example in 5G discussions is sharing of a given physical network to simultaneously run mMTC, eMBB, and URLLC. NOMA with network slicing has been under focus recently. 
	\par
\begin{figure*}[t]
		\centering	\includegraphics[width=5.4in,height=3.1in]{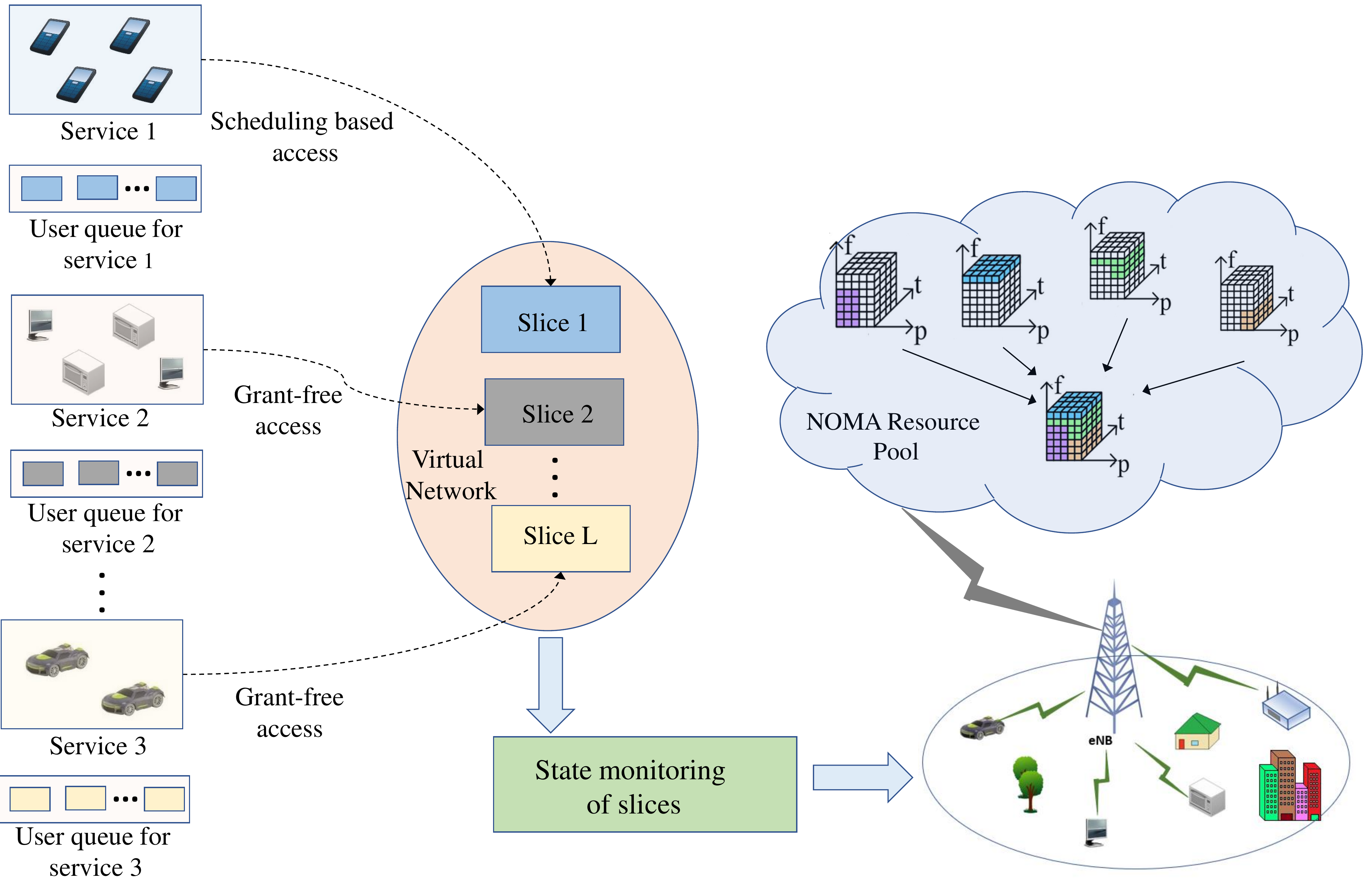}
		\caption{Grant-free UL NOMA with network slicing}
		\label{fig23}
	\end{figure*}

	In \cite{kazmi2019network}, vital challenges of resource management pertaining to network slicing using the NOMA-based scheme are highlighted. In this context, efficient solutions for resource management in network slicing for NOMA-based scheme are provided. Moreover, a slice-based virtual resource scheduling scheme with NOMA
	to enhance the QoS of the system is proposed in \cite{tang2018adaptive}. While network slicing with NOMA has been explored to some extent in the existing literature, these works only focus on the scheduling based access. Therefore, in order to provide a unified framework with grant-free/contention-based transmission support, a desperate need of the hour is to devise grant-free UL NOMA based solutions using network slicing. An example scenario is further shown in Fig. \ref{fig23}.
	\subsubsection{Resource pool partitioning for grant-free UL NOMA}
\begin{figure}[!t]
		\centering
		\includegraphics[width=2.5in,height=1.2in]{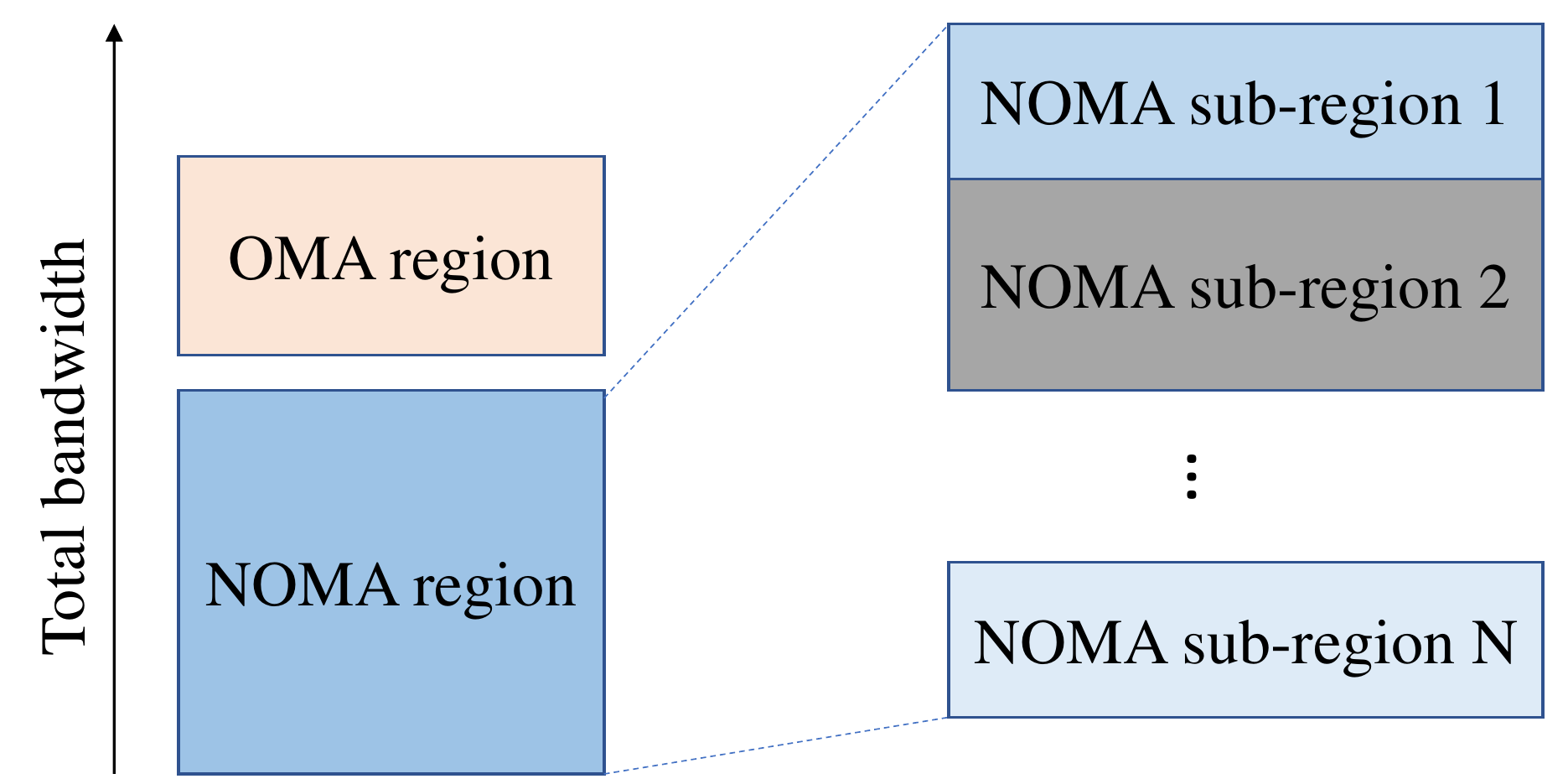}
		\caption{NOMA resource pool partitioning}
		\label{fig24}
	\end{figure}

	Similar to network slicing, depending on applications/services, packet payload sizes from multiple users could be different. To allow efficient resource use for different payload sizes and reduce eNB receiver complexity, multiple NOMA sub-regions can be defined within one NOMA resource pool, where each NOMA sub-region may be tailored for one particular MCS, transmission block size or coverage enhancement level if supported for mMTC \cite{Intel2016Grant2}. Moreover, some OMA based regions can be defined/reserved for priority use cases, as shown in Fig. \ref{fig24}.
	\subsubsection{Switching between grant-free and grant-based}
	For a unified network based operation, using dynamic grant to override grant-free transmissions can enable switching between grant-based and grant-free transmission and provide flexibility to the eNB scheduler for handling urgent events or reconfiguring resources. For example, users may be able to access different type of services, e.g., URLLC and eMBB, and switching between grant-based and grant-free may be useful in this context \cite{Samsung2016Discussion}. 
	\subsection{Integration of grant-free UL NOMA with other technologies}
	The integration of scheduling based NOMA with other technologies is supported through various studies. However, when it comes to grant-free NOMA, there is hardly any such integration found in the existing literature. The performance of grant-free UL NOMA can be further improved by its integration with other cutting edge technologies.
	\subsubsection{Relaying based grant-free UL NOMA}
\begin{figure}[t]
		\centering
		\includegraphics[width=3.3in,height=3.0in]{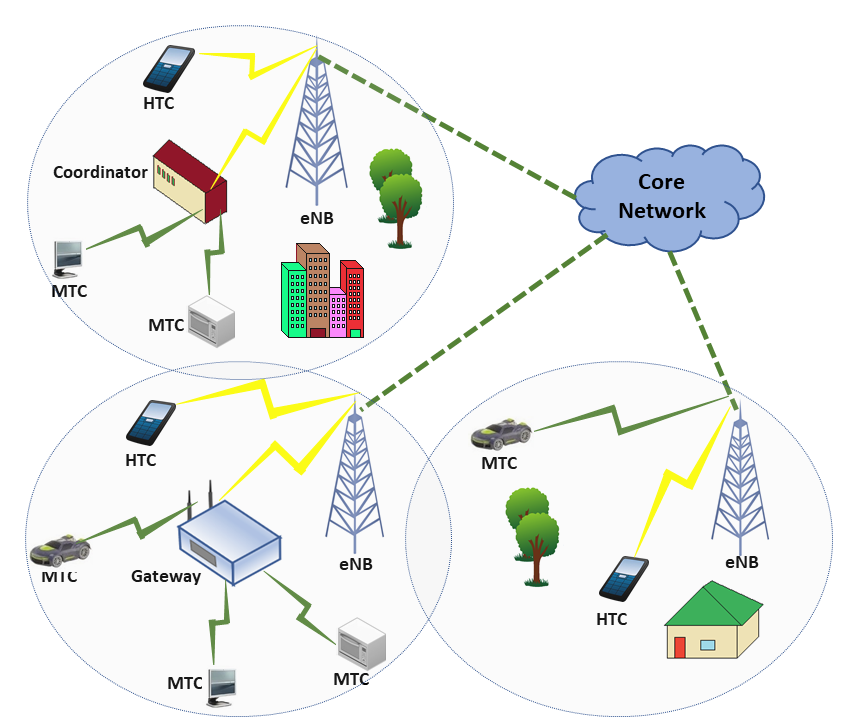}
		\caption{mMTC access scenarios}
		\label{fig25}
	\end{figure}

	According to the European telecommunications standards institute MTC architecture, and developments by 3GPP on mMTC \cite{zheng2014challenges,zheng2015radio,3gpp2016access}, three basic mMTC scenarios have been identified as shown in Fig. \ref{fig25}, and are defined as:
	\begin{itemize}
		\item \textbf{Direct Access:} A MTCD can access the eNB without any intermediate device, also referred to as a direct 3GPP connection \cite{3gpp2016access}. This is the simplest access method, but may lead to traffic congestion and excessive signaling overhead when number of MTCDs becomes very high.
		\item \textbf{Gateway Access:} A MTCD can obtain cellular connectivity through a M2M gateway, which is a dedicated device for data relaying between eNB and a group of MTCDs, and does not generate its own traffic. This is also termed as an indirect 3GPP connection \cite{3gpp2016access}. 
		\item \textbf{Coordinator Access:} A group/cluster of MTCDs obtain cellular connectivity through a coordinator (temporary M2M gateway), which itself is also a MTCD with its own traffic. 
	\end{itemize}
	Direct access is the point of focus in most of the existing literature as it is simple, but may lead to worse traffic congestion if the number of MTCDs is very high. Moreover, gateway and coordinator access are also critical, as group data transmission from a dedicated gateway or a temporary coordinator may reduce the overall power consumption of all the MTCDs and extend their service life, which is a key goal in mMTC/IoT. Some additional complex access scenarios might apply e.g., when a personal area network of MTCDs (a person wearing several smart wearables) attempts a gateway/coordinator access, or when the MTCDs and the relay/coordinator belong to different subscribers \cite{3gpp2016access}. The network should also aim to provide service continuity for devices switching between various access types (e.g., gateway to direct access or vice versa), their continuous authorization, and flexibility of choosing a radio access technology (licensed or unlicensed) \cite{3gpp2016access}. 
%	Furthermore, as most of the existing works focus on grant-free/contention-based transmissions for UL mMTC using direct access scenario only with no intermediate device between users and eNB, research on using the relaying based scenarios is significantly important.
	\par
	The poor channel quality of some distant-from-eNB users is also a reason behind the importance of gateway and coordinator access. In this context, the relaying based access can be further studied by considering half/full-duplex relaying mechanisms for grant-free UL NOMA transmissions. Furthermore, as the relays normally operate in either decode-and-forward or amplify-and-forward modes \cite{han2018joint,zhang2018performance}, their use in relaying based grant-free UL NOMA can be further investigated.
	\subsubsection{MIMO-NOMA}
	Scheduling based NOMA with multiple-input multiple-output (MIMO) has gained significant research interest recently, as MIMO provides an additional spatial degree of freedom. MIMO can be applied either for enhancing the reliability of data by introducing diversity, or enhancing the per-user capacity through spatial multiplexing. Some works on MIMO-NOMA are discussed in \cite{zeng2019energy,ding2016general}. However, the spatial diversity achieved through MIMO can also be used in facilitating grant-free UL NOMA transmissions, where the channel gains of different MIMO links can facilitate user separation, as can be seen in spatial domain multiple access (SDMA), where unique user-specific channel impulse responses are used for user multiplexing.
	\subsubsection{NOMA with energy harvesting}
	IoT devices are power-limited and maintaining a long life of these devices is of prime importance. In this context, NOMA can be incorporated to enable simultaneous wireless information and power transfer (SWIPT). Through SWIPT-NOMA, the massive connectivity is realizable, while providing energy harvesting opportunities to the IoT devices for a long battery life. Some NOMA based energy harvesting solutions are provided in \cite{ni2019analysis}. While all these works are focused on scheduling based NOMA schemes where power split ratios or time slots for information and energy transfer are predefined, the use of grant-free UL NOMA needs to be studied. 
	\subsection{New IoT use cases}
	Recently, some other use cases of IoT have also been identified/agreed e.g., vehicle-to-everything (V2X) and unmanned aerial vehicle (UAV). Hence, the use of UL grant-free NOMA should be considered for these scenarios. 
	\subsubsection{NOMA for V2X}
	The LTE network has recently been considered as a promising candidate to support V2X services. However, with a massive number of devices accessing the network, conventional OFDM-based LTE network faces congestion and access delay issues due to inefficient orthogonal resource allocation. Hence, the use of NOMA in supporting cellular V2X services to achieve low latency and high reliability is crucial \cite{di2017noma,di2017v2x}.
	\subsubsection{NOMA for UAV}
	UAV-enabled DL/UL wireless system wherein a UAV serves as flying BS to communicate with multiple ground users has gained much research interest recently. On the other side, the scenarios where multiple flying UAVs (as users) are connected to a ground-based BS are also equally important. In this context, the use of NOMA can increase the connectivity density manifold. Some recent investigations on NOMA-based UAV communications are provided in \cite{mei2019uplink}. However, these works focus on scheduling based NOMA, and can be extended to grant-free UL transmissions.
	\section{Conclusion}
	It is agreed that 5G should support autonomous/grant-free/contention-based UL transmission for mMTC. In this context, unlike the existing works, this article provides a comprehensive survey of NOMA from a grant-free connectivity perspective. Various candidate techniques for UL NOMA are explained first with their categorization into different groups. Furthermore, the design of these schemes to meet the grant-free requirements is then discussed in detail. Some, existing grant-free NOMA techniques are introduced alongwith the mechanism of blind MUD at the receiver. Moreover, the related research/practical challenges are comprehensively discussed and solutions are highlighted. In the end, some possible future directions are also provided. 
	% if have a single appendix:
	%\appendix[Proof of the Zonklar Equations]
	% or
	%\appendix  % for no appendix heading
	% do not use \section anymore after \appendix, only \section*
	% is possibly needed
	
	% use appendices with more than one appendix
	% then use \section to start each appendix
	% you must declare a \section before using any
	% \subsection or using \label (\appendices by itself
	% starts a section numbered zero.)
	%

	%\appendices
	%\section{Proof of the First Zonklar Equation}
	%Appendix one text goes here.
	
	% you can choose not to have a title for an appendix
	% if you want by leaving the argument blank
	
	% use section* for acknowledgment
	%\section*{Acknowledgment}
	%This 

	% Can use something like this to put references on a page
	% by themselves when using endfloat and the captionsoff option.
	\ifCLASSOPTIONcaptionsoff
	\newpage
	\fi

	%\IEEEtriggeratref{215}
	\bibliographystyle{IEEEtran}
	\bibliography{references}

% Generated by IEEEtran.bst, version: 1.14 (2015/08/26)
\begin{thebibliography}{100}
\providecommand{\url}[1]{#1}
\csname url@samestyle\endcsname
\providecommand{\newblock}{\relax}
\providecommand{\bibinfo}[2]{#2}
\providecommand{\BIBentrySTDinterwordspacing}{\spaceskip=0pt\relax}
\providecommand{\BIBentryALTinterwordstretchfactor}{4}
\providecommand{\BIBentryALTinterwordspacing}{\spaceskip=\fontdimen2\font plus
\BIBentryALTinterwordstretchfactor\fontdimen3\font minus
  \fontdimen4\font\relax}
\providecommand{\BIBforeignlanguage}[2]{{%
\expandafter\ifx\csname l@#1\endcsname\relax
\typeout{** WARNING: IEEEtran.bst: No hyphenation pattern has been}%
\typeout{** loaded for the language `#1'. Using the pattern for}%
\typeout{** the default language instead.}%
\else
\language=\csname l@#1\endcsname
\fi
#2}}
\providecommand{\BIBdecl}{\relax}
\BIBdecl

\bibitem{al2015internet}
A.~Al-Fuqaha, M.~Guizani, M.~Mohammadi, M.~Aledhari, and M.~Ayyash, ``Internet
  of things: A survey on enabling technologies, protocols, and applications,''
  \emph{IEEE Communications Surveys \& Tutorials}, vol.~17, no.~4, pp.
  2347--2376, 2015.

\bibitem{zanella2014internet}
A.~Zanella, N.~Bui, A.~Castellani, L.~Vangelista, and M.~Zorzi, ``Internet of
  things for smart cities,'' \emph{IEEE Internet of Things journal}, vol.~1,
  no.~1, pp. 22--32, 2014.

\bibitem{da2014internet}
L.~Da~Xu, W.~He, and S.~Li, ``Internet of things in industries: A survey,''
  \emph{IEEE Transactions on industrial informatics}, vol.~10, no.~4, pp.
  2233--2243, 2014.

\bibitem{lien2011toward}
S.-Y. Lien, K.-C. Chen, and Y.~Lin, ``Toward ubiquitous massive accesses in
  {3GPP} machine-to-machine communications,'' \emph{IEEE Communications
  Magazine}, vol.~49, no.~4, 2011.

\bibitem{wp5d2015imt}
{ITU-R}, ``{IMT} vision--framework and overall objectives of the future
  development of {IMT} for 2020 and beyond,'' \emph{International
  Telecommunication Union, Geneva, Switzerland, ITU Recommendation M 2083},
  2015.

\bibitem{zheng2014challenges}
K.~Zheng \emph{et~al.}, ``Challenges of massive access in highly dense
  {LTE}-advanced networks with machine-to-machine communications,'' \emph{IEEE
  Wireless Communications}, vol.~21, no.~3, pp. 12--18, 2014.

\bibitem{zheng2012radio}
K.~Zheng, F.~Hu, W.~Wang, W.~Xiang, and M.~Dohler, ``Radio resource allocation
  in {LTE}-advanced cellular networks with {M2M} communications,'' \emph{IEEE
  communications Magazine}, vol.~50, no.~7, 2012.

\bibitem{Ericsson2016cellular}
Ericsson, ``Cellular networks for massive {IoT},'' \emph{tech. rep. Uen 284
  23-3278}, Jan. 2016.

\bibitem{3gpp2019medical}
{3GPP}, ``Study on communication services for critical medical applications
  (release 17),'' \emph{3GPP TR 22.826 V1.0.0}, May 2019.

\bibitem{3gpp2019vehical}
------, ``Study on enhancement of {3GPP} support for {5G} {V2X} services
  (release 16),'' \emph{3GPP TR 22.886 V16.2.0}, December 2018.

\bibitem{wu2011m2m}
G.~Wu, S.~Talwar, K.~Johnsson, N.~Himayat, and K.~D. Johnson, ``{M2M}: From
  mobile to embedded internet,'' \emph{IEEE Communications Magazine}, vol.~49,
  no.~4, 2011.

\bibitem{jain2012machine}
P.~Jain, P.~Hedman, and H.~Zisimopoulos, ``Machine type communications in
  {3GPP} systems,'' \emph{IEEE Communications Magazine}, vol.~50, no.~11, pp.
  28--35, 2012.

\bibitem{etsi2011machine}
V.~ETSI, ``Machine-to-machine communications ({M2M}): Functional
  architecture,'' \emph{Int. Telecommun. Union, Geneva, Switzerland, Tech. Rep.
  TS}, vol. 102, p. 690, 2011.

\bibitem{cisco2018cisco}
{Cisco, VNI}, ``Cisco visual networking index: Forecast and trends,
  2017--2022,'' \emph{White Paper}, 2018.

\bibitem{Ericsson2016mobility}
Ericsson, ``Ericsson mobility report,'' \emph{tech. rep.}, Nov. 2015.

\bibitem{dai2015non}
L.~Dai \emph{et~al.}, ``Non-orthogonal multiple access for {5G}: solutions,
  challenges, opportunities, and future research trends,'' \emph{IEEE
  Communications Magazine}, vol.~53, no.~9, pp. 74--81, 2015.

\bibitem{tao2015survey}
Y.~{Tao}, L.~{Liu}, S.~{Liu}, and Z.~{Zhang}, ``A survey: Several technologies
  of non-orthogonal transmission for {5G},'' \emph{China Communications},
  vol.~12, no.~10, pp. 1--15, Oct 2015.

\bibitem{wang2016analysis}
Y.~{Wang}, B.~{Ren}, S.~{Sun}, S.~{Kang}, and X.~{Yue}, ``Analysis of
  non-orthogonal multiple access for {5G},'' \emph{China Communications},
  vol.~13, no. Supplement2, pp. 52--66, N 2016.

\bibitem{islam2016power}
S.~R. Islam, N.~Avazov, O.~A. Dobre, and K.-S. Kwak, ``Power-domain
  non-orthogonal multiple access ({NOMA}) in {5G} systems: Potentials and
  challenges,'' \emph{IEEE Communications Surveys \& Tutorials}, vol.~19,
  no.~2, pp. 721--742, 2016.

\bibitem{yang2017uplink2}
S.~Yang, P.~Chen, L.~Liang, J.~Zhu, and X.~She, ``Uplink multiple access
  schemes for {5G}: A survey,'' \emph{ZTE Communications}, vol.~15, no.~S1, pp.
  31--40, 2017.

\bibitem{ding2017survey}
Z.~Ding \emph{et~al.}, ``A survey on non-orthogonal multiple access for {5G}
  networks: Research challenges and future trends,'' \emph{IEEE Journal on
  Selected Areas in Communications}, vol.~35, no.~10, pp. 2181--2195, Oct 2017.

\bibitem{cai2017modulation}
Y.~Cai, Z.~Qin, F.~Cui, G.~Y. Li, and J.~A. McCann, ``Modulation and multiple
  access for {5G} networks,'' \emph{IEEE Communications Surveys \& Tutorials},
  vol.~20, no.~1, pp. 629--646, 2017.

\bibitem{liu2017nonorthogonal}
Y.~Liu \emph{et~al.}, ``Nonorthogonal multiple access for {5G} and beyond,''
  \emph{Proceedings of the IEEE}, vol. 105, no.~12, pp. 2347--2381, 2017.

\bibitem{basharat2018survey}
M.~Basharat, W.~Ejaz, M.~Naeem, A.~M. Khattak, and A.~Anpalagan, ``A survey and
  taxonomy on nonorthogonal multiple-access schemes for {5G} networks,''
  \emph{Transactions on Emerging Telecommunications Technologies}, vol.~29,
  no.~1, p. e3202, 2018.

\bibitem{ding2018embracing}
Z.~Ding, M.~Xu, Y.~Chen, M.~Peng, and H.~V. Poor, ``Embracing
  non-orthogonalmultiple access in future wireless networks,'' \emph{Frontiers
  of Information Technology \& Electronic Engineering}, vol.~19, no.~3, pp.
  322--339, 2018.

\bibitem{dai2018survey}
L.~Dai \emph{et~al.}, ``A survey of non-orthogonal multiple access for {5G},''
  \emph{IEEE communications surveys \& tutorials}, vol.~20, no.~3, pp.
  2294--2323, 2018.

\bibitem{ye2018uplink}
N.~Ye, H.~Han, L.~Zhao, and A.-h. Wang, ``Uplink nonorthogonal multiple access
  technologies toward {5G}: a survey,'' \emph{Wireless Communications and
  Mobile Computing}, vol. 2018, 2018.

\bibitem{maraqa2019survey}
O.~Maraqa, A.~S. Rajasekaran, S.~Al-Ahmadi, H.~Yanikomeroglu, and S.~M. Sait,
  ``A survey of rate-optimal power domain {NOMA} schemes for enabling
  technologies of future wireless networks,'' \emph{arXiv preprint
  arXiv:1909.08011}, 2019.

\bibitem{ding2017application}
Z.~Ding \emph{et~al.}, ``Application of non-orthogonal multiple access in {LTE}
  and {5G} networks,'' \emph{IEEE Communications Magazine}, vol.~55, no.~2, pp.
  185--191, February 2017.

\bibitem{docomo2014docomo}
{NTT DOCOMO}, ``5{G} radio access: Requirements, concept and technologies,''
  \emph{White Paper, Jul (2014)}, 2014.

\bibitem{abramson1970aloha}
N.~Abramson, ``The {ALOHA SYSTEM}: another alternative for computer
  communications,'' in \emph{Proceedings of the November 17-19, 1970, fall
  joint computer conference}.\hskip 1em plus 0.5em minus 0.4em\relax ACM, 1970,
  pp. 281--285.

\bibitem{roberts1975aloha}
L.~G. Roberts, ``{ALOHA} packet system with and without slots and capture,''
  \emph{ACM SIGCOMM Computer Communication Review}, vol.~5, no.~2, pp. 28--42,
  1975.

\bibitem{3GPPGP100892}
{3GPP}, ``{RACH} capacity evaluation for {MTC},'' \emph{GP-100892, TSG GERAN
  No. 46}, 2010.

\bibitem{shirvanimoghaddam2015probabilistic}
M.~Shirvanimoghaddam, Y.~Li, M.~Dohler, B.~Vucetic, and S.~Feng,
  ``Probabilistic rateless multiple access for machine-to-machine
  communication,'' \emph{IEEE Transactions on Wireless Communications},
  vol.~14, no.~12, pp. 6815--6826, 2015.

\bibitem{chen2018ultra}
H.~Chen \emph{et~al.}, ``Ultra-reliable low latency cellular networks: Use
  cases, challenges and approaches,'' \emph{IEEE Communications Magazine},
  vol.~56, no.~12, pp. 119--125, 2018.

\bibitem{sesia2011lte}
S.~Sesia, I.~Toufik, and M.~Baker, \emph{{LTE}-the {UMTS} long term evolution:
  From theory to practice}.\hskip 1em plus 0.5em minus 0.4em\relax John Wiley
  \& Sons, 2011.

\bibitem{dahlman20103g}
E.~Dahlman, S.~Parkvall, J.~Skold, and P.~Beming, \emph{{3G} evolution: {HSPA}
  and {LTE} for mobile broadband}.\hskip 1em plus 0.5em minus 0.4em\relax
  Academic press, 2010.

\bibitem{khan2018priority}
N.~Khan, J.~Mi{\v{s}}i{\'c}, and V.~Mi{\v{s}}i{\'c}, ``Priority-based
  machine-to-machine overlay network over {LTE} for a smart city,''
  \emph{Journal of Sensor and Actuator Networks}, vol.~7, no.~3, p.~27, 2018.

\bibitem{wiriaatmadja2015hybrid}
D.~T. Wiriaatmadja and K.~W. Choi, ``Hybrid random access and data transmission
  protocol for machine-to-machine communications in cellular networks,''
  \emph{IEEE Transactions on Wireless Communications}, vol.~14, no.~1, pp.
  33--46, 2015.

\bibitem{shariatmadari2015machine}
H.~Shariatmadari \emph{et~al.}, ``Machine-type communications: current status
  and future perspectives toward {5G} systems,'' \emph{IEEE Communications
  Magazine}, vol.~53, no.~9, pp. 10--17, 2015.

\bibitem{durisi2016toward}
G.~Durisi, T.~Koch, and P.~Popovski, ``Toward massive, ultrareliable, and
  low-latency wireless communication with short packets,'' \emph{Proceedings of
  the IEEE}, vol. 104, no.~9, pp. 1711--1726, 2016.

\bibitem{lien2011cooperative}
S.-Y. Lien, T.-H. Liau, C.-Y. Kao, and K.-C. Chen, ``Cooperative access class
  barring for machine-to-machine communications,'' \emph{IEEE Transactions on
  Wireless Communications}, vol.~11, no.~1, pp. 27--32, 2011.

\bibitem{cheng2011prioritized}
J.-P. Cheng, C.-h. Lee, and T.-M. Lin, ``Prioritized random access with dynamic
  access barring for {RAN} overload in {3GPP LTE-A} networks,'' in \emph{2011
  IEEE GLOBECOM Workshops (GC Wkshps)}.\hskip 1em plus 0.5em minus 0.4em\relax
  IEEE, 2011, pp. 368--372.

\bibitem{wu2013fasa}
H.~Wu, C.~Zhu, R.~J. La, X.~Liu, and Y.~Zhang, ``{FASA}: Accelerated {S-ALOHA}
  using access history for event-driven {M2M} communications,'' \emph{IEEE/ACM
  Transactions on Networking (ToN)}, vol.~21, no.~6, pp. 1904--1917, 2013.

\bibitem{3GPP2011Studyx}
{3GPP}, ``Study on {RAN} improvements for machine type communications,''
  \emph{Technical Report {3GPP TS 37.868 V11.0}}, 2011.

\bibitem{lo2011enhanced}
A.~Lo, Y.~W. Law, M.~Jacobsson, and M.~Kucharzak, ``Enhanced {LTE}-advanced
  random-access mechanism for massive machine-to-machine ({M2M})
  communications,'' in \emph{27th World Wireless Research Forum (WWRF)
  Meeting}.\hskip 1em plus 0.5em minus 0.4em\relax WWRF27-WG4-08,, 2011, pp.
  1--5.

\bibitem{3gpp2013accesse}
{3GPP}, ``Evolved universal terrestrial radio access ({E-UTRA}): Physical
  channels and modulation,'' \emph{3GPP TS 36.211 V14.11.0}, June 2019.

\bibitem{Ericsson2016NRPrach}
{Ericsson}, ``{NR PRACH} preamble design,'' \emph{document R1-1611904, 3GPP
  TSG-RAN WG1 {87}}, November 2016.

\bibitem{ZTE2016Considerations1}
{ZTE, ZTE Microelectronics}, ``Considerations on the preamble design for
  grant-free non-orthogonal {MA},'' \emph{R1-1608955, 3GPP TSG RAN WG1 Meeting
  {86b}}, October 2016.

\bibitem{Huawei2016Evaluation1}
{Huawei, HiSilicon}, ``Evaluation on {CP} types for {UL} transmissions,''
  \emph{R1-1611198, 3GPP TSG RAN WG1 Meeting {87}}, November 2016.

\bibitem{Ericsson2016nrrandom}
{Ericsson}, ``{NR} random-access response design,'' \emph{R1-1611911, 3GPP TSG
  RAN WG1 Meeting {87}}, November 2016.

\bibitem{Qualcomm2016RACH1}
{Qualcomm Incorporated}, ``{RACH} timeline considerations,'' \emph{R1-1612035,
  3GPP TSG RAN WG1 Meeting {87}}, November 2016.

\bibitem{ZTE2016On2}
{ZTE Corporation, ZTE Microelectronics}, ``On 2-step random access procedure,''
  \emph{R1-1608969, 3GPP TSG RAN WG1 Meeting {86b}}, October 2016.

\bibitem{Nokia2016Random12}
{Nokia, Alcatel-Lucent Shanghai Bell}, ``Random access principles for new
  radio,'' \emph{R1-1609737, 3GPP TSG RAN WG1 Meeting {86b}}, October 2016.

\bibitem{shafi20175g}
M.~Shafi \emph{et~al.}, ``{5G}: A tutorial overview of standards, trials,
  challenges, deployment, and practice,'' \emph{IEEE Journal on Selected Areas
  in Communications}, vol.~35, no.~6, pp. 1201--1221, 2017.

\bibitem{li20145g}
Q.~C. Li, H.~Niu, A.~T. Papathanassiou, and G.~Wu, ``{5G} network capacity: Key
  elements and technologies,'' \emph{IEEE Vehicular Technology Magazine},
  vol.~9, no.~1, pp. 71--78, 2014.

\bibitem{docomo2016initial}
{NTT DOCOMO}, ``Initial views and evaluation results on non-orthogonal multiple
  access for {NR},'' \emph{R1-165175, 3GPP TSG RAN WG1 Meeting {85}}, May 2016.

\bibitem{ding2016impact}
Z.~Ding, P.~Fan, and H.~V. Poor, ``Impact of user pairing on {5G} nonorthogonal
  multiple-access downlink transmissions.'' \emph{IEEE Trans. Vehicular
  Technology}, vol.~65, no.~8, pp. 6010--6023, 2016.

\bibitem{wang2006comparison}
P.~Wang, J.~Xiao, and P.~Li, ``Comparison of orthogonal and non-orthogonal
  approaches to future wireless cellular systems,'' \emph{IEEE Vehicular
  Technology Magazine}, vol.~1, no.~3, pp. 4--11, 2006.

\bibitem{zhang2016layered}
L.~Zhang \emph{et~al.}, ``Layered-division-multiplexing: Theory and practice,''
  \emph{IEEE Transactions on Broadcasting}, vol.~62, no.~1, pp. 216--232, 2016.

\bibitem{3gpp2015access}
{3GPP}, ``Study on downlink multiuser superposition transmission ({MUST}) for
  {LTE} (release 13),'' \emph{3GPP TR 36.859 V13.0.0}, December 2015.

\bibitem{3GPP2016WF}
------, ``{WF} on common features and general framework of {MA} schemes,''
  \emph{document R1-1610956, 3GPP TSG-RAN WG1 {86b}}, October 2016.

\bibitem{China2016Classification}
{China Telecom}, ``Classification of candidate {UL} non-orthogonal {MA}
  schemes,'' \emph{document R1-167445, 3GPP TSG-RAN WG1 {86}}, August 2016.

\bibitem{ETSI2016Final}
{ETSI}, ``Final report of {3GPP TSG RAN WG1} 84bis v1.0.0,'' \emph{document
  R1-165448, 3GPP TSG-RAN WG1 {84b}}, May 2016.

\bibitem{ETSI2016Final2}
------, ``Final report of {3GPP TSG RAN WG1} 86 v1.0.0,'' \emph{document
  R1-1608562, 3GPP TSG-RAN WG1 {86}}, October 2016.

\bibitem{hoshyar2008novel}
R.~Hoshyar, F.~P. Wathan, and R.~Tafazolli, ``Novel low-density signature for
  synchronous {CDMA} systems over {AWGN} channel,'' \emph{IEEE Transactions on
  Signal Processing}, vol.~56, no.~4, pp. 1616--1626, 2008.

\bibitem{guo2008multiuser}
D.~Guo and C.-C. Wang, ``Multiuser detection of sparsely spread {CDMA},''
  \emph{IEEE journal on selected areas in communications}, vol.~26, no.~3, pp.
  421--431, 2008.

\bibitem{van2009multiple}
J.~Van De~Beek and B.~M. Popovic, ``Multiple access with low-density
  signatures,'' in \emph{GLOBECOM 2009-2009 IEEE Global Telecommunications
  Conference}.\hskip 1em plus 0.5em minus 0.4em\relax IEEE, 2009, pp. 1--6.

\bibitem{hoshyar2010lds}
R.~Hoshyar, R.~Razavi, and M.~Al-Imari, ``{LDS-OFDM} an efficient multiple
  access technique,'' in \emph{2010 IEEE 71st Vehicular Technology
  Conference}.\hskip 1em plus 0.5em minus 0.4em\relax IEEE, 2010, pp. 1--5.

\bibitem{al2011subcarrier}
M.~Al-Imari, M.~A. Imran, R.~Tafazolli, and D.~Chen, ``Subcarrier and power
  allocation for {LDS-OFDM} system,'' in \emph{2011 IEEE 73rd Vehicular
  Technology Conference (VTC Spring)}.\hskip 1em plus 0.5em minus 0.4em\relax
  IEEE, 2011, pp. 1--5.

\bibitem{wen2016non}
L.~Wen, ``Non-orthogonal multiple access schemes for future cellular systems.''
  Ph.D. dissertation, University of Surrey, 2016.

\bibitem{nikopour2013sparse}
H.~Nikopour and H.~Baligh, ``Sparse code multiple access,'' in \emph{2013 IEEE
  24th Annual International Symposium on Personal, Indoor, and Mobile Radio
  Communications (PIMRC)}.\hskip 1em plus 0.5em minus 0.4em\relax IEEE, 2013,
  pp. 332--336.

\bibitem{zhang2014sparse}
S.~Zhang, X.~Xu, L.~Lu, Y.~Wu, G.~He, and Y.~Chen, ``Sparse code multiple
  access: An energy efficient uplink approach for {5G} wireless systems,'' in
  \emph{2014 IEEE Global Communications Conference}.\hskip 1em plus 0.5em minus
  0.4em\relax IEEE, 2014, pp. 4782--4787.

\bibitem{huawei2016sparse}
{Huawei, HiSilicon}, ``Sparse code multiple access ({SCMA}) for {5G} radio
  transmission,'' \emph{document R1-162155, 3GPP TSG-RAN WG1 {84b}}, April
  2016.

\bibitem{taherzadeh2014scma}
M.~Taherzadeh, H.~Nikopour, A.~Bayesteh, and H.~Baligh, ``{SCMA} codebook
  design,'' in \emph{2014 IEEE 80th Vehicular Technology Conference
  (VTC2014-Fall)}.\hskip 1em plus 0.5em minus 0.4em\relax IEEE, 2014, pp. 1--5.

\bibitem{wu2015iterative}
Y.~Wu, S.~Zhang, and Y.~Chen, ``Iterative multiuser receiver in sparse code
  multiple access systems,'' in \emph{2015 IEEE International Conference on
  Communications (ICC)}.\hskip 1em plus 0.5em minus 0.4em\relax IEEE, 2015, pp.
  2918--2923.

\bibitem{xiao2015iterative}
B.~Xiao \emph{et~al.}, ``Iterative detection and decoding for {SCMA} systems
  with {LDPC} codes,'' in \emph{2015 International Conference on Wireless
  Communications \& Signal Processing (WCSP)}.\hskip 1em plus 0.5em minus
  0.4em\relax IEEE, 2015, pp. 1--5.

\bibitem{du2016fast}
Y.~Du, B.~Dong, Z.~Chen, J.~Fang, and X.~Wang, ``A fast convergence multiuser
  detection scheme for uplink {SCMA} systems,'' \emph{IEEE Wireless
  Communications Letters}, vol.~5, no.~4, pp. 388--391, 2016.

\bibitem{mu2015fixed}
H.~Mu, Z.~Ma, M.~Alhaji, P.~Fan, and D.~Chen, ``A fixed low complexity message
  pass algorithm detector for up-link {SCMA} system,'' \emph{IEEE Wireless
  Communications Letters}, vol.~4, no.~6, pp. 585--588, 2015.

\bibitem{bayesteh2015low}
A.~Bayesteh, H.~Nikopour, M.~Taherzadeh, H.~Baligh, and J.~Ma, ``Low complexity
  techniques for {SCMA} detection,'' in \emph{2015 IEEE Globecom Workshops (GC
  Wkshps)}.\hskip 1em plus 0.5em minus 0.4em\relax IEEE, 2015, pp. 1--6.

\bibitem{dai2014successive}
X.~Dai \emph{et~al.}, ``Successive interference cancelation amenable multiple
  access ({SAMA}) for future wireless communications,'' in \emph{2014 IEEE
  International Conference on Communication Systems}.\hskip 1em plus 0.5em
  minus 0.4em\relax IEEE, 2014, pp. 222--226.

\bibitem{catt2016pdma}
{CATT}, ``Candidate solution for new multiple access,'' \emph{document
  R1-163383, 3GPP TSG-RAN WG1 {84}}, April 2016.

\bibitem{zeng2015pattern}
J.~Zeng, B.~Li, X.~Su, L.~Rong, and R.~Xing, ``Pattern division multiple access
  ({PDMA}) for cellular future radio access,'' in \emph{2015 international
  conference on wireless communications \& signal processing (WCSP)}.\hskip 1em
  plus 0.5em minus 0.4em\relax IEEE, 2015, pp. 1--5.

\bibitem{kang2015pattern}
S.~Kang, X.~Dai, and B.~Ren, ``Pattern division multiple access for {5G},''
  \emph{Telecommun. Netw. Technol.}, vol.~5, no.~5, pp. 43--47, 2015.

\bibitem{chen2016pattern}
S.~Chen \emph{et~al.}, ``Pattern division multiple access—a novel
  nonorthogonal multiple access for fifth-generation radio networks,''
  \emph{IEEE Transactions on Vehicular Technology}, vol.~66, no.~4, pp.
  3185--3196, 2016.

\bibitem{dai2018pattern}
X.~Dai, Z.~Zhang, B.~Bai, S.~Chen, and S.~Sun, ``Pattern division multiple
  access: A new multiple access technology for {5G},'' \emph{IEEE Wireless
  Communications}, vol.~25, no.~2, pp. 54--60, 2018.

\bibitem{ren2016advanced}
B.~Ren \emph{et~al.}, ``Advanced {IDD} receiver for {PDMA} uplink system,'' in
  \emph{2016 IEEE/CIC International Conference on Communications in China
  (ICCC)}.\hskip 1em plus 0.5em minus 0.4em\relax IEEE, 2016, pp. 1--6.

\bibitem{Fujitsu2016ldssve}
{3GPP}, ``Initial {LLS} results for {UL} non-orthogonal multiple access,''
  \emph{document R1-164329, 3GPP TSG-RAN WG1 {85}}, May 2016.

\bibitem{yuan2016multi}
Z.~Yuan \emph{et~al.}, ``Multi-user shared access for internet of things,'' in
  \emph{2016 IEEE 83rd Vehicular Technology Conference (VTC Spring)}.\hskip 1em
  plus 0.5em minus 0.4em\relax IEEE, 2016, pp. 1--5.

\bibitem{yuan2015multi}
Z.~Yuan, G.~Yu, and W.~Li, ``Multi-user shared access for {5G},''
  \emph{Telecommun. Network Technology}, vol.~5, no.~5, pp. 28--30, 2015.

\bibitem{lg2014ncma}
{LG Electronics}, ``Considerations on {DL/UL} multiple access for {NR},''
  \emph{document R1-162517, 3GPP TSG-RAN WG1 {84b}}, April 2014.

\bibitem{hu2013new}
H.~Hu and J.~Wu, ``New constructions of codebooks nearly meeting the {Welch}
  bound with equality,'' \emph{IEEE Transactions on Information Theory},
  vol.~60, no.~2, pp. 1348--1355, 2013.

\bibitem{nokia2016noca}
{Nokia, Alcatel-Lucent Shanghai Bell}, ``Non-orthogonal multiple access for new
  radio,'' \emph{document R1-165019, 3GPP TSG-RAN WG1 {85}}, May 2016.

\bibitem{intel2016fds}
{Intel Corporation}, ``Multiple access schemes for new radio interface,''
  \emph{document R1-162385, 3GPP TSG-RAN WG1 {84b}}, April 2016.

\bibitem{MediaTek2016goca}
{MediaTek Inc}, ``New uplink non-orthogonal multiple access schemes for {NR},''
  \emph{document R1-167535, 3GPP TSG-RAN WG1 {86}}, Aug 2016.

\bibitem{benjebbour2013concept}
A.~Benjebbour \emph{et~al.}, ``Concept and practical considerations of
  non-orthogonal multiple access ({NOMA}) for future radio access,'' in
  \emph{2013 International Symposium on Intelligent Signal Processing and
  Communication Systems}.\hskip 1em plus 0.5em minus 0.4em\relax IEEE, 2013,
  pp. 770--774.

\bibitem{higuchi2015non}
K.~Higuchi and A.~Benjebbour, ``Non-orthogonal multiple access ({NOMA}) with
  successive interference cancellation for future radio access,'' \emph{IEICE
  Transactions on Communications}, vol.~98, no.~3, pp. 403--414, 2015.

\bibitem{benjebbour2015non1}
A.~Benjebbour, K.~Saito, A.~Li, Y.~Kishiyama, and T.~Nakamura, ``Non-orthogonal
  multiple access ({NOMA}): Concept, performance evaluation and experimental
  trials,'' in \emph{2015 International Conference on Wireless Networks and
  Mobile Communications (WINCOM)}.\hskip 1em plus 0.5em minus 0.4em\relax IEEE,
  2015, pp. 1--6.

\bibitem{benjebbour2015noma2}
A.~Benjebbour \emph{et~al.}, ``{NOMA}: From concept to standardization,'' in
  \emph{2015 IEEE Conference on Standards for Communications and Networking
  (CSCN)}.\hskip 1em plus 0.5em minus 0.4em\relax IEEE, 2015, pp. 18--23.

\bibitem{yan2015receiver}
C.~Yan \emph{et~al.}, ``Receiver design for downlink non-orthogonal multiple
  access ({NOMA}),'' in \emph{2015 IEEE 81st vehicular technology conference
  (VTC Spring)}.\hskip 1em plus 0.5em minus 0.4em\relax IEEE, 2015, pp. 1--6.

\bibitem{ding2014performance}
Z.~Ding, Z.~Yang, P.~Fan, and H.~V. Poor, ``On the performance of
  non-orthogonal multiple access in {5G} systems with randomly deployed
  users,'' \emph{IEEE signal processing letters}, vol.~21, no.~12, pp.
  1501--1505, 2014.

\bibitem{shahab2016user}
M.~B. Shahab, M.~Irfan, M.~F. Kader, and S.~Young~Shin, ``User pairing schemes
  for capacity maximization in non-orthogonal multiple access systems,''
  \emph{Wireless Communications and Mobile Computing}, vol.~16, no.~17, pp.
  2884--2894, 2016.

\bibitem{shahab2016power}
M.~B. Shahab, M.~F. Kader, and S.~Y. Shin, ``On the power allocation of
  non-orthogonal multiple access for {5G} wireless networks,'' in \emph{2016
  International Conference on Open Source Systems \& Technologies
  (ICOSST)}.\hskip 1em plus 0.5em minus 0.4em\relax IEEE, 2016, pp. 89--94.

\bibitem{shahab2018user}
M.~B. Shahab and S.~Y. Shin, ``User pairing and power allocation for
  non-orthogonal multiple access: Capacity maximization under data reliability
  constraints,'' \emph{Physical Communication}, vol.~30, pp. 132--144, 2018.

\bibitem{zhang2016uplink}
N.~Zhang, J.~Wang, G.~Kang, and Y.~Liu, ``Uplink nonorthogonal multiple access
  in {5G} systems,'' \emph{IEEE Communications Letters}, vol.~20, no.~3, pp.
  458--461, 2016.

\bibitem{al2014uplink}
M.~Al-Imari, P.~Xiao, M.~A. Imran, and R.~Tafazolli, ``Uplink non-orthogonal
  multiple access for {5G} wireless networks,'' in \emph{2014 11th
  international symposium on wireless communications systems (ISWCS)}.\hskip
  1em plus 0.5em minus 0.4em\relax IEEE, 2014, pp. 781--785.

\bibitem{sheng2017novel}
Z.~Sheng, X.~Su, and X.~Zhang, ``A novel power allocation method for
  non-orthogonal multiple access in cellular uplink network,'' in \emph{2017
  International Conference on Intelligent Environments (IE)}.\hskip 1em plus
  0.5em minus 0.4em\relax IEEE, 2017, pp. 157--159.

\bibitem{ali2016dynamic}
M.~S. Ali, H.~Tabassum, and E.~Hossain, ``Dynamic user clustering and power
  allocation for uplink and downlink non-orthogonal multiple access ({NOMA})
  systems,'' \emph{IEEE access}, vol.~4, pp. 6325--6343, 2016.

\bibitem{yang2016general}
Z.~Yang, Z.~Ding, P.~Fan, and N.~Al-Dhahir, ``A general power allocation scheme
  to guarantee quality of service in downlink and uplink {NOMA} systems,''
  \emph{IEEE transactions on wireless communications}, vol.~15, no.~11, pp.
  7244--7257, 2016.

\bibitem{ding2015cooperative}
Z.~Ding, M.~Peng, and H.~V. Poor, ``Cooperative non-orthogonal multiple access
  in {5G} systems,'' \emph{IEEE Communications Letters}, vol.~19, no.~8, pp.
  1462--1465, 2015.

\bibitem{zhang2016full}
Z.~Zhang, Z.~Ma, M.~Xiao, Z.~Ding, and P.~Fan, ``Full-duplex
  device-to-device-aided cooperative non-orthogonal multiple access,''
  \emph{IEEE Transactions on Vehicular Technology}, vol.~66, no.~5, pp.
  4467--4471, 2016.

\bibitem{zhong2016non}
C.~Zhong and Z.~Zhang, ``Non-orthogonal multiple access with cooperative
  full-duplex relaying,'' \emph{IEEE Communications Letters}, vol.~20, no.~12,
  pp. 2478--2481, 2016.

\bibitem{shahab2018time}
M.~B. Shahab and S.~Y. Shin, ``Time shared half/full-duplex cooperative {NOMA}
  with clustered cell edge users,'' \emph{IEEE Communications Letters},
  vol.~22, no.~9, pp. 1794--1797, 2018.

\bibitem{ding2015application}
Z.~Ding, F.~Adachi, and H.~V. Poor, ``The application of {MIMO} to
  non-orthogonal multiple access,'' \emph{IEEE Transactions on Wireless
  Communications}, vol.~15, no.~1, pp. 537--552, 2015.

\bibitem{qualcomm2016RSMA}
{Qualcomm}, ``{RSMA},'' \emph{document R1-164688, 3GPP TSG-RAN WG1 {85}}, May
  2016.

\bibitem{qualcom2016rsma1}
------, ``{RSMA},'' \emph{document R1-164688, TSG-RAN WG1 {85}}, May 2016.

\bibitem{qualcom2016rsma2}
------, ``Candidate {NR} multiple access schemes,'' \emph{document R1-163510,
  TSG-RAN WG1 {84b}}, April 2016.

\bibitem{etri2016lssa}
{ETRI}, ``Low code rate and signature based multiple access scheme for new
  radio,'' \emph{document R1-164869, TSG-RAN WG1 {85}}, May 2016.

\bibitem{ping2006interleave}
L.~Ping, L.~Liu, K.~Wu, and W.~K. Leung, ``Interleave division
  multiple-access,'' \emph{IEEE transactions on wireless communications},
  vol.~5, no.~4, pp. 938--947, 2006.

\bibitem{nokia2016idma}
{Nokia, Alcatel-Lucent Shanghai Bell}, ``Performance of interleave division
  mul- tiple access ({IDMA}) in combination with {OFDM} family waveforms,''
  \emph{document R1-165021, 3GPP TSG-RAN WG1 {85}}, May 2016.

\bibitem{samsung2016igma}
{Samsung}, ``Non-orthogonal multiple access candidate for {NR},''
  \emph{document R1-163992, 3GPP TSG-RAN WG1 {85}}, May 2016.

\bibitem{bana2001space}
S.~V. Bana and P.~Varaiya, ``Space division multiple access ({SDMA}) for robust
  ad hoc vehicle communication networks,'' in \emph{ITSC 2001. 2001 IEEE
  Intelligent Transportation Systems. Proceedings (Cat. No. 01TH8585)}.\hskip
  1em plus 0.5em minus 0.4em\relax IEEE, 2001, pp. 962--967.

\bibitem{fang2016lattice}
D.~Fang \emph{et~al.}, ``Lattice partition multiple access: A new method of
  downlink non-orthogonal multiuser transmissions,'' in \emph{2016 IEEE Global
  Communications Conference (GLOBECOM)}.\hskip 1em plus 0.5em minus 0.4em\relax
  IEEE, 2016, pp. 1--6.

\bibitem{naim2014building}
M.~A. Naim, J.~P. Fonseka, and E.~M. Dowling, ``A building block approach for
  designing multilevel coding schemes,'' \emph{IEEE Communications Letters},
  vol.~19, no.~1, pp. 2--5, 2014.

\bibitem{docomo2016uplink}
{NTT DOCOMO}, ``Uplink multiple access schemes for {NR},'' \emph{R1-165174,
  3GPP TSG-RAN WG1 Meeting {85}}, May 2016.

\bibitem{shirvanimoghaddam2017massive2}
M.~Shirvanimoghaddam, M.~Dohler, and S.~J. Johnson, ``Massive multiple access
  based on superposition raptor codes for cellular {M2M} communications,''
  \emph{IEEE Transactions on Wireless Communications}, vol.~16, no.~1, pp.
  307--319, 2017.

\bibitem{Intel2016Grant}
{Intel Corporation}, ``Grant-less and non-orthogonal {UL} transmissions in
  {NR},'' \emph{document R1-167698, 3GPP TSG-RAN WG1 {86}}, August 2016.

\bibitem{Huawei2016Discussion}
{Huawei, HiSilicon}, ``Discussion on grant-free transmission,''
  \emph{R1-166095, 3GPP TSG-RAN WG1 Meeting {86}}, August 2016.

\bibitem{zte2016grant}
{ZTE}, ``Grant-based and grant-free multiple access for {mMTC},''
  \emph{R1-164268, 3GPP TSG-RAN WG1 Meeting {85}}, May 2016.

\bibitem{lenovo2016uplink}
{Lenovo}, ``Uplink grant-free access for {5G mMTC},'' \emph{R1-1609398, 3GPP
  TSG-RAN WG1 Meeting {86b}}, October 2016.

\bibitem{catt2016consideration}
{CATT}, ``Consideration on grant-free transmission,'' \emph{R1-1608757, 3GPP
  TSG-RAN WG1 Meeting {86b}}, October 2016.

\bibitem{docomo2016discussion}
{DOCOMO}, ``Discussion on multiple access for {UL mMTC},'' \emph{document
  R1-167392, 3GPP TSG-RAN WG1 {86}}, August 2016.

\bibitem{zte2016discussion}
{ZTE}, ``Discussion on grant-free concept for {UL mMTC},'' \emph{document
  R1-166405, 3GPP TSG-RAN WG1 {86}}, August 2016.

\bibitem{au2014uplink}
K.~Au \emph{et~al.}, ``Uplink contention based {SCMA} for {5G} radio access,''
  in \emph{2014 IEEE Globecom workshops (GC wkshps)}.\hskip 1em plus 0.5em
  minus 0.4em\relax IEEE, 2014, pp. 900--905.

\bibitem{nokia2016basic}
{Nokia, Alcatel Lucent Shanghai Bell}, ``Basic principles of contention-based
  access,'' \emph{document R1-167252, 3GPP TSG-RAN WG1 {86}}, August 2016.

\bibitem{qualcom2016RSMA}
{Qualcomm}, ``{RSMA} and {SCMA} comparison,'' \emph{document R1-166358, 3GPP
  TSG-RAN WG1 {86}}, August 2016.

\bibitem{bayesteh2014blind}
A.~Bayesteh, E.~Yi, H.~Nikopour, and H.~Baligh, ``Blind detection of {SCMA} for
  uplink grant-free multiple-access,'' in \emph{2014 11th international
  symposium on wireless communications systems (ISWCS)}.\hskip 1em plus 0.5em
  minus 0.4em\relax IEEE, 2014, pp. 853--857.

\bibitem{Huawei2016LLS}
{Huawei, HiSilicon}, ``{LLS} results for uplink multiple access,''
  \emph{document R1-164037, 3GPP TSG-RAN WG1 {85}}, May 2016.

\bibitem{zte2016discussion1}
{ZTE}, ``Discussion on multiple access for new radio interface,''
  \emph{document R1-162226, 3GPP TSG-RAN WG1 {84b}}, April 2016.

\bibitem{zte2016contention}
------, ``Contention-based non-orthogonal multiple access for {UL mMTC},''
  \emph{document R1-164269, 3GPP TSG-RAN WG1 {85}}, May 2016.

\bibitem{zte2016System2}
------, ``Link-level performance evaluation for {MUSA},'' \emph{document
  R1-1608953, 3GPP TSG-RAN WG1 {86b}}, October 2016.

\bibitem{zte2016System}
------, ``System level performance evaluation for {MUSA},'' \emph{document
  R1-1608952, 3GPP TSG-RAN WG1 {86b}}, October 2016.

\bibitem{zte2016receiver}
------, ``Receiver implementation for {MUSA},'' \emph{document R1-164270, 3GPP
  TSG-RAN WG1 {85}}, August 2016.

\bibitem{zte2016receiver2}
------, ``Receiver details and link performance for {MUSA},'' \emph{document
  R1-166404, 3GPP TSG-RAN WG1 {85}}, August 2016.

\bibitem{balevi2018aloha}
E.~Balevi, F.~T. Al~Rabee, and R.~D. Gitlin, ``{ALOHA-NOMA} for massive
  machine-to-machine {IoT} communication,'' in \emph{2018 IEEE International
  Conference on Communications (ICC)}.\hskip 1em plus 0.5em minus 0.4em\relax
  IEEE, 2018, pp. 1--5.

\bibitem{choi2017nomaal}
J.~Choi, ``{NOMA-based} random access with multichannel {ALOHA},'' \emph{IEEE
  Journal on Selected Areas in Communications}, vol.~35, no.~12, pp.
  2736--2743, 2017.

\bibitem{elkourdi2018enabling}
M.~Elkourdi, A.~Mazin, E.~Balevi, and R.~D. Gitlin, ``Enabling slotted
  {Aloha-NOMA} for massive {M2M} communication in {IoT} networks,'' in
  \emph{2018 IEEE 19th Wireless and Microwave Technology Conference
  (WAMICON)}.\hskip 1em plus 0.5em minus 0.4em\relax IEEE, 2018, pp. 1--4.

\bibitem{qualcomm2016candidate}
{Qualcomm}, ``Candidate {NR} multiple access schemes,'' \emph{document
  R1-163510, 3GPP TSG-RAN WG1 {84b}}, April 2016.

\bibitem{shirvanimoghaddam2017massive}
M.~Shirvanimoghaddam, M.~Dohler, and S.~J. Johnson, ``Massive non-orthogonal
  multiple access for cellular {IoT}: Potentials and limitations,'' \emph{IEEE
  Communications Magazine}, vol.~55, no.~9, pp. 55--61, 2017.

\bibitem{shirvanimoghaddam2017fundamental}
M.~Shirvanimoghaddam, M.~Condoluci, M.~Dohler, and S.~J. Johnson, ``On the
  fundamental limits of random non-orthogonal multiple access in cellular
  massive {IoT},'' \emph{IEEE Journal on Selected Areas in Communications},
  vol.~35, no.~10, pp. 2238--2252, 2017.

\bibitem{abbas2018novel}
R.~Abbas, M.~Shirvanimoghaddam, Y.~Li, and B.~Vucetic, ``A novel analytical
  framework for massive grant-free {NOMA},'' \emph{IEEE Transactions on
  Communications}, vol.~67, no.~3, pp. 2436--2449, 2018.

\bibitem{abbas2018multi}
------, ``A multi-layer grant-free {NOMA} scheme for short packet
  transmissions,'' in \emph{2018 IEEE Global Communications Conference
  (GLOBECOM)}.\hskip 1em plus 0.5em minus 0.4em\relax IEEE, 2018, pp. 1--6.

\bibitem{hong2014sparsity}
J.-P. Hong, W.~Choi, and B.~D. Rao, ``Sparsity controlled random multiple
  access with compressed sensing,'' \emph{IEEE Transactions on Wireless
  Communications}, vol.~14, no.~2, pp. 998--1010, 2014.

\bibitem{fazel2013random}
F.~Fazel, M.~Fazel, and M.~Stojanovic, ``Random access compressed sensing over
  fading and noisy communication channels,'' \emph{IEEE Transactions on
  Wireless Communications}, vol.~12, no.~5, pp. 2114--2125, 2013.

\bibitem{alam2018survey}
M.~Alam and Q.~Zhang, ``A survey: Non-orthogonal multiple access with
  compressed sensing multiuser detection for {mMTC},'' \emph{arXiv preprint
  arXiv:1810.05422}, 2018.

\bibitem{monsees2014reliable}
F.~Monsees, C.~Bockelmann, and A.~Dekorsy, ``Reliable activity detection for
  massive machine to machine communication via multiple measurement vector
  compressed sensing,'' in \emph{2014 IEEE Globecom Workshops (GC
  Wkshps)}.\hskip 1em plus 0.5em minus 0.4em\relax IEEE, 2014, pp. 1057--1062.

\bibitem{monsees2015compressive}
F.~Monsees, M.~Woltering, C.~Bockelmann, and A.~Dekorsy, ``Compressive sensing
  multi-user detection for multicarrier systems in sporadic machine type
  communication,'' in \emph{2015 IEEE 81st Vehicular Technology Conference (VTC
  Spring)}.\hskip 1em plus 0.5em minus 0.4em\relax IEEE, 2015, pp. 1--5.

\bibitem{abebe2015compressive}
A.~T. Abebe and C.~G. Kang, ``Compressive sensing-based random access with
  multiple-sequence spreading for {MTC},'' in \emph{2015 IEEE Globecom
  Workshops (GC Wkshps)}.\hskip 1em plus 0.5em minus 0.4em\relax IEEE, 2015,
  pp. 1--6.

\bibitem{wang2015compressive}
B.~Wang, L.~Dai, Y.~Yuan, and Z.~Wang, ``Compressive sensing based multi-user
  detection for uplink grant-free non-orthogonal multiple access,'' in
  \emph{2015 IEEE 82nd Vehicular Technology Conference (VTC2015-Fall)}.\hskip
  1em plus 0.5em minus 0.4em\relax IEEE, 2015, pp. 1--5.

\bibitem{needell2009cosamp}
D.~Needell and J.~A. Tropp, ``{CoSaMP}: Iterative signal recovery from
  incomplete and inaccurate samples,'' \emph{Applied and computational harmonic
  analysis}, vol.~26, no.~3, pp. 301--321, 2009.

\bibitem{wang2016joint}
B.~Wang, L.~Dai, T.~Mir, and Z.~Wang, ``Joint user activity and data detection
  based on structured compressive sensing for {NOMA},'' \emph{IEEE
  Communications Letters}, vol.~20, no.~7, pp. 1473--1476, 2016.

\bibitem{tan2016compressive}
J.~Tan, W.~Ding, F.~Yang, C.~Pan, and J.~Song, ``Compressive sensing based
  time-frequency joint non-orthogonal multiple access,'' in \emph{2016 IEEE
  International Symposium on Broadband Multimedia Systems and Broadcasting
  (BMSB)}.\hskip 1em plus 0.5em minus 0.4em\relax IEEE, 2016, pp. 1--4.

\bibitem{shim2012multiuser}
B.~Shim and B.~Song, ``Multiuser detection via compressive sensing,''
  \emph{IEEE Communications Letters}, vol.~16, no.~7, pp. 972--974, 2012.

\bibitem{wang2016dynamic}
B.~Wang, L.~Dai, Y.~Zhang, T.~Mir, and J.~Li, ``Dynamic compressive
  sensing-based multi-user detection for uplink grant-free {NOMA},'' \emph{IEEE
  Communications Letters}, vol.~20, no.~11, pp. 2320--2323, 2016.

\bibitem{vaswani2016recursive}
N.~Vaswani and J.~Zhan, ``Recursive recovery of sparse signal sequences from
  compressive measurements: A review,'' \emph{IEEE Transactions on Signal
  Processing}, vol.~64, no.~13, pp. 3523--3549, 2016.

\bibitem{chen2017sparsity}
G.~Chen, J.~Dai, K.~Niu, and C.~Dong, ``Sparsity-inspired sphere decoding
  ({SI-SD}): A novel blind detection algorithm for uplink grant-free sparse
  code multiple access,'' \emph{IEEE Access}, vol.~5, pp. 19\,983--19\,993,
  2017.

\bibitem{liu2017blind}
J.~Liu, G.~Wu, S.~Li, and O.~Tirkkonen, ``Blind detection of uplink grant-free
  {SCMA} with unknown user sparsity,'' in \emph{2017 IEEE International
  Conference on Communications (ICC)}.\hskip 1em plus 0.5em minus 0.4em\relax
  IEEE, 2017, pp. 1--6.

\bibitem{abebe2017comprehensive}
A.~T. Abebe and C.~G. Kang, ``Comprehensive grant-free random access for
  massive and low latency communication,'' in \emph{2017 IEEE International
  Conference on Communications (ICC)}.\hskip 1em plus 0.5em minus 0.4em\relax
  IEEE, 2017, pp. 1--6.

\bibitem{polyanskiy2017perspective}
Y.~Polyanskiy, ``A perspective on massive random-access,'' in \emph{2017 IEEE
  International Symposium on Information Theory (ISIT)}.\hskip 1em plus 0.5em
  minus 0.4em\relax IEEE, 2017, pp. 2523--2527.

\bibitem{ordentlich2017low}
O.~Ordentlich and Y.~Polyanskiy, ``Low complexity schemes for the random access
  gaussian channel,'' in \emph{2017 IEEE International Symposium on Information
  Theory (ISIT)}.\hskip 1em plus 0.5em minus 0.4em\relax IEEE, 2017, pp.
  2528--2532.

\bibitem{nazer2011compute}
B.~Nazer and M.~Gastpar, ``Compute-and-forward: Harnessing interference through
  structured codes,'' \emph{IEEE Transactions on Information Theory}, vol.~57,
  no.~10, pp. 6463--6486, 2011.

\bibitem{bar1993forward}
I.~Bar-David, E.~Plotnik, and R.~Rom, ``Forward collision resolution-a
  technique for random multiple-access to the adder channel,'' \emph{IEEE
  transactions on information theory}, vol.~39, no.~5, pp. 1671--1675, 1993.

\bibitem{yang2017non}
T.~Yang, L.~Yang, Y.~J. Guo, and J.~Yuan, ``A non-orthogonal multiple-access
  scheme using reliable physical-layer network coding and cascade-computation
  decoding,'' \emph{IEEE Transactions on Wireless Communications}, vol.~16,
  no.~3, pp. 1633--1645, 2017.

\bibitem{yang2016multiuser}
L.~Yang, T.~Yang, Y.~Xie, J.~Yuan, and J.~An, ``Multiuser decoding scheme for
  {K}-user fading multiple-access channel based on physical-layer network
  coding,'' \emph{IEEE Communications Letters}, vol.~20, no.~5, pp. 1046--1049,
  2016.

\bibitem{goseling2015random}
J.~Goseling, M.~Gastpar, and J.~H. Weber, ``Random access with physical-layer
  network coding,'' \emph{IEEE Transactions on Information Theory}, vol.~61,
  no.~7, pp. 3670--3681, 2015.

\bibitem{du2018block}
Y.~Du \emph{et~al.}, ``Block-sparsity-based multiuser detection for uplink
  grant-free {NOMA},'' \emph{IEEE Transactions on Wireless Communications},
  vol.~17, no.~12, pp. 7894--7909, 2018.

\bibitem{ye2019deep}
N.~Ye \emph{et~al.}, ``Deep learning aided grant-free {NOMA} toward reliable
  low-latency access in tactile internet of things,'' \emph{IEEE Transactions
  on Industrial Informatics}, vol.~15, no.~5, pp. 2995--3005, 2019.

\bibitem{ding2019sparsity}
T.~Ding, X.~Yuan, and S.~C. Liew, ``Sparsity learning based multiuser detection
  in grant-free massive-device multiple access,'' \emph{IEEE Transactions on
  Wireless Communications}, 2019.

\bibitem{bacstanlar2014introduction}
Y.~Ba{\c{s}}tanlar and M.~{\"O}zuysal, ``Introduction to machine learning,'' in
  \emph{miRNomics: MicroRNA Biology and Computational Analysis}.\hskip 1em plus
  0.5em minus 0.4em\relax Springer, 2014, pp. 105--128.

\bibitem{o2016convolutional}
T.~J. O’Shea, J.~Corgan, and T.~C. Clancy, ``Convolutional radio modulation
  recognition networks,'' in \emph{International conference on engineering
  applications of neural networks}.\hskip 1em plus 0.5em minus 0.4em\relax
  Springer, 2016, pp. 213--226.

\bibitem{o2016learning}
T.~J. O'Shea, K.~Karra, and T.~C. Clancy, ``Learning to communicate: Channel
  auto-encoders, domain specific regularizers, and attention,'' in \emph{2016
  IEEE International Symposium on Signal Processing and Information Technology
  (ISSPIT)}.\hskip 1em plus 0.5em minus 0.4em\relax IEEE, 2016, pp. 223--228.

\bibitem{nachmani2016learning}
E.~Nachmani, Y.~Be'ery, and D.~Burshtein, ``Learning to decode linear codes
  using deep learning,'' in \emph{2016 54th Annual Allerton Conference on
  Communication, Control, and Computing (Allerton)}.\hskip 1em plus 0.5em minus
  0.4em\relax IEEE, 2016, pp. 341--346.

\bibitem{gruber2017deep}
T.~Gruber, S.~Cammerer, J.~Hoydis, and S.~ten Brink, ``On deep learning-based
  channel decoding,'' in \emph{2017 51st Annual Conference on Information
  Sciences and Systems (CISS)}.\hskip 1em plus 0.5em minus 0.4em\relax IEEE,
  2017, pp. 1--6.

\bibitem{borgerding2016onsager}
M.~Borgerding and P.~Schniter, ``Onsager-corrected deep learning for sparse
  linear inverse problems,'' in \emph{2016 IEEE Global Conference on Signal and
  Information Processing (GlobalSIP)}.\hskip 1em plus 0.5em minus 0.4em\relax
  IEEE, 2016, pp. 227--231.

\bibitem{samuel2017deep}
N.~Samuel, T.~Diskin, and A.~Wiesel, ``Deep {MIMO} detection,'' in \emph{2017
  IEEE 18th International Workshop on Signal Processing Advances in Wireless
  Communications (SPAWC)}.\hskip 1em plus 0.5em minus 0.4em\relax IEEE, 2017,
  pp. 1--5.

\bibitem{jeon2017blind}
Y.-S. Jeon, S.-N. Hong, and N.~Lee, ``Blind detection for {MIMO} systems with
  low-resolution {ADCs} using supervised learning,'' in \emph{2017 IEEE
  International Conference on Communications (ICC)}.\hskip 1em plus 0.5em minus
  0.4em\relax IEEE, 2017, pp. 1--6.

\bibitem{farsad2017detection}
N.~Farsad and A.~Goldsmith, ``Detection algorithms for communication systems
  using deep learning,'' \emph{arXiv preprint arXiv:1705.08044}, 2017.

\bibitem{o2016unsupervised}
T.~J. O'Shea, J.~Corgan, and T.~C. Clancy, ``Unsupervised representation
  learning of structured radio communication signals,'' in \emph{2016 First
  International Workshop on Sensing, Processing and Learning for Intelligent
  Machines (SPLINE)}.\hskip 1em plus 0.5em minus 0.4em\relax IEEE, 2016, pp.
  1--5.

\bibitem{goodfellow2016deep}
I.~Goodfellow, Y.~Bengio, and A.~Courville, \emph{Deep learning}.\hskip 1em
  plus 0.5em minus 0.4em\relax MIT press, 2016.

\bibitem{pan2009survey}
S.~J. Pan and Q.~Yang, ``A survey on transfer learning,'' \emph{IEEE
  Transactions on knowledge and data engineering}, vol.~22, no.~10, pp.
  1345--1359, 2009.

\bibitem{mitzenmacher2009survey}
M.~Mitzenmacher \emph{et~al.}, ``A survey of results for deletion channels and
  related synchronization channels,'' \emph{Probability Surveys}, vol.~6, pp.
  1--33, 2009.

\bibitem{dorner2017deep}
S.~D{\"o}rner, S.~Cammerer, J.~Hoydis, and S.~ten Brink, ``Deep learning based
  communication over the air,'' \emph{IEEE Journal of Selected Topics in Signal
  Processing}, vol.~12, no.~1, pp. 132--143, 2017.

\bibitem{liu2019uav}
Y.~Liu \emph{et~al.}, ``{UAV} communications based on non-orthogonal multiple
  access,'' \emph{IEEE Wireless Communications}, vol.~26, no.~1, pp. 52--57,
  2019.

\bibitem{gui2018deep}
G.~Gui, H.~Huang, Y.~Song, and H.~Sari, ``Deep learning for an effective
  nonorthogonal multiple access scheme,'' \emph{IEEE Transactions on Vehicular
  Technology}, vol.~67, no.~9, pp. 8440--8450, 2018.

\bibitem{kim2018deep}
M.~Kim, N.-I. Kim, W.~Lee, and D.-H. Cho, ``Deep learning-aided {SCMA},''
  \emph{IEEE Communications Letters}, vol.~22, no.~4, pp. 720--723, 2018.

\bibitem{awan2018detection}
D.~A. Awan, R.~L. Cavalcante, M.~Yukawa, and S.~Stanczak, ``Detection for
  {5G-NOMA}: An online adaptive machine learning approach,'' in \emph{2018 IEEE
  International Conference on Communications (ICC)}.\hskip 1em plus 0.5em minus
  0.4em\relax IEEE, 2018, pp. 1--6.

\bibitem{liu2018deep}
M.~Liu, T.~Song, and G.~Gui, ``Deep cognitive perspective: Resource allocation
  for {NOMA-based} heterogeneous {IoT} with imperfect {SIC},'' \emph{IEEE
  Internet of Things Journal}, vol.~6, no.~2, pp. 2885--2894, 2018.

\bibitem{nguyen2018power}
K.~Nguyen \emph{et~al.}, ``Power allocations in cache-aided {NOMA} systems:
  Optimization and deep learning approaches,'' \emph{submitted to IEEE Journal
  of Selected Topics in Signal Processing (J-STSP)}, 2018.

\bibitem{luo2019deep}
J.~Luo \emph{et~al.}, ``A deep learning-based approach to power minimization in
  multi-carrier {NOMA} with {SWIPT},'' \emph{IEEE Access}, vol.~7, pp.
  17\,450--17\,460, 2019.

\bibitem{cui2018unsupervised}
J.~Cui, Z.~Ding, P.~Fan, and N.~Al-Dhahir, ``Unsupervised machine
  learning-based user clustering in millimeter-wave-noma systems,'' \emph{IEEE
  Transactions on Wireless Communications}, vol.~17, no.~11, pp. 7425--7440,
  2018.

\bibitem{gallager1985perspective}
R.~Gallager, ``A perspective on multiaccess channels,'' \emph{IEEE Transactions
  on information Theory}, vol.~31, no.~2, pp. 124--142, 1985.

\bibitem{cover2012elements}
T.~M. Cover and J.~A. Thomas, \emph{Elements of information theory}.\hskip 1em
  plus 0.5em minus 0.4em\relax John Wiley \& Sons, 2012.

\bibitem{molavianjazi2015second}
E.~MolavianJazi and J.~N. Laneman, ``A second-order achievable rate region for
  gaussian multi-access channels via a central limit theorem for functions,''
  \emph{IEEE Transactions on Information Theory}, vol.~61, no.~12, pp.
  6719--6733, 2015.

\bibitem{molavianjazi2012random}
------, ``A random coding approach to gaussian multiple access channels with
  finite blocklength,'' in \emph{2012 50th Annual Allerton Conference on
  Communication, Control, and Computing (Allerton)}.\hskip 1em plus 0.5em minus
  0.4em\relax IEEE, 2012, pp. 286--293.

\bibitem{jazi2012simpler}
E.~M. Jazi and J.~N. Laneman, ``Simpler achievable rate regions for multiaccess
  with finite blocklength,'' in \emph{2012 IEEE International Symposium on
  Information Theory Proceedings}.\hskip 1em plus 0.5em minus 0.4em\relax IEEE,
  2012, pp. 36--40.

\bibitem{truong2017gaussian}
L.~V. Truong and V.~Y. Tan, ``On the gaussian {MAC} with stop-feedback,'' in
  \emph{2017 IEEE International Symposium on Information Theory (ISIT)}.\hskip
  1em plus 0.5em minus 0.4em\relax IEEE, 2017, pp. 2303--2307.

\bibitem{huang2012finite}
Y.-W. Huang and P.~Moulin, ``Finite blocklength coding for multiple access
  channels,'' in \emph{2012 IEEE International Symposium on Information Theory
  Proceedings}.\hskip 1em plus 0.5em minus 0.4em\relax IEEE, 2012, pp.
  831--835.

\bibitem{scarlett2014second}
J.~Scarlett, A.~Martinez, and A.~G. i~F{\`a}bregas, ``Second-order rate region
  of constant-composition codes for the multiple-access channel,'' \emph{IEEE
  Transactions on Information Theory}, vol.~61, no.~1, pp. 157--172, 2014.

\bibitem{chen2017capacity}
X.~Chen, T.-Y. Chen, and D.~Guo, ``Capacity of gaussian many-access channels,''
  \emph{IEEE Transactions on Information Theory}, vol.~63, no.~6, pp.
  3516--3539, 2017.

\bibitem{effros2018random}
M.~Effros, V.~Kostina, and R.~C. Yavas, ``Random access channel coding in the
  finite blocklength regime,'' in \emph{2018 IEEE International Symposium on
  Information Theory (ISIT)}.\hskip 1em plus 0.5em minus 0.4em\relax IEEE,
  2018, pp. 1261--1265.

\bibitem{Nokia2016Preamble}
{Nokia, Alcatel-Lucent Shanghai Bell}, ``Preamble transmission procedures for
  the {mMTC} uplink,'' \emph{document R1-1609653, 3GPP TSG-RAN WG1 {86b}},
  October 2016.

\bibitem{Samsung2016Discussion}
{Samsung}, ``Discussion on grant-free/contention-based non-orthogonal multiple
  access,'' \emph{document R1-166752, 3GPP TSG-RAN WG1 {86}}, August 2016.

\bibitem{Samsung2016Non}
------, ``Non-orthogonal multiple access candidate for {NR},'' \emph{document
  R1-163992, 3GPP TSG-RAN WG1 {85}}, May 2016.

\bibitem{Huawei2016Advanced}
{Huawei, HiSilicon}, ``Advanced multi-user detectors for grant-free
  transmissions,'' \emph{document R1-1608855, 3GPP TSG-RAN WG1 {86b}}, October
  2016.

\bibitem{chen2019noncoherent}
H.~Chen, Z.~Dong, and B.~Vucetic, ``Noncoherent and non-orthogonal massive
  {SIMO} for critical industrial {IoT} communications,'' \emph{arXiv preprint
  arXiv:1903.01650}, 2019.

\bibitem{liu2018sparse}
L.~Liu \emph{et~al.}, ``Sparse signal processing for grant-free massive
  connectivity: A future paradigm for random access protocols in the internet
  of things,'' \emph{IEEE Signal Processing Magazine}, vol.~35, no.~5, pp.
  88--99, 2018.

\bibitem{senel2018joint}
K.~Senel and E.~G. Larsson, ``Joint user activity and non-coherent data
  detection in {mMTC}-enabled massive {MIMO} using machine learning
  algorithms,'' in \emph{WSA 2018; 22nd International ITG Workshop on Smart
  Antennas}.\hskip 1em plus 0.5em minus 0.4em\relax VDE, 2018, pp. 1--6.

\bibitem{Nokia2016Collision}
{Nokia, Alcatel-Lucent Shanghai Bell}, ``Collision handling for grant-free,''
  \emph{document R1-1609648, 3GPP TSG-RAN WG1 {86b}}, October 2016.

\bibitem{Huawei2016Solutions}
{Huawei, HiSilicon}, ``Solutions for collisions of {MA} signatures,''
  \emph{document R1-1608860, 3GPP TSG-RAN WG1 {86b}}, October 2016.

\bibitem{Sony2016Non}
{Sony}, ``Non-orthogonal multiple access for uplink,'' \emph{document
  R1-166651, 3GPP TSG-RAN WG1 {86}}, August 2016.

\bibitem{Huawei2016The}
{Huawei, HiSilicon}, ``The retransmission and {HARQ} schemes for grant-free,''
  \emph{document R1-1608859, 3GPP TSG-RAN WG1 {86b}}, October 2016.

\bibitem{ZTE2016Discussion3}
{ZTE}, ``Discussion on grant-free transmission based on sensing,''
  \emph{document R1-1609801, 3GPP TSG-RAN WG1 {86b}}, October 2016.

\bibitem{kazmi2019network}
S.~A. Kazmi, L.~U. Khan, N.~H. Tran, and C.~S. Hong, ``Network slicing: Radio
  resource allocation using non-orthogonal multiple access,'' in \emph{Network
  Slicing for 5G and Beyond Networks}.\hskip 1em plus 0.5em minus 0.4em\relax
  Springer, 2019, pp. 69--89.

\bibitem{tang2018adaptive}
L.~Tang, Q.~Tan, Y.~Shi, C.~Wang, and Q.~Chen, ``Adaptive virtual resource
  allocation in {5G} network slicing using constrained markov decision
  process,'' \emph{IEEE Access}, vol.~6, pp. 61\,184--61\,195, 2018.

\bibitem{Intel2016Grant2}
{Intel}, ``Grant-free {UL} transmissions in {NR},'' \emph{R1-1609499, 3GPP
  TSG-RAN WG1 Meeting {86b}}, October 2016.

\bibitem{zheng2015radio}
K.~{Zheng}, F.~{Hu}, W.~{Wang}, W.~{Xiang}, and M.~{Dohler}, ``Radio resource
  allocation in {LTE}-advanced cellular networks with {M2M} communications,''
  \emph{IEEE Communications Magazine}, vol.~50, no.~7, pp. 184--192, July 2012.

\bibitem{3gpp2016access}
{3GPP}, ``Feasibility study on new services and markets technology enablers for
  massive internet of things (release 14),'' \emph{3GPP TR 22.861 V14.1.0},
  September 2016.

\bibitem{han2018joint}
S.~Han, X.~Xu, L.~Zhao, and X.~Tao, ``Joint time and power allocation for
  uplink cooperative non-orthogonal multiple access based massive machine-type
  communication network,'' \emph{International Journal of Distributed Sensor
  Networks}, vol.~14, no.~5, p. 1550147718778215, 2018.

\bibitem{zhang2018performance}
Y.~Zhang, Z.~Yang, Y.~Feng, and S.~Yan, ``Performance analysis of a novel
  uplink cooperative {NOMA} system with full-duplex relaying,'' \emph{IET
  Communications}, vol.~12, no.~19, pp. 2408--2417, 2018.

\bibitem{zeng2019energy}
M.~Zeng, W.~Hao, O.~A. Dobre, and H.~V. Poor, ``Energy-efficient power
  allocation in uplink {mmWave} massive {MIMO} with {NOMA},'' \emph{IEEE
  Transactions on Vehicular Technology}, vol.~68, no.~3, pp. 3000--3004, 2019.

\bibitem{ding2016general}
Z.~Ding, R.~Schober, and H.~V. Poor, ``A general {MIMO} framework for {NOMA}
  downlink and uplink transmission based on signal alignment,'' \emph{IEEE
  Transactions on Wireless Communications}, vol.~15, no.~6, pp. 4438--4454,
  2016.

\bibitem{ni2019analysis}
Z.~Ni, Z.~Chen, Q.~Zhang, and C.~Zhou, ``Analysis of {RF} energy harvesting in
  uplink-{NOMA IoT}-based network,'' \emph{arXiv preprint arXiv:1907.11647},
  2019.

\bibitem{di2017noma}
B.~Di, L.~Song, Y.~Li, and G.~Y. Li, ``{NOMA}-based low-latency and
  high-reliable broadcast communications for {5G V2X} services,'' in
  \emph{GLOBECOM 2017-2017 IEEE Global Communications Conference}.\hskip 1em
  plus 0.5em minus 0.4em\relax IEEE, 2017, pp. 1--6.

\bibitem{di2017v2x}
B.~Di, L.~Song, Y.~Li, and Z.~Han, ``{V2X} meets {NOMA}: Non-orthogonal
  multiple access for {5G}-enabled vehicular networks,'' \emph{IEEE Wireless
  Communications}, vol.~24, no.~6, pp. 14--21, 2017.

\bibitem{mei2019uplink}
W.~Mei and R.~Zhang, ``Uplink cooperative {NOMA} for cellular-connected
  {UAV},'' \emph{IEEE Journal of Selected Topics in Signal Processing},
  vol.~13, no.~3, pp. 644--656, 2019.

\end{thebibliography}

	%\begin{IEEEbiography}[{\includegraphics[width=1in,height=1.25in,clip,keepaspectratio]{mshell}}]{Michael Shell}

\end{document}